\documentclass[proof]{pasj01}
\Received{}%{yyyy/mm/dd}
\Accepted{}%{yyyy/mm/dd}
\usepackage{graphicx}
\usepackage{multirow}
\usepackage{url}

\newcommand{\hi}{H{\sc i}}
\newcommand{\hii}{H$_{2}$}
\newcommand{\ICO}{$I_{\rm{CO}}$}
\newcommand{\IHI}{$I_{\rm{H{\textsc{i}}}}$}
\newcommand{\aco}{$\alpha_{\rm CO}$}
\newcommand{\xco}{$X_{\rm CO}$}
\newcommand{\acomw}{$\alpha_{\rm CO}$(MW)}

\newcommand{\sigatom}{$\Sigma_{\rm{atom}}$}

\newcommand{\sigmol}{$\Sigma_{\rm{mol}}$}
\newcommand{\sigdust}{$\Sigma_{\rm{dust}}$}
\newcommand{\sigSFR}{$\Sigma_{\rm{SFR}}$}
\newcommand{\massunit}{$M_\odot$ pc$^{-2}$}
\newcommand{\acounit}{$M_\odot$ pc$^{-2}$ (K km s$^{-1}$)$^{-1}$}
\newcommand{\xcounit}{cm$^{-2}$ (K km s$^{-1}$)$^{-1}$}

\begin{document}

\title{CO Multi-line Imaging of Nearby Galaxies (COMING). XII. CO-to-H$_{2}$ Conversion Factor and Dust-to-Gas Ratio}
\author{Atsushi YASUDA$^{1, \ast}$, Nario KUNO$^{1, 2, 3}$, Kazuo SORAI$^{1, 2, 4, 5}$, Kazuyuki MURAOKA$^{6}$, Yusuke MIYAMOTO$^{7}$, Hiroyuki KANEKO$^{7, 8, 9}$, Yoshiyuki YAJIMA$^{5,10}$, Takahiro TANAKA$^{1}$, Kana MOROKUMA-MATSUI$^{11}$, Tsutomu T. TAKEUCHI$^{12}$, and Masato I. N. KOBAYASHI$^{13, 14}$}
\altaffiltext{1}{Graduate School of Pure and Applied Sciences, University of Tsukuba, 1-1-1 Tennodai, Tsukuba, Ibaraki 305-8571, Japan}
\altaffiltext{2}{Tomonaga Center for the History of the Universe, University of Tsukuba, 1-1-1 Tennodai, Tsukuba, Ibaraki 305-8571, Japan}
\altaffiltext{3}{Department of Physics, School of Science and Technology, Kwansei Gakuin University, 2-1 Gakuen, Sanda, Hyogo 669-1337, Japan}
\altaffiltext{4}{Department of Physics, Faculty of Science, Hokkaido University, Kita 10, Nishi 8, Kita-ku, Sapporo, Hokkaido 060-0810, Japan}
\altaffiltext{5}{Department of Cosmosciences, Graduate School of Science, Hokkaido University, Kita 10, Nishi 8, Kita-ku, Sapporo, Hokkaido 060-0810, Japan}
\altaffiltext{6}{Department of Physics, Graduate School of Science, Osaka Metropolitan University, 1-1 Gakuen-cho, Naka-ku, Sakai, Osaka 599-8531, Japan}
\altaffiltext{7}{Department of Electrical, Electronic and Computer Engineering, Faculty of Engineering, Fukui University of Technology, 3-6-1 Gakuen, Fukui, Fukui 910-8505, Japan}
\altaffiltext{8}{Graduate School of Education, Joetsu University of Education, 1, Yamayashiki-machi, Joetsu, Niigata, 943-8512, Japan}
\altaffiltext{9}{Center for Astronomy, Ibaraki University, 2-1-1 Bunkyo, Mito, Ibaraki, 310-8512, Japan}
\altaffiltext{10}{Present address: NEC Corporation, 1753, Shimonumabe, Nakahara-ku, Kawasaki, Kanagawa, 211-8666 Japan}
\altaffiltext{11}{Institute of Astronomy, Graduate School of Science, The University of Tokyo, 2-21-1 Osawa, Mitaka, Tokyo 181-0015, Japan}
\altaffiltext{12}{Division of Particle and Astrophysical Science, Nagoya University, Furo-cho, Chikusa-ku, Nagoya, Aichi 464-8602, Japan}
\altaffiltext{13}{Division of Science, National Astronomical Observatory of Japan, Osawa 2-21-1, Mitaka, Tokyo 181-8588, Japan}
\altaffiltext{14}{I. Physikalisches Institut, Universit\"{a}t zu K\"{o}ln, Z\"{u}lpicher Str 77, D-50937 K\"{o}ln, Germany}

\altaffiltext{}{}

\email{yasuda.atsushi.mp@alumni.tsukuba.ac.jp}


\maketitle

\begin{abstract}
We simultaneously measured the spatially-resolved CO-to-\hii\ conversion factor (\aco) and dust-to-gas ratio (DGR) in nearby galaxies on a kiloparsec scale. In this study, we used $^{12}$CO($J=1-0$) data obtained by the Nobeyama 45-m radio telescope with \hi\ and dust mass surface densities. We obtained the values of global \aco\ and DGR in 22 nearby spiral galaxies, with averages of $2.66 \pm 1.36$ \acounit and $0.0052 \pm 0.0026$, respectively. Furthermore, the radial variations of \aco\ and DGR in four barred spiral galaxies (IC\,342, NGC\,3627, NGC\,5236, and NGC\,6946) were obtained by dividing them into the inner and outer regions with a boundary of $0.2R_{25}$, where $R_{25}$ is the isophotal radius at 25 mag arcsec$^{-2}$ in the $B$ band. The averages of \aco\ and DGR in the inner region ($\leq 0.2R_{25}$) are $0.36 \pm 0.08$ \acounit\ and $0.0199 \pm 0.0058$, while those in the outer region ($> 0.2R_{25}$) are $1.49 \pm 0.76$ \acounit\ and $0.0084 \pm 0.0037$, respectively. The value of \aco\ in the outer region is $2.3$ to $5.3$ times larger than that of the inner region. When separated into the inner and outer regions, we find that \aco\ and DGR correlate with the metallicity and the star formation rate surface density. The value of \aco\ derived in this study tends to be smaller than those obtained in previous studies for the Milky Way and nearby star-forming galaxies. This fact can be attributed to our measurements being biased toward the inner region; we measured \aco\ at 0.85 and 0.76 times smaller in radius than the previous works for nearby star-forming galaxies and the Milky Way, respectively.
\end{abstract}

%%%%%%%%%%%%%%%%%%%%
%% Introduction
%%%%%%%%%%%%%%%%%%%%
\section{Introduction} \label{introduction}
Stars are formed in dense and cold molecular clouds, which mainly consist of molecular hydrogen (\hii). It is important to measure the precise amount and distribution of \hii\ gas to understand star formation in galaxies. However, \hii\ molecules cannot emit line emission at the low temperature of molecular clouds, $\sim$10 K, because they do not have permanent dipole moments. This indicates that \hii\ molecules in molecular clouds cannot be directly observed. Carbon monoxide (CO) is the most common tracer of \hii\ gas mass in molecular clouds. $^{12}$CO($J=1-0$) is easily excited in molecular clouds by frequent collisions with \hii\ molecules because the energy gap of $^{12}$CO rotational transition between $J=0$ to $J=1$ is low ($\Delta E/k_{\rm{B}}\sim5.5$ K, where $\Delta E$ is the energy gap between $J=0$ to $J=1$ and $k_{\rm{B}}$ is the Boltzmann constant). Thus, $^{12}$CO($J=1-0$) emission is widely used as a cold \hii\ gas tracer. The \hii\ column density ($N_{\rm{H_{2}}}$) is often derived from the integrated intensity of $^{12}$CO($J=1-0$) (\ICO) by using the empirical CO-to-\hii\ conversion factor (\xco) with the following equation:

\begin{equation}
  \left(\frac{N_{\mathrm{H_{2}}}}{\mathrm{cm^{-2}}} \right) = \left[\frac{X_{\mathrm{CO}}}{\mathrm{cm^{-2}\ (K\ km\ s^{-1})^{-1}}} \right] \times \left[\frac{I_{\mathrm{CO}}}{\mathrm{K\ km\ s^{-1}}} \right].
\end{equation}

\noindent
The mass surface density of molecular gas (\sigmol) is also derived using the CO-to-\hii\ conversion factor of \aco\ including a factor of 1.36 for helium using the following equations:

\begin{equation}
  \left(\frac{\Sigma_{\mathrm{mol}}}{M_{\odot}\ \mathrm{pc^{-2}}} \right) = \left[\frac{\alpha_{\mathrm{CO}}}{M_\odot\ \mathrm{pc^{-2}\ (K\ km\ s^{-1})^{-1}}} \right] \times \left[\frac{I_{\mathrm{CO}}}{\mathrm{K\ km\ s^{-1}}} \right].
\end{equation}

\noindent
The measurements of \aco\ or \xco\ in galaxies are necessary to estimate accurate molecular gas mass and understand star formation in galaxies.

The Galactic conversion factor $\alpha_{\rm{CO}}(\rm{MW})=4.35$ \acounit\ or $X_{\rm{CO}}(\rm{MW})=2.0\times10^{20}$ \xcounit\ (e.g., \cite{Bolatto2013}) has been widely used to estimate the molecular gas mass of galaxies. On the other hand, studies have reported that \aco\ depends on various properties of the interstellar medium (ISM). For example, the CO-to-\hii\ conversion factor in galaxies with low metallicity tends to be larger than that with high metallicity (e.g., \cite{Arimoto1996}, \cite{Gratier2010}, \cite{Leroy2011}, \cite{Narayanan2012}). Ultra-luminous infrared galaxies and merger galaxies with active star formation tend to have a lower CO-to-\hii\ conversion factor than the galactic value (e.g., \cite{Downes1998}, \cite{Narayanan2011}, \cite{Papadopoulos2012}). These studies suggest that \aco\ would be different among galaxies and within a galaxy depending on the physical properties of the ISM or local environments within a galaxy. Therefore, spatially-resolved CO mapping data of various galaxies is essential to investigate the variety of \aco\ among different galaxies and within a particular galaxy.

\citet{Sandstrom2013} (hereafter S13) statistically investigated \aco\ in 26 nearby star-forming galaxies using the data of HERA CO-Line Extragalactic Survey (HERACLES, \cite{Leroy2009}), which was the $^{12}$CO($J=2-1$) mapping survey of nearby galaxies with the IRAM 30-m telescope. The mean $R_{25}$, where $R_{25}$ is the which is the isophotal radius at 25 mag arcsec$^{-2}$ in the $B$ band, of 26 galaxies is 4.4$^{\prime}$. S13 measured \aco\ by using CO integrated intensity \ICO\ (K km s$^{-1}$), atomic gas mass surface density \sigatom\ (\massunit), and dust mass surface density \sigdust\ (\massunit) by using the following equation:

\begin{eqnarray}
  \frac{\Sigma_{\mathrm{dust}}}{\mathrm{DGR}} &=& \Sigma_\mathrm{atom} + \Sigma_\mathrm{mol} \nonumber \\
  &=& \Sigma_\mathrm{atom} + \alpha_{\mathrm{CO}}I_{\mathrm{CO}},
  \label{eq:alpha_DGR}
\end{eqnarray}

\noindent
where DGR is the mass surface density ratio of interstellar dust to that of interstellar gas. The total gas mass surface density (= \sigatom\ + \sigmol) includes a factor of 1.36 for helium. S13 derived \aco\ considering the following assumptions: (1) dust and gas are mixed well, (2) DGR does not depend on the gas phase, namely, atomic or molecular gas, (3) the fraction of mass in the ionized gas is negligible, (4) the line ratio of $^{12}$CO($J=2-1$) to $^{12}$CO($J=1-0$), $R_{21}$, is constant for the HERACLES samples ($R_{21}=0.7$; e.g., \cite{Rosolowsky2015}), and (5) DGR is constant on $\sim$ kiloparsec scale. Furthermore, their \aco\ and DGR were measured in each hexagonal region called ``solution pixel'' on kiloparsec scales. They measured \aco\ at which the scatter of DGR was minimized in each solution pixel. Finally, they reported that the average \aco\ for their samples was 3.1 \acounit\ with a standard deviation of 0.3 dex.

Because the energy of $J=2$ from the ground level of $^{12}$CO is 16.6 K, which is higher than the temperature of typical molecular clouds ($\sim$10 K; \cite{Scoville_Sanders1987}), $R_{21}$ depends on the excitation condition of molecular gas. Studies have reported that $R_{21}$ is not constant within a galaxy and that $R_{21}=0.7$ is inappropriate in certain nearby galaxies. \citet{Koda2012} investigated the variation of $R_{21}$ within the nearby spiral galaxy M\,51 by using $^{12}$CO data observed by the Combined Array for Research in Millimeter Astronomy (CARMA), the Nobeyama 45-m radio telescope, and the IRAM 30-m telescope. They obtained that $R_{21}$ in the interarm regions is typically $< 0.7$ (often 0.4 -- 0.6) and the rise of $R_{21}$ in the spiral arms (often 0.8 -- 1.0) and central region ($\sim$0.8 -- 1.0). \citet{Yajima2021} presented the variations of $R_{21}$ among 24 nearby galaxies by using $^{12}$CO data observed by the Nobeyama 45-m radio telescope and the IRAM 30-m telescope. They reported that the median of $R_{21}$ for 24 nearby galaxies is 0.61, and the weighted mean of $R_{21}$ by $^{12}$CO($J=1-0$) integrated intensity is 0.66 with a standard deviation of 0.19. Considering the variation of $R_{21}$ from 0.7 within and among galaxies reported in these studies, the assumption of a constant $R_{21}$ could impose additional uncertainty on the \aco\ estimation. Therefore, statistical studies of \aco\ without the assumption of $R_{21}$ are required. Although it is necessary to measure \aco\ in nearby galaxies by using $^{12}$CO($J=1-0$) data without additional uncertainties, statistical studies regarding \aco\ using $^{12}$CO($J=1-0$) mapping data are limited. Therefore, we measured \aco\ in nearby galaxies using the data of the large $^{12}$CO($J=1-0$) mapping surveys of nearby galaxies and investigated the relation between \aco\ and various properties of the ISM. We used the spatially-resolved $^{12}$CO($J=1-0$) data of CO multi-line imaging of nearby galaxies (COMING, \cite{Sorai2019}) obtained with the multi-beam receiver FOur-beam REceiver System (FOREST, \cite{Minamidani2016}) on the Nobeyama 45-m radio telescope. For COMING, the mapping region covers 70\% of the optical disk diameter with a kiloparsec scale resolution. We also used the spatially-resolved $^{12}$CO($J=1-0$) data from the Nobeyama CO Atlas of nearby galaxies (CO Atlas, \cite{Kuno2007}). Fifteen out of 40 galaxies have larger optical diameter than the mean optical diameter of the samples of S13 ($=\timeform{4.4'}$). They are also sufficient to derive \aco\ in nearby galaxies. We measured \aco\ in nearby galaxies for the sample galaxies of the COMING and the CO Atlas by modifying the method of S13 using $^{12}$CO($J=1-0$) instead of $^{12}$CO($J=2-1$).

We explain the sample selection and the method for measuring \aco\ and DGR in section 2. In section 3, we present the derivation of resolved \ICO, \sigatom, and \sigdust. Details regarding the ancillary data for metallicity and SFR used in section 6 are presented in section 4 followed by the results of the \aco\ and DGR measurements in our sample galaxies in section 5. In subsection 6.1, we discuss the correlation of the \aco\ and DGR values with metallicity and SFR. In subsection 6.2, we compare the value of \aco\ obtained in this study with the standard value in the Milky Way \acomw\ and \aco\ derived by S13 and discuss the cause of the difference between them.

%%%%%%%%%%%%%%%%%%%%
%% Method
%%%%%%%%%%%%%%%%%%%%
\section{Sample selection and measurement method of \aco\ and DGR} \label{method}

%%%%%%%%%%%%%%%%%%%%
%% sample selection
%%%%%%%%%%%%%%%%%%%%
\subsection{Sample selection}
We set two criteria for sample selection. First, because our method to measure \aco\ requires infrared and \hi\ data, in addition to CO, we selected the galaxies of COMING and CO Atlas whose archival data of \hi\ and infrared wavelengths is available. For the \hi\ data, we used the archival data of the \hi\ Nearby Galaxies Survey (THINGS, \cite{Walter2008}) and the VLA Imaging of Virgo in Atomic gas (VIVA, \cite{Chung2009}). For the galaxies that were not included in the THINGS and the VIVA, we obtained the data from the NRAO Science Data Archive and performed data reduction. For IC\,342, we used \hi\ data of \citet{Crosthwaite2000} observed by the Very Large Array (VLA). For the infrared data, we used the archival data of the DustPedia project (\cite{Davies2017}), which was the legacy project of the $Herschel$ Space Observatory and releases photometric data across 42 UV-microwave bands for 875 galaxies. Second, we excluded galaxies that were categorized as edge-on in table 1 of \citet{Sorai2019} because we were investigating the radial variation of \aco\ and DGR within the galaxies. We also selected three interacting galaxy pairs from the COMING samples: NGC\,4298/NGC\,4302, VV\,219 (NGC\,4567/NGC\,4568), and Arp\,116 (NGC\,4647/NGC\,4649). We excluded NGC\,4302 because this galaxy was classified as edge-on. We divided NGC\,4567 and NGC\,4568 by using the filamentary structure described in \citet{Kaneko2018} as the boundary between NGC\,4567 and NGC\,4568 and measured \aco\ and DGR in each galaxy. We excluded NGC\,4649 because CO emission from this galaxy was very faint and we could not detect it significantly. Finally, we analyzed 40 nearby galaxies listed in table \ref{tab:galaxy_list}.

%%%%%%%%%%%%%%%%%%%%
%% alpha and DGR measurement method
%%%%%%%%%%%%%%%%%%%%
\subsection{Method}
As explained in section \ref{introduction}, we used the integrated intensity of $^{12}$CO($J=1-0$) (\ICO), the atomic gas mass surface density (\sigatom), and the dust mass surface density (\sigdust) to obtain \aco\ and DGR. The grid size and beam size of all data were matched to the data with the largest grid and beam sizes among $^{12}$CO($J=1-0$), \hi\, and the infrared data we used. The adopted grid size and beam size for the measurements of \aco\ and DGR are listed in table \ref{tab:HI_CO_list}.

We determined \aco\ and DGR simultaneously by using the $\chi^{2}$ fitting described in the following equations:

\begin{equation}
\Sigma_{\mathrm{dust}, \mathrm {model}} \equiv \mathrm{DGR} \times (\Sigma_{\mathrm{atom, obs}}+\alpha_{\mathrm{CO}}I_{\mathrm{CO, obs}}),
\label{eq:dust_model}
\end{equation}

\begin{equation}
\chi^{2} \equiv \sum_{j,\ k} \frac{\left(\Sigma_{\mathrm{dust}, \mathrm{obs} (j,\ k)}-\Sigma_{\mathrm{dust}, \mathrm {model} (j,\ k)}\right)^{2}}{\Sigma_{\mathrm{dust}, \mathrm {model} (j,\ k)}}.
\label{eq:chi}
\end{equation}

\begin{figure}[htbp]
    \begin{center}
    \includegraphics[width=\linewidth]{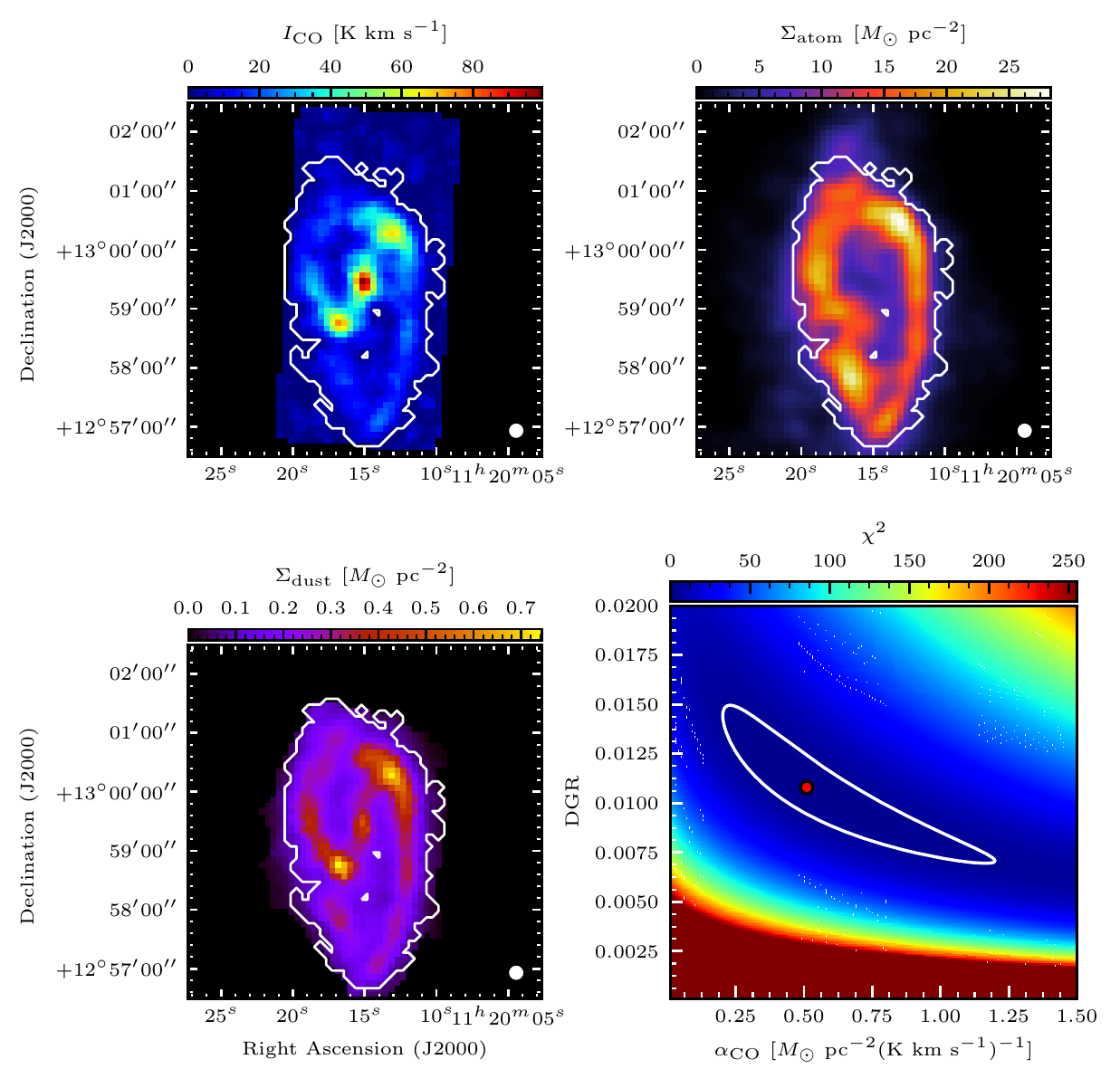}        
    \end{center}
  \caption{\aco\ and DGR measurement in NGC\,3627. The top left panel indicates the observed integrated intensity of $^{12}$CO($J=1-0$) ($I_{\rm{CO, obs}}$). The top right panel indicates the observed atomic gas mass surface density ($\Sigma_{\rm{atom, obs}}$). The bottom left panel indicates the observed dust mass surface density ($\Sigma_{\rm{dust, obs}}$). The white contour of these panels indicates the region used for the measurement of \aco\ and DGR. The bottom right panel indicates the result of $\chi^{2}$ fitting, and the red point indicates the optimal values of \aco\ and DGR, which minimizes the $\chi^{2}$ value ($\equiv \chi^{2}_{\rm{min}}$) described in equation (\ref{eq:chi}). The white contour in the bottom right panel indicates the 1$\sigma$ range with $\chi^{2} = \chi^{2}_{\rm{min}} + 2.3$.}
 \label{fig:alpha_dgr_search}
\end{figure}

\noindent
We calculated the dust mass surface density ($\Sigma_{\rm{dust, model}}$) in each pixel ($j,\ k$) by using equation (\ref{eq:dust_model}), changing \aco\ and DGR as free parameters. $I_{\rm{CO, obs}}$, $\Sigma_{\rm{atom, obs}}$, and $\Sigma_{\rm{dust, obs}}$ are the observed values and corrected for the inclination of a galactic disk by multiplying $\cos{i}$, where $i$ is the inclination angle listed in table \ref{tab:galaxy_list}. Thus, we calculated the value of $\chi^{2}$ by using equation (\ref{eq:chi}) and the observed dust mass surface density ($\Sigma_{\rm{dust, obs}}$) and $\Sigma_{\rm{dust, model}}$ in the region where $I_\mathrm{CO} > 3 \Delta I_\mathrm{CO}$, $I_{\rm{H{\textsc{i}}}} > 3\Delta I_{\rm{H{\textsc{i}}}}$, and the infrared intensity $>$ 3$\times$rms noise of emission-free regions in an infrared intensity map. $\Delta  I_\mathrm{CO}$ and $\Delta I_{\rm{H{\textsc{i}}}}$ are the error of integrated intensity of CO and \hi, respectively. The derivations of these physical quantities are described in section \ref{Derivation_PP}. The $\chi^{2}$ fitting assumes that \aco\ and DGR are constant over the region where \aco\ and DGR are measured. We find the optimal values of \aco\ and DGR which minimize the value of $\chi^2$ ($\equiv \chi^{2}_{\rm{min}}$), indicating the minimized difference between $\Sigma_{\rm{dust, obs}}$ and $\Sigma_{\rm{dust, model}}$. As an example, figure \ref{fig:alpha_dgr_search} presents the result of NGC\,3627. The top left, top right, and bottom left panels are maps of $I_{\rm{CO, obs}}$, $\Sigma_{\rm{atom, obs}}$, and $\Sigma_{\rm{dust, obs}}$, respectively. The measurement regions of \aco\ and DGR are shown as the white contour in these panels. The bottom right panel shows the calculated result of equation (\ref{eq:chi}), and the red point in the panel presents \aco\ and DGR, which minimizes the $\chi^{2}$ value for optimization. The calculated results of equation (\ref{eq:chi}) for other galaxies are shown in the Appendix \ref{maps_PP}.

The optimal values of \aco\ and DGR were searched within the ranges of $\alpha_{\rm{CO}}=0.01-50.0$ \acounit\ with a step of 0.01 \acounit and $\mathrm{DGR} = 0.0001 - 0.15$ with a step of 0.0001, respectively. We determined the maximum of the ranges of \aco\ (= 50.0 \acounit) and DGR ($= 0.15$) by referring to the observational studies of \aco, DGR, and metallicity in nearby galaxies including our samples. In the measurement regions of our samples, the range of normalized metallicity ($Z^{\prime} \equiv Z/Z_\odot$) estimated by the metallicity ($Z$) from \citet{Pilyugin2014} presented in subsection \ref{metallicity} and the one in the solar neighborhood ($Z_\odot$) from \citet{Pilyugin2006} is 0.6 -- 2.9.

\begin{figure}
    \begin{center}
    \includegraphics[scale=1.5]{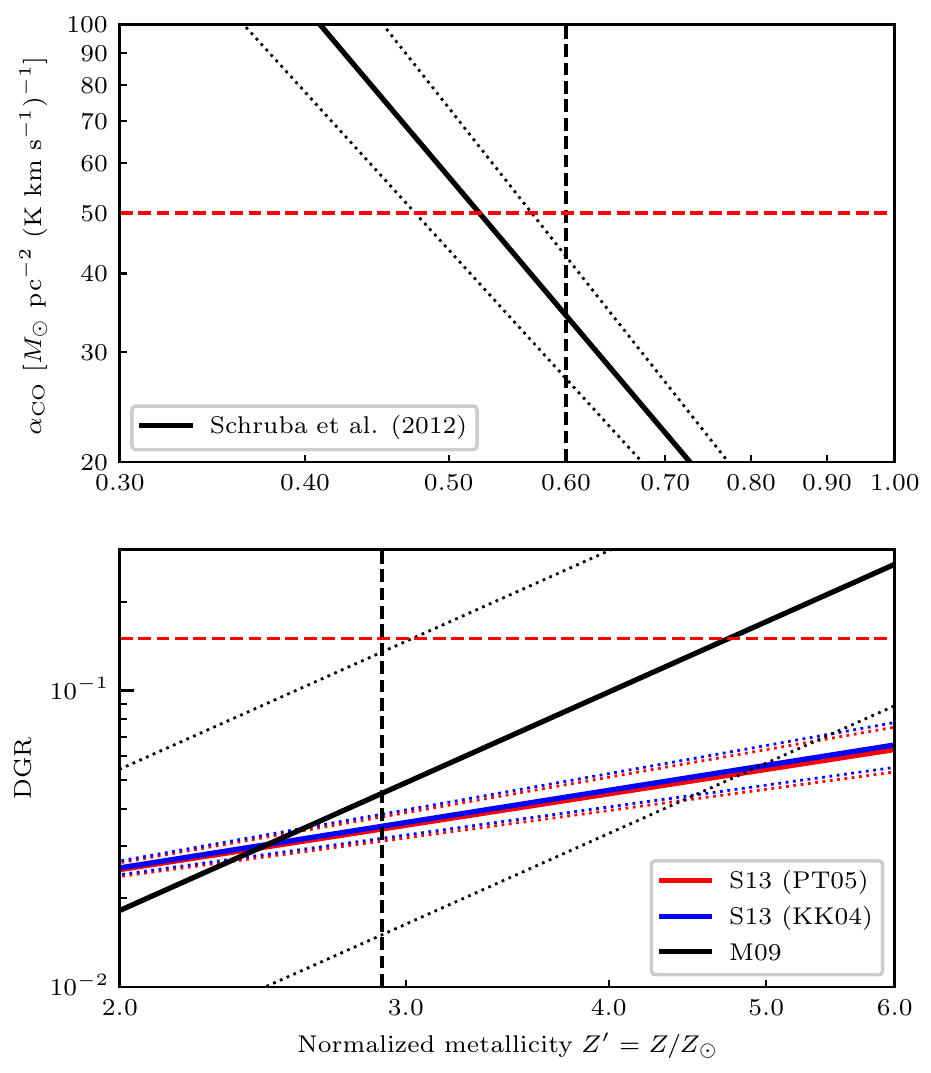}        
    \end{center}
  \caption{$Top\ panel$: \aco\ as a function of normalized metallicity in nearby galaxies investigated by \citet{Schruba2012} (solid black). The dashed vertical black line and the dashed horizontal red line indicate $Z^{\prime} = 0.6$ and \aco\ = 50.0 \acounit, respectively. The black dotted lines indicate the range of uncertainty. $Bottom\ panel$: DGR as a function of normalized metallicity in nearby galaxies investigated by \citet{Mu2009} (M09; solid black) and S13 with metallicity derived from \citet{Moustakas2010} with the oxygen abundance calibrated by \citet{Kobulnicky2004} (KK04; solid bule) and \citet{Pilyugin2005} (PT05; solid red). The dashed vertical black line and the dashed horizontal red line indicate $Z^{\prime} = 2.9$ and $\mathrm{DGR} = 0.15$, respectively. The dotted red line and the dotted black line indicate the range of uncertainty. The dotted black line indicate the variations of a factor of three around the solid black line.}
 \label{fig:aco_DGR_metallicity}
\end{figure}

The top panel of figure \ref{fig:aco_DGR_metallicity} shows \aco\ as a function of $Z^{\prime}$ investigated by several studies of nearby galaxies. For the range of our \aco, we referred to the result listed in the top row of table 7 from \citet{Schruba2012}, who measured global \aco\ derived from $\tau_\mathrm{dep} \times \mathrm{SFR} / L_\mathrm{CO}$, where $\tau_\mathrm{dep}$ is a constant \hii\ depletion time and $L_\mathrm{CO}$ is a observed CO luminosity, in dwarf and spiral galaxies. 22 out of 40 samples in this study overlap with the samples from \citet{Schruba2012}. Considering the relation between the result of \citet{Schruba2012} and the minimum $Z^{\prime}\ (=0.6)$ as shown in the top panel of figure \ref{fig:aco_DGR_metallicity}, we suspected that \aco\ of our samples were lower than 50 \acounit.

The bottom panel of figure \ref{fig:aco_DGR_metallicity} shows DGR as a function of $Z^{\prime}$ investigated by several studies of nearby galaxies. For the range of our DGR, we referred to the results of \citet{Mu2009} (hereafter M09) and S13. M09 measured the radial variations of DGR for 12 spiral galaxies by combining \hi\ profiles from the THINGS with CO profiles from the literature and using the metallicity-dependent CO-to-\hii\ conversion factor. Seven out of 12 samples of M09 overlap with our samples. They obtained the linear fitted relation between the measured DGR and metallicity of \citet{Moustakas2010} with the oxygen abundance calibrated by \citet{Kobulnicky2004} (hereafter KK04) while they did not evaluate the error of fitting. We investigated the relation between the metallicity measured by using values listed in table 8 of \citet{Moustakas2010} and the radial DGR values, which were measured in galaxies whose CO and \hi\ data were available and listed in table 4 of M09. We found that 74\% ($\approx 1 \sigma$) of their radial DGR followed their linear fit within a factor of three. Therefore, we defined the range within a factor of three around the linear fit of M09 as the error of the linear fit. S13 also obtained the linear fitted relation between DGR and metallicity with the oxygen abundance calibrated by KK04 and \citet{Pilyugin2005} (hereafter PT05) in eight spiral galaxies overlapping with our samples. Considering the relation between their results and the maximum $Z^{\prime}\ (=2.9)$ as shown in the bottom panel of figure \ref{fig:aco_DGR_metallicity}, we suspected that DGR of our samples was lower than 0.15.

%%%%%%%%%%%%%%%%%%%%
%% Uncertainties of alpha and DGR
%%%%%%%%%%%%%%%%%%%%
\subsection{Uncertainties of \aco\ and DGR}
To evaluate the uncertainties of \aco\ and DGR, we adopted a confidence interval of parameters denoted as $\Delta \chi^{2} \equiv \chi^{2} - \chi^{2}_{\rm{min}}$ for the $\chi^{2}$ fitting. If there are two free parameters used in the $\chi^{2}$ fitting, the 1$\sigma$ confidence interval of the parameters is $\Delta \chi^{2} = 2.3$. Therefore, the uncertainties of \aco\ and DGR are estimated from the minimum and maximum values of \aco\ and DGR with $\chi^{2} = \chi^{2}_{\rm{min}} + 2.3$. The white contour shown in the bottom right panel of figure \ref{fig:alpha_dgr_search} is an example of the curve with $\chi^{2} = \chi^{2}_{\rm{min}} + 2.3$. If $\chi^{2}$ of \aco\ and DGR extends outside of the parameter search space, we exclude the galaxy from the analysis of \aco\ and DGR.

Our method determining the uncertainties in \aco\ and DGR is different from that of S13. The uncertainties of \aco\ and DGR in S13 were determined by Monte Carlo tests and bootstrapping iterations. Our method can measure the uncertainties of \aco\ and DGR in a simple way compared to the S13 method using the $\chi^{2}$ method.

%%%%%%%%%%%%%%%%%%%%
%% Data
%%%%%%%%%%%%%%%%%%%%
\section{Derivation of physical quantities} \label{Derivation_PP}

%%%%%%%%%%%%%%%%%%%%
%% CO integrated intensity
%%%%%%%%%%%%%%%%%%%%
\subsection{CO integrated intensity}\label{I_CO}
We used the $^{12}$CO($J=1-0$) mapping data obtained in the COMING project (\cite{Sorai2019}) and the archival data of the CO Atlas (\cite{Kuno2007}) project to measure \ICO. The original angular resolutions of COMING and CO Atlas are \timeform{17"} and \timeform{15"}, respectively. The pixel size of COMING and the CO Atlas archival data are \timeform{6"} and \timeform{1"}, respectively. We used the auto-reduction system COMING-ART for the data reduction of COMING. COMING-ART contains the procedures for flagging bad data, baseline subtraction, and basket-weaving. Details regarding data reduction are described in \citet{Sorai2019}.

The integrated intensity of CO (\ICO) is evaluated as follows:

\begin{equation}
\left( \frac{I_\mathrm{CO}}{\mathrm{K\ km\ s^{-1}}} \right) = \int T_\mathrm{MB} dv = \sum_{k=1} T_\mathrm{MB} \Delta v_\mathrm{ch},
\label{eq:ICO}
\end{equation}

\noindent
where $k$ is the number of emission channels, $\Delta v_{\rm{ch}}$ is the velocity width of one channel (= 10 km s$^{-1}$). \citet{Sorai2019} determined the emission channels of the COMING cube data by identifying the channels above 2.5 $\sigma$ in the smoothed cube by $3 \times 3$ pixels in space. We also determined the emission channels of CO Atlas cube data adopting the same method.

The error of \ICO ($\Delta I_{\rm{CO}}$) is evaluated as follows:

\begin{equation}
\left(\frac{\Delta I_\mathrm{CO}}{\mathrm{K\ km\ s^{-1}}} \right) = T_{\mathrm{rms}}\sqrt{\Delta v_{\mathrm{signal}} \Delta v_{\mathrm{ch}}},
\label{eq:Delta_I}
\end{equation}

\noindent
where $T_{\rm{rms}}$ is the rms noise temperature on the $T_{\rm{MB}}$ scale (K) and $\Delta v_{\rm{signal}}$ is the total velocity width of identified emission channels ($= 10 \times k$ km s$^{-1}$). $T_{\rm{rms}}$ was calculated in the baseline range. For the COMING cube data, 200 km s$^{-1}$ at the edges of the observed velocity range was defined as the baseline channels. For the CO Atlas cube data, the baseline range was determined by eye. To measure \aco\ and DGR in the region where $^{12}$CO($J=1-0$) emission is significantly detected, we masked the pixels where $I_{\rm{CO}} \leq 3\Delta I_{\rm{CO}}$. The distribution of \ICO\ in individual galaxies is shown in the Appendix \ref{maps_PP}.

%%%%%%%%%%%%%%%%%%%%
%% Atomic gas mass surface density
%%%%%%%%%%%%%%%%%%%%

\subsection{Atomic gas mass surface density} \label{HI_mass}
For galaxies in the THINGS and the VIVA projects, and IC\,342 (\cite{Crosthwaite2000}), we measured \sigatom\ by using their publicly available \hi\ integrated intensity map. For galaxies whose data were obtained from the NRAO Science Data Archive, \sigatom\ was derived by integrating over the entire velocity range of each spectrum. Assuming that the optical depth of \hi\ gas is thin, \sigatom\ is calculated by the following equation:

\begin{equation}
\left(\frac{\Sigma_\mathrm{atom}}{M_\odot\ \mathrm{pc^{-2}}} \right) = 0.02 \left[ \frac{I_{\mathrm{H{\textsc{i}}}}}{\mathrm{K\ km\ s^{-1}}} \right],
  \label{eq:HI_cd}
\end{equation}

\noindent
where $I_{\mathrm{H{\textsc{i}}}}$ is the \hi\ integrated intensity. This equation includes a factor of 1.36 for helium. We masked the pixel where $I_{\rm{H{\textsc{i}}}} \leq 3\Delta I_{\rm{H{\textsc{i}}}}$ to measure \aco\ and DGR in the region where \hi\ emission is significantly detected. For the THINGS samples, \citet{Walter2008} reported that the typical sensitivity of \sigatom\ was $\sim$0.5 \massunit, which corresponds to $\Delta I_{\rm{H{\textsc{i}}}} = 0.5/0.02 = 25\ \mathrm{K\ km\ s^{-1}}$, at $\timeform{30"}$ resolution. Therefore, we convolved the publicly available \hi\ integrated intensity map of the THINGS with a Gaussian kernel to match $\timeform{30"}$ resolution and masked the pixel where \IHI\ of the convolved map was lower than $3 \Delta I_{\mathrm{H{\textsc{i}}}} = 75\ \mathrm{K\ km\ s^{-1}}$. For the samples from the NRAO Science Data Archive, we determined the emission channels by identifying the channels with $3\sigma$ in the smoothed cube convolved to twice the original beam size. We made a map of \IHI\ by using the identified \hi\ emission channels and equation (\ref{eq:ICO}) and measured $\Delta$\IHI\ by using equation (\ref{eq:Delta_I}). $T_\mathrm{rms}$ of \hi\ data was calculated in the velocity ranges which were not identified as \hi\ emission channel. To identify \hi\ emission channels and derived maps of \IHI\ and $\Delta$\IHI\ for the VIVA samples, we adopted the same method as samples of the NRAO Science Data Archive to archival cleaned cube data of VIVA. For IC\,342, we masked pixels of \sigatom\ with negative value because there was no archival cube data for \hi\ and we could not evaluate $\Delta I_{\rm{H{\textsc{i}}}}$ from equation (\ref{eq:Delta_I}). The distributions of \sigatom\ in individual galaxies are shown in the Appendix \ref{maps_PP}.

%%%%%%%%%%%%%%%%%%%%
%% Dust mass surface density
%%%%%%%%%%%%%%%%%%%%
\subsection{Dust mass surface density}\label{dust_sed}
Our samples mainly comprise star-forming galaxies and the peak of the spectral energy distribution (SED) of the cold dust component in a star-forming galaxy is located in the wavelength range of 100 to 200 $\mu$m (e.g., \cite{Boselli2003}; \cite{Dale2012}). Therefore, we measured the dust mass $M_{\rm{dust}}$ by using the infrared data of DustPedia (\cite{Clark2018}) from the archived PACS 70, 100, 160 $\mu$m, and SPIRE 250 $\mu$m photometry. $M_{\rm{dust}}$ can be derived from the dust temperature and infrared flux density (e.g., \cite{Hildebrand1983}, \cite{Gall2011}) using the following equation:

\begin{equation}
\left(\frac{M_{\mathrm{dust}}}{M_{\odot}} \right) =\frac{S_{\nu} D^{2}}{\kappa_{\nu} B_{\nu}(T_{\mathrm{dust}})},
\label{eq:dust_mass}
\end{equation}

\noindent
where $S_{\nu}$ is the infrared flux density at a frequency $\nu$, $\kappa_{\nu}$ is the dust absorption coefficient, $D$ is the distance to a galaxy, and $B_{\nu}(T_{\rm{dust}})$ is the Planck function as a function of dust temperature ($T_{\rm{dust}}$). Finally, \sigdust\ is derived by converting $M_{\rm{dust}}$ to the mass surface density in units of $M_{\odot}\ \rm{pc^{-2}}$ in each pixel. We used the infrared flux density at 250 $\mu$m to measure \sigdust\ because the angular resolution (original FWHM = \timeform{18"}; \cite{Clark2018}) is similar to that of the $^{12}$CO($J=1-0$) data of COMING. We adopted the dust absorption coefficient\footnote{https://www.astro.princeton.edu/~draine/dust/dustmix.html} at 250 $\mu$m of 3.98 cm$^{2}$ g$^{-1}$ derived from the model using the dust size distribution for $R_{\rm{V}}=3.1$ in the Milky Way (\cite{Weingartner2001}).

We can obtain the distribution of $T_{\rm{dust}}$ by the SED fitting at each pixel in the infrared maps. We adopted the SED model of Casey (2012) (hereafter C12) to obtain $T_{\rm{dust}}$. C12 described the SED model ($S_{\lambda}$), in which a modified blackbody (i.e., greybody) dominates at wavelengths $>$ 50 $\mu$m, whereas a power law dominates at wavelengths $<$ 50 $\mu$m. The model is described as follows:

\begin{equation}
\left(\frac{S_{\lambda}}{\mathrm{Jy}} \right) = N_{\mathrm{bb}} \frac{\left(1-e^{-\left(\frac{\lambda_{0}}{\lambda}\right)^{\beta}}\right)\left(\frac{c}{\lambda}\right)^{3}}{e^{h c / \lambda k T}-1}+N_{\mathrm{pl}} \lambda^{\alpha} e^{-\left(\frac{\lambda}{\lambda_{c}}\right)^{2}},
\label{eq:sed_model}
\end{equation}

\begin{equation}
N_{\mathrm{pl}} = N_{\mathrm{bb}} \frac{\left(1-e^{-\left(\frac{\lambda_{0}}{\lambda}\right)^{\beta}}\right)\lambda_{c}^{3}}{e^{h c / \lambda_{c} k T}-1}.
\label{eq:Npl}
\end{equation}

\noindent
$N_{\rm{bb}}$, $T$, $\beta$, and $\alpha$ are free parameters; $N_{\rm{bb}}$ is the normalization of the greybody term, $T$ is the galaxy's characteristic cold dust temperature ($=T_{\rm{dust}}$), $\beta$ is the emissivity, and $\alpha$ is the mid-infrared power-law slope. $\lambda_{0}$ is the wavelength at which the optical depth is at unity ($\equiv$ 200 $\mu$m). $N_{\rm{pl}}$ is the normalization of the power-law term. $\lambda_{c}$ is the wavelength where the mid-infrared power law turns over and described by $\alpha$ and $T$ as follows:

\begin{equation}
\lambda_{c} = [(b_{1} + b_{2} \alpha)^{-2} + (b_{3} + b_{4} \alpha) \times T]^{-1},
\label{eq:lamda_c}
\end{equation}

\noindent
where $b_{1} = 26.68$, $b_{2} = 6.246$, $b_{3}=1.905 \times 10^{-4}$, and $b_{4}=7.243\times10^{-5}$. Details regarding $S_{\lambda}$ are described in table 1 of C12, which suggests if there are more than three independent photometric points at $\lambda \geq 200\ \mu\rm{m}$, $\beta$ should be maintained as a free parameter; otherwise, the fixed $\beta=1.5$ should be adopted. They also suggested that if the photometric points are more than three at $\lambda \leq 70\ \mu\rm{m}$, $\alpha$ should be maintained as a free parameter; otherwise, a fixed $\alpha = 2.0$ should be adopted. In our study, the photometric points are less than three at $\lambda \leq 70\ \mu\rm{m}$ and $\lambda \geq 200\ \mu\rm{m}$. Therefore, we applied fixed $\beta=1.5$ and $\alpha=2.0$ for all samples listed in table \ref{tab:galaxy_list}.

We performed the SED fitting with the C12 SED model as follows:

\begin{enumerate}
\item We calculated the rms noise $\rm{RMS}_{\lambda}$ of the background in the 70, 100, 160, and 250 $\mu$m data for the emission-free region.
\item If the infrared flux density at a pixel ($j,\ k$) is lower than $3 \times \rm{RMS}_{\lambda}$ at one or more wavelengths, the pixel ($j,\ k$) is not used for the SED fitting.
\item We determined the optimal $N_{\rm{bb}}$ and $T$ of the SED model $S_{\lambda}$ by fitting the observed data using the non-linear least squares in each pixel and obtained the distribution of the dust temperature $T$ ($=T_{\rm{dust}}$).
\end{enumerate}

The distributions of \sigdust\ and $T_{\rm{dust}}$ in individual galaxies are shown in the Appendix \ref{maps_PP}.

%%%%%%%%%%%%%%%%%%%%
%% Ancillary data
%%%%%%%%%%%%%%%%%%%%
\section{Ancillary data}
Metallicity affects the abundance of interstellar dust, CO and \hii\ (e.g., \cite{Arimoto1996}, \cite{Leroy2011}, \cite{RemyRuyer2014}). Therefore, \aco\ and DGR are expected to be related to metallicity. Furthermore, \aco\ depends on the gas density and temperature of molecular clouds (e.g., \cite{Bolatto2013}), which are related to the star-formation activity of molecular clouds. Therefore, we discuss the dependence of \aco\ and DGR on metallicity and SFR.

%%%%%%%%%%%%%%%%%%%%
%% Star formation rate
%%%%%%%%%%%%%%%%%%%%
\subsection{Star formation rate}
We obtained the SFR surface density (\sigSFR) to discuss the correlation with \aco\ and DGR in subsection \ref{alpha_dgr_SFR_metal}. \sigSFR\ is measured by using far-ultraviolet (FUV) and mid-infrared 24 $\mu$m as a tracer of SFR with the following equation (\cite{Leroy2008}):

\begin{equation}
\left( \frac{\Sigma_{\mathrm{SFR}}}{M_{\odot}\ \mathrm{yr^{-1}\ kpc^{-2}}} \right) = \left( 8.1 \times 10^{-2} I_{\mathrm{FUV}} + 3.2 \times 10^{-3} I_{24 \mu \mathrm{m}} \right) \times \cos i,
\label{eq:SFR}
\end{equation}

\noindent
where $I_{\rm{FUV}}$ and $I_{24\mu \rm{m}}$ are the intensities of FUV and 24 $\mu$m, in units of MJy sr$^{-1}$, respectively, and $\cos i$ corrects the inclination angle $i$ of a galaxy. We obtained the distribution of \sigSFR\ in individual galaxies by using the archival data of the FUV and 24 $\mu$m from the DustPedia project. \citet{Clark2018} reported that the DustPedia acquired the UV data observed by the GALaxy Evolution eXplorer (GALEX; \cite{Morrissey2007}) from the GR6/7 data release (\cite{Bianchi2014}) and the 24 $\mu$m observed by the $Spitzer$ Space Telescope and provided from the MIPS Local Galaxies Program (\cite{Bendo2012}). The original FWHM of the FUV and 24 $\mu$m from DustPedia are \timeform{5.3"} and \timeform{6"}, respectively. The original grid sizes of these wavelengths are \timeform{3.2"} and \timeform{2.4"}--\timeform{2.6"}, respectively. For NGC\ 628, we used the 24 $\mu$m data observed by the Local Volume Legacy (\cite{Dale2009}), whose beam size and grid size were \timeform{6"} and \timeform{1.5"}, respectively, owing to the lack of data of 24 $\mu$m in the DustPedia project. The maps of FUV and 24 $\mu$m were regridded and convolved with a Gaussian kernel to match the spatial resolution listed in table \ref{tab:HI_CO_list}.

%%%%%%%%%%%%%%%%%%%%
%% Metallicity
%%%%%%%%%%%%%%%%%%%%
\subsection{Metallicity} \label{metallicity}
We discuss the correlation of metallicity with \aco\ and DGR. We used the metallicity reported by \citet{Pilyugin2014}. They investigated the distributions of the abundance of gas phase oxygen and nitrogen across the optical disks from 3740 spectra of the H{\sc ii} regions in 130 nearby late-type galaxies by combining the line intensities of [O\,{\sc ii}]$\lambda$3727+$\lambda$3729, [O\,{\sc iii}]$\lambda$5007, [N\,{\sc ii}]$\lambda$6584, [S, {\sc ii}]$\lambda$6717, and [S\,{\sc ii}]$\lambda$6731 normalized by H$\beta$. They modeled the radial distribution of the oxygen abundance ($12+\log (\rm{O/H})$) within the isophotal radius in each galaxy as follows:

\begin{equation}
12+\log (\mathrm{O} / \mathrm{H})=12+\log (\mathrm{O} / \mathrm{H})_{R_{0}}+C_{\mathrm{O} / \mathrm{H}} \times\left(R / R_{25}\right),
\label{eq:metal}
\end{equation}

\noindent
where $12+\log \mathrm{(O/H)}_{R_{0}}$ is the oxygen abundance at the galactic center $R_{0}$, $C_{\rm{O} / \rm{H}}$ is the slope of the oxygen abundance gradient, and $R / R_{25}$ is the fractional radius as a function of $R_{25}$. Each pixels were assigned a value of $12+\log (\rm{O/H})$ based on equation (\ref{eq:metal}) and the values of $12+\log \mathrm{(O/H)}_{R_{0}}$ and $C_{\rm{O/H}}$ from \citet{Pilyugin2014}.

%%%%%%%%%%%%%%%%%%%%
%% Results
%%%%%%%%%%%%%%%%%%%%
\section{Results}\label{Result}

%%%%%%%%%%%%%%%%%%%%
%% alpha and DGR in nearby galaxies
%%%%%%%%%%%%%%%%%%%%
\subsection{Global \aco\ and DGR} \label{result_alpha_dgr}
The regions where \aco\ and DGR were measured in each galaxy are shown in the Appendix \ref{maps_PP}. In 18 out of 40 galaxies, the 1$\sigma$ range of \aco\ of DGR extended outside the parameter search regions of \aco\ and DGR. These galaxies do not have a sufficient number of pixels for the $\chi^{2}$ fitting owing to the low signal-to-noise ratio of the CO data or the small contrast of CO/\hi\ to derive \aco\ and DGR using this method. Small contrast in the CO/\hi\ ratio (\ICO/\sigatom\ $\approx$ constant) causes the degeneracy of \aco\ and DGR in equation (\ref{eq:dust_model}) and increases the 1$\sigma$ range of \aco\ and DGR. We did not use these galaxies for subsequent analysis. Finally, we obtained \aco\ and DGR for 22 spiral galaxies. We define these \aco\ and DGR as ``global'' \aco\ and DGR, \aco(global) and DGR(global). These values and the largest radius of the measurement regions in individual galaxies are listed in table \ref{tab:alpha_DGR_measure}. Among the interacting galaxies listed in table \ref{tab:galaxy_list}, we obtained \aco(global) and DGR(global) in the spiral galaxy NGC\,4568 only. The mean radius of the measurement regions of \aco(global) and DGR(global) for the 22 spiral galaxies is 0.38$R_{25}$. The average and standard deviation of \aco(global) and DGR(global) for the 22 spiral galaxies are $2.66 \pm 1.36$ \acounit and $0.0052 \pm 0.0026$, respectively. The values of \aco(global) in 20 out of 22 spiral galaxies are lower than the standard values in the Milky Way $\alpha_{\rm{CO}}(\rm{MW})=4.35$ \acounit\ in \citet{Bolatto2013}. The comparison of \aco(global) with \acomw\ is discussed in sub-subsection \ref{comp_with_MW}.

\begin{figure}[htbp]
    \begin{center}
    \includegraphics[scale=1.3]{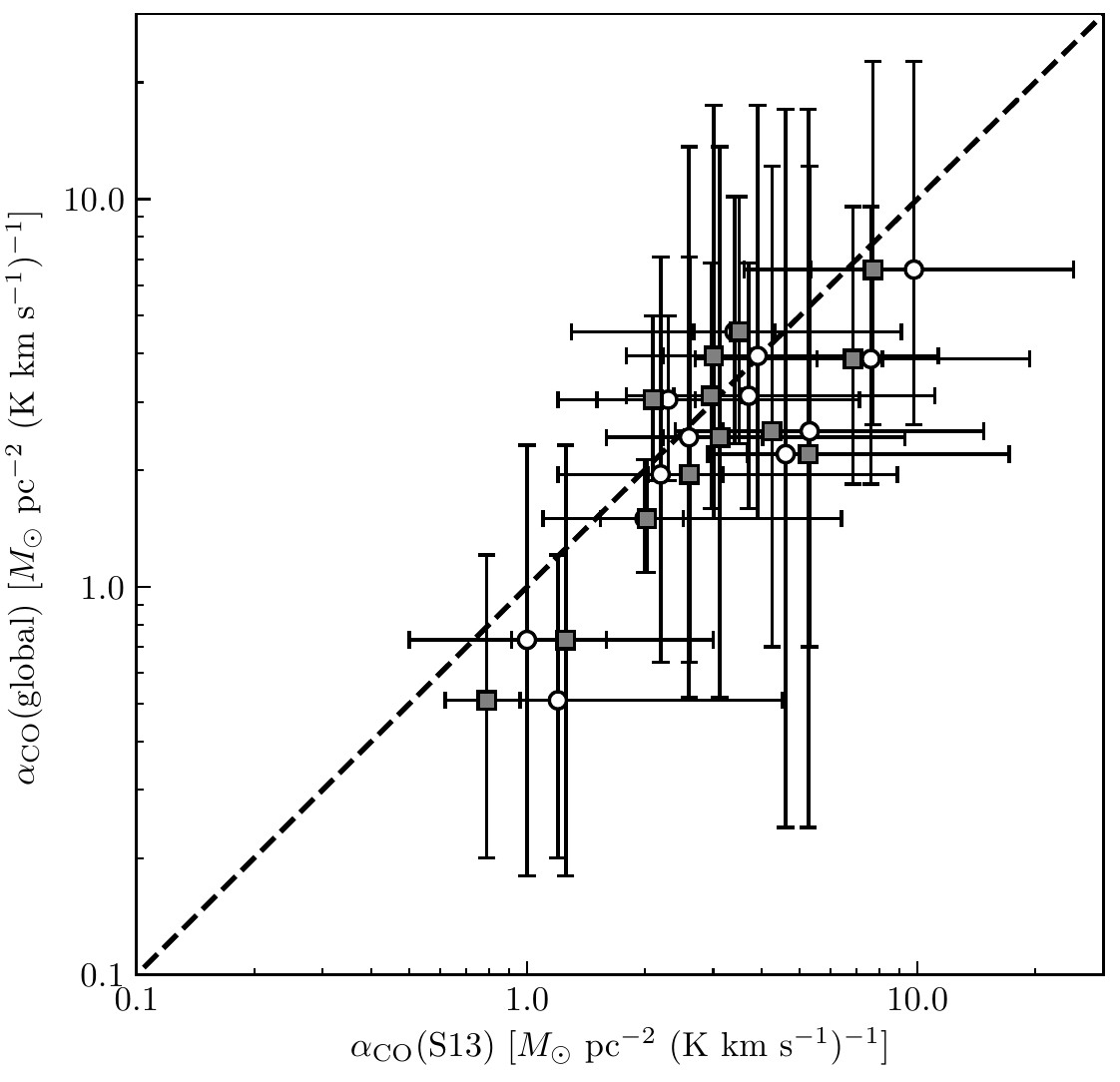}
    \end{center}
 \caption{The relation between our \aco(global) and mean \aco\ of individual galaxy measured by S13, \aco(S13). The white circles of \aco(S13) indicate the values of \aco(S13) listed in the second column of table \ref{tab:alpha_R21}. The gray squares of \aco(S13) indicate \aco(S13) corrected for $R_{21}$ by \aco(S13)$\times (\overline{R_{21}}/0.7)$, where $\overline{R_{21}}$ is the mean $R_{21}$ measured by \citet{Yajima2021}. The error bars of \aco(global) indicate the minimum and maximum values of \aco\ at $\chi^{2}=\chi^{2}_{\mathrm{min}}+2.3$. The error bars of \aco(S13) plotted as white circles and gray squares indicate the standard deviation listed in table 4 of S13 and the error of \aco(S13)$\times (\overline{R_{21}}/0.7)$ considering the standard deviation of $\overline{R_{21}}$ in \citet{Yajima2021}, respectively. The dashed black line indicates \aco(global) = \aco(S13).}
 \label{fig:comp_S13_aco}
\end{figure}

Thirteen out of 22 galaxies are in common with samples of S13 as shown in figure \ref{fig:comp_S13_aco}. For the 13 galaxies, table \ref{tab:alpha_R21} presents the values of \aco(global) in this study and the mean values of \aco\ obtained by S13 in individual galaxies (\aco(S13)). The values of \aco(global) in 10 out of 13 galaxies are lower than those of S13. S13 adopted the constant $R_{21}(=0.7)$ to measure \aco(S13). \citet{Yajima2021} reported the mean $R_{21}$ weighted by the integrated intensity of $^{12}$CO($J=1-0$) ($\overline{R_{21}}$) in nearby galaxies. $\overline{R_{21}}$ was derived using the COMING and the CO Atlas for $^{12}$CO($J=1-0$), and the HERACLES for $^{12}$CO($J=2-1$). The range of $\overline{R_{21}}$ in 13 galaxies is 0.46--0.88. Figure \ref{fig:comp_S13_aco} also shows \aco(S13) corrected for $R_{21}$ by \aco(S13)$\times (\overline{R_{21}}/0.7)$ (gray squares). We discuss the difference between \aco(global) and \aco(S13) in sub-subsection \ref{comp_S13}: (1) total dust mass, (2) CO($J=2-1$)/CO($J=1-0$) line ratio $R_{21}$, and (3) difference in the data sampling area.

\subsection{Radial variations of \aco\ and DGR} \label{radial_alpha_dgr}
\begin{figure}[htbp]
 \begin{center}
  \includegraphics[width=\linewidth]{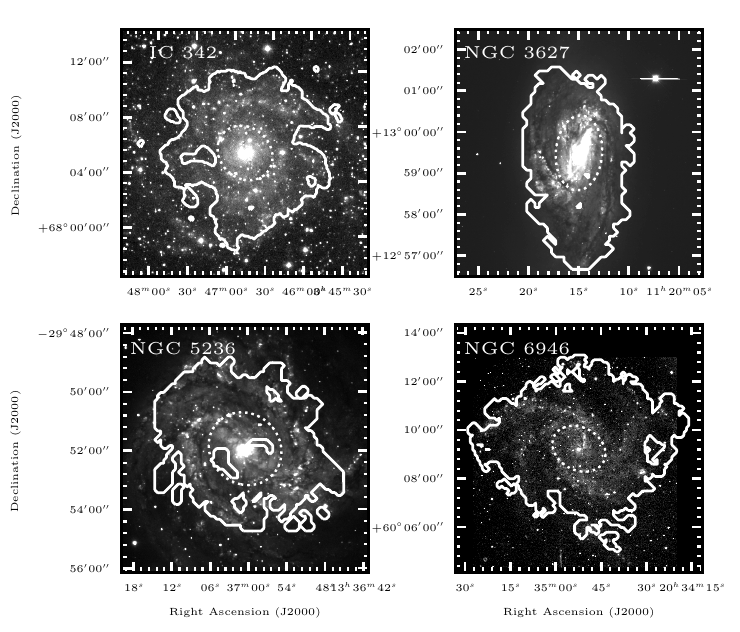}
 \end{center}
 \caption{The boundary of the inner region ($r \leq 0.2 R_{25}$) and the outer region ($r > 0.2R_{25}$) where radial variations of \aco\ and DGR were measured in four galaxies. The dotted white ellipse indicates the boundary at $r=0.2R_{25}$. The white contour indicates the region used for the measurement of \aco\ and DGR. The gray images are obtained from the NED Database (NGC\,3627: $B$ band image from the fifth delivery of the Spitzer Data from the $Spitzer$ Infrared Nearby Galaxies Survey (SINGS, \cite{Moustakas2010}), NGC\,5236: $B$ band image from \cite{David2014}; NGC\,6946: $B$ band image from \cite{Knapen2004}, IC\,342: 645 nm image from the Digitized Sky Survey).}
 \label{fig:R25_Vband}
\end{figure}

We investigated the radial variation of \aco\ and DGR for the galaxies whose \aco(global) and DGR(global) are discussed in subsection \ref{result_alpha_dgr}. To obtain an adequate amount of pixels used for the $\chi^{2}$ fitting and sufficient contrast of the CO/\hi\ ratio, we divided the region within a galaxy only into inner and outer regions. We adopted 0.2$R_{25}$ as the boundary of the inner and outer regions, as we could obtain \aco\ and DGR in both the inner and outer regions in the largest number of galaxies with the boundary. We define \aco\ and DGR in the inner region as \aco(inner) and DGR(outer) and those in the outer region as \aco(outer) and DGR(outer). Finally, we obtained \aco\ and DGR in the inner and outer regions of the following four barred spiral galaxies: IC\,342, NGC\,3627, NGC\,5236, and NGC\,6946, as listed in table \ref{tab:alpha_DGR_radial}. The boundary of 0.2$R_{25}$ in each four galaxies is shown in figure \ref{fig:R25_Vband}. The inclination and the position angle were corrected. In these galaxies, the inner region includes the center and the bar structure, while the outer region includes the spiral arms. The average and standard deviation of \aco(inner) and DGR(inner) are $0.36 \pm 0.08$ \acounit\ and $0.0199 \pm 0.0058$, while \aco(outer) and DGR(outer) are $1.49 \pm 0.76$ \acounit\ and $0.0084 \pm 0.0037$, respectively. In each galaxy, while \aco(outer) is 2.3 to 5.3 times larger than \aco(inner), DGR(outer) is 0.3 to 0.6 times smaller than that DGR(inner). Moreover, the mean values of \aco(inner) and \aco(outer) are smaller than the mean \aco(global) within the whole samples ($=2.66$ \acounit). The four galaxies with \aco(inner) and \aco(outer) also have the lower \aco(global) than the mean \aco(global). Therefore, our measurement of the radial variation of \aco\ is biased towards the galaxies with small \aco(global) and this causes the small \aco(inner) and \aco(outer).

The radial variation of \aco\ and DGR in galaxies has also been reported in previous studies. \citet{Cormier2018} found that \xco\ has lower values in the center of eight nearby spiral galaxies. \citet{Nakai1995} measured \xco\ in M\,51 and reported $X_{\rm{CO}}/10^{20}=0.75$ \xcounit\ ($\alpha_{\rm{CO}} = 1.6$ \acounit)\ in the inner region ($R < $ \timeform{1.2'}) and $X_{\rm{CO}}/10^{20}=1-4$ \xcounit\ ($\alpha_{\rm{CO}} = 2.2 - 8.7$ \acounit) in the outer region ($R >$ \timeform{2'}). S13 found the decrease in DGR with a galactocentric radius in nearby star-forming galaxies. \citet{Giannetti2017} demonstrated the radial variation of the gas-to-dust ratio ($= \mathrm{DGR}^{-1}$) as a function of the galactocentric radius between $\sim$ 2 kpc and $\sim$ 20 kpc from the center of the Milky Way by the following equation:

\begin{equation}
\log (\mathrm{DGR}^{-1}) = \left(0.087[^{+0.045}_{-0.025}] \pm 0.007 \right) R_{\mathrm{GC}} + \left(1.44[^{-0.45}_{+0.21}] \pm 0.03 \right),
\label{eq:gdr_giannetti}
\end{equation}

\noindent
where $R_{\rm{GC}}$ is the galactocentric radius in units of kpc.

\subsection{Relations with environmental properties} \label{correlation_Z_SFR}

\begin{figure}[htbp]
    \begin{center}
    \includegraphics[width=\linewidth]{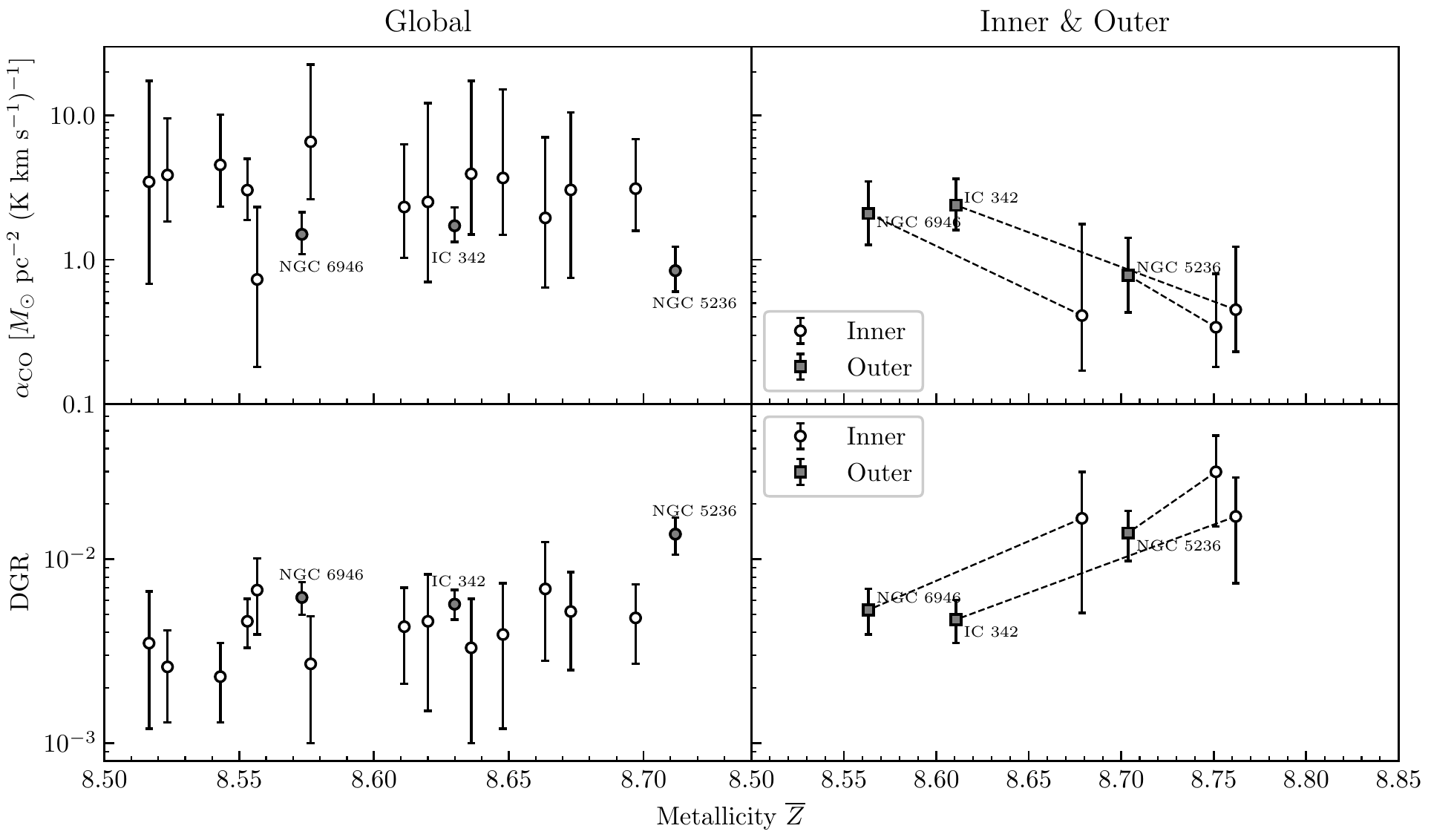}
    \end{center}
 \caption{Relations of metallicity with \aco\ (top panels) and DGR (bottom panels). The left panels indicate global values listed in table \ref{tab:alpha_DGR_measure}. The gray circles indicate IC\,342, NGC\,5236, and NGC\,6946 whose \aco\ and DGR in the inner and outer regions were measured. The right panels indicate the values in the inner and outer regions listed in table \ref{tab:alpha_DGR_radial}. $\overline{Z}$ is the average $12+\log(\rm{O/H})$ over pixels where \aco\ and DGR are measured in the global region (left panels), the inner region, and the outer region (right panels), respectively. The error bars of \aco\ and DGR indicate the minimum and maximum values of \aco\ and DGR at $\chi^{2}=\chi^{2}_{\mathrm{min}}+2.3$.}
 \label{fig:alpha_dgr_metal}
\end{figure}

We investigated the correlations of \aco\ and DGR with metallicity and SFR. Figure \ref{fig:alpha_dgr_metal} presents the relations of \aco\ and DGR with metallicity. \aco\ and DGR in the left panels are the global values listed in table \ref{tab:alpha_DGR_measure}. The metallicity in the left panels of figure \ref{fig:alpha_dgr_metal} is the average $12+\log(\rm{O/H})$ over pixels where \aco(global) and DGR(global) are measured. We define this metallicity as $\overline{Z}$(global). The results of NGC\,3147, NGC\,3627, NGC\,4527, NGC\,4536, NGC\,4568, and NGC\,5713 are not plotted owing to the lack of metallicity data in \citet{Pilyugin2014}. The Spearman's rank correlation coefficient of \aco(global) and DGR(global) with $\overline{Z}$(global) in the left panels of figure \ref{fig:alpha_dgr_metal} are 0.15 and 0.19, respectively. The right panels of figure \ref{fig:alpha_dgr_metal} present \aco\ and DGR in the inner and outer regions measured listed in table \ref{tab:alpha_DGR_radial} as a function of metallicity which is the average $12+\log(\rm{O/H})$ over pixels where \aco\ and DGR in the inner and outer regions are measured. We define the metallicity in the inner region as $\overline{Z}$(inner) and that in the outer region as $\overline{Z}$(outer). The Spearman's rank correlation coefficients of \aco\ and DGR with $\overline{Z}$ in the inner and outer regions are $-0.66$ and 0.83, respectively. We can find a clearer correlation of metallicity with \aco\ and DGR in the inner and outer regions than \aco(global) and DGR(global).

\begin{figure*}
    \begin{center}
    \includegraphics[width=\linewidth]{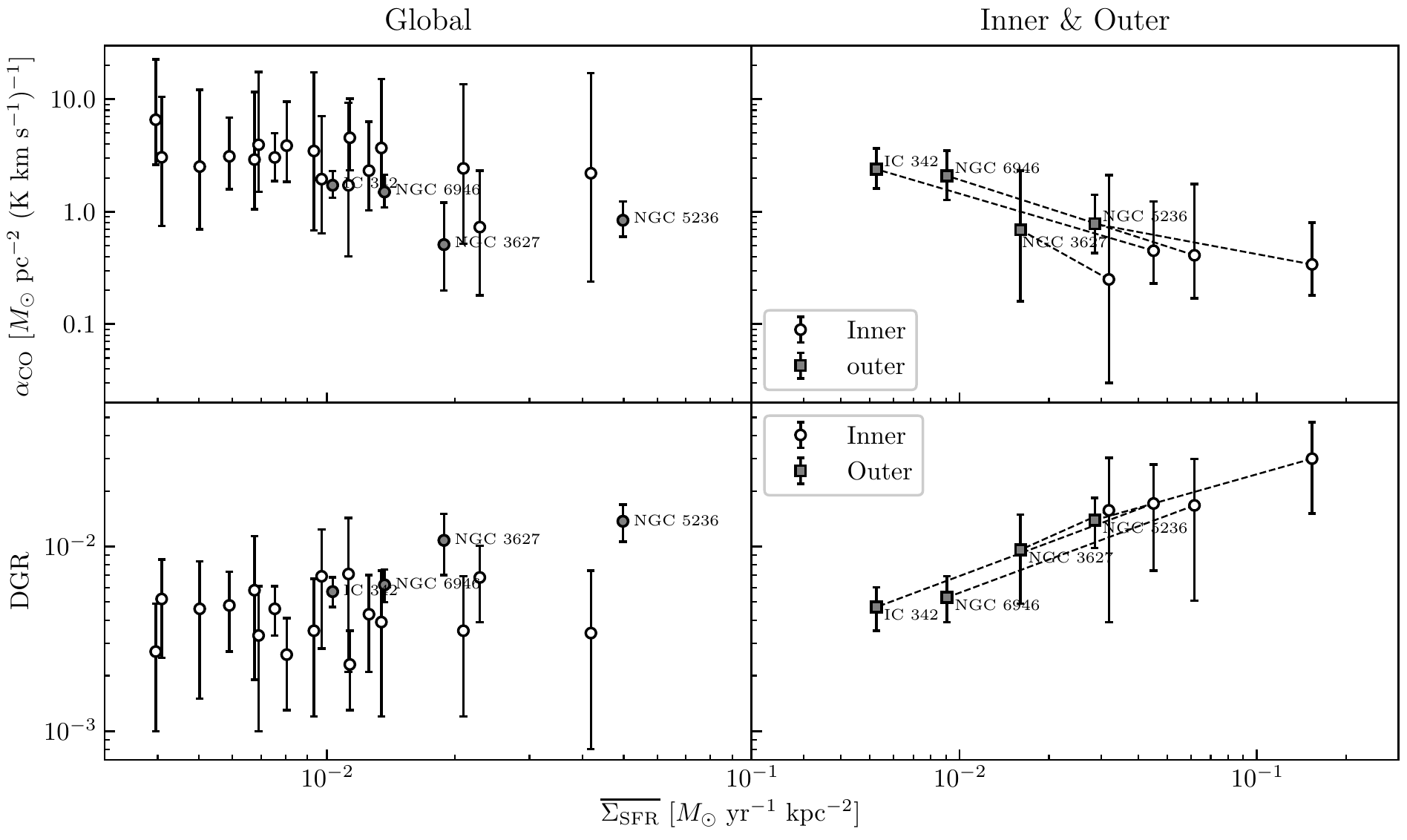}
    \end{center}
 \caption{Relations of \sigSFR\ with \aco\ (top panels) and DGR (bottom panels). The left panels indicate global values listed in table \ref{tab:alpha_DGR_measure}. The gray circles indicate IC\,342, NGC\,3627, NGC\,5236, and NGC\,6946 whose \aco\ and DGR in the inner and outer regions were measured. The right panels indicate the values in the inner and outer regions listed in table \ref{tab:alpha_DGR_radial}. $\overline{\Sigma_{\rm{SFR}}}$ is the average \sigSFR\ over pixels where \aco\ and DGR are measured in the global region (left panels) and the radial region (right panels), respectively. The error bars of \aco\ and DGR indicate the minimum and maximum values of \aco\ and DGR.}
 \label{fig:alpha_DGR_SFR}
\end{figure*}

Figure \ref{fig:alpha_DGR_SFR} presents the relations of \aco\ and DGR with \sigSFR. \aco\ and DGR in the left panels are the global values listed in table \ref{tab:alpha_DGR_measure}. NGC\,4527 is not plotted in the left panels of figure \ref{fig:alpha_DGR_SFR} due to the lack of 24 $\mu$m data. The SFR surface density in the left panels of figure \ref{fig:alpha_DGR_SFR} is the average \sigSFR\ over pixels where \aco(global) and DGR(global) are measured. We define this average \sigSFR\ as $\overline{\Sigma_{\rm{SFR}}}$(global). The Spearman's rank correlation coefficient of \aco(global) and DGR(global) with $\overline{\Sigma_{\rm{SFR}}}$(global) are $-0.53$ and 0.29, respectively. The right panels of figure \ref{fig:alpha_DGR_SFR} present \aco\ and DGR in the inner and outer regions measured listed in table \ref{tab:alpha_DGR_radial} as a function of SFR surface density which is the average \sigSFR\ over pixels where \aco\ and DGR in the inner and outer regions are measured. We define the average \sigSFR\ in the inner region as $\overline{\Sigma_{\rm{SFR}}}$(inner) and that in the outer region as $\overline{\Sigma_{\rm{SFR}}}$(outer). The Spearman's rank correlation coefficients of \aco\ and DGR with $\overline{\Sigma_{\rm{SFR}}}$ in the inner and outer regions are $-0.83$ and 0.98, respectively. As with metallicity, a more apparent correlation for the radial values can be observed compared to the global values. We discuss the difference of these correlations between global values and radial values in subsection \ref{alpha_dgr_SFR_metal}.

%%%%%%%%%%%%%%%%%%%%
%% Discussion
%%%%%%%%%%%%%%%%%%%%
\section{Discussion}\label{discussion}

%%%%%%%%%%%%%%%%%%%%
%% Correlations of \aco\ and DGR with environmental properties
%%%%%%%%%%%%%%%%%%%%
\subsection{Dependence of \aco\ and DGR on environmental properties} \label{alpha_dgr_SFR_metal}
We discuss the dependence of \aco\ and DGR on the metallicity obtained from figure \ref{fig:alpha_dgr_metal}. According to theoretical models (e.g., \cite{Bolatto1999}), in the low metallicity region, CO photodissociation is promoted owing to the decrease in interstellar dust and CO abundance. Although \hii\ is also dissociated by FUV, it is more likely to exist in the outer layer of a molecular cloud than CO owing to its stronger self-shielding. Observations indicate that low metallicity galaxies have a large \aco. For example, \citet{Arimoto1996} measured \xco\ as a function of metallicity for nearby spiral and dwarf irregular galaxies and reported a linear relationship. \citet{Leroy2011} measured \aco\ in local group galaxies, such as M\,31, M\,33, the large magellanic cloud (LMC), small magellanic cloud (SMC), and NGC\,6822, utilizing the same relation between \aco\ and DGR described in equation (\ref{eq:alpha_DGR}) used in this study. They found that low metallicity galaxies, such as NGC\,6822 and SMC, whose metallicities are $12+\rm{log(O/H)} \lesssim 8.0 - 8.2$, have a large \aco\ ($\sim$ 30 and 70), while the \aco\ of M\,31, M\,33, and LMC, whose metallicities are $12+\rm{log(O/H)} > 8.2$, are nearly constant ($\alpha_{\rm{CO}} \approx 6$) within a factor of two. They also found that DGR has a nearly linear relationship with metallicity in their samples. Large values of \xco\ in SMC have also been reported in other previous studies (e.g., \cite{Israel1997}; \cite{Leroy2007}).

\begin{figure*}
    \begin{center}
    \includegraphics[width=\linewidth]{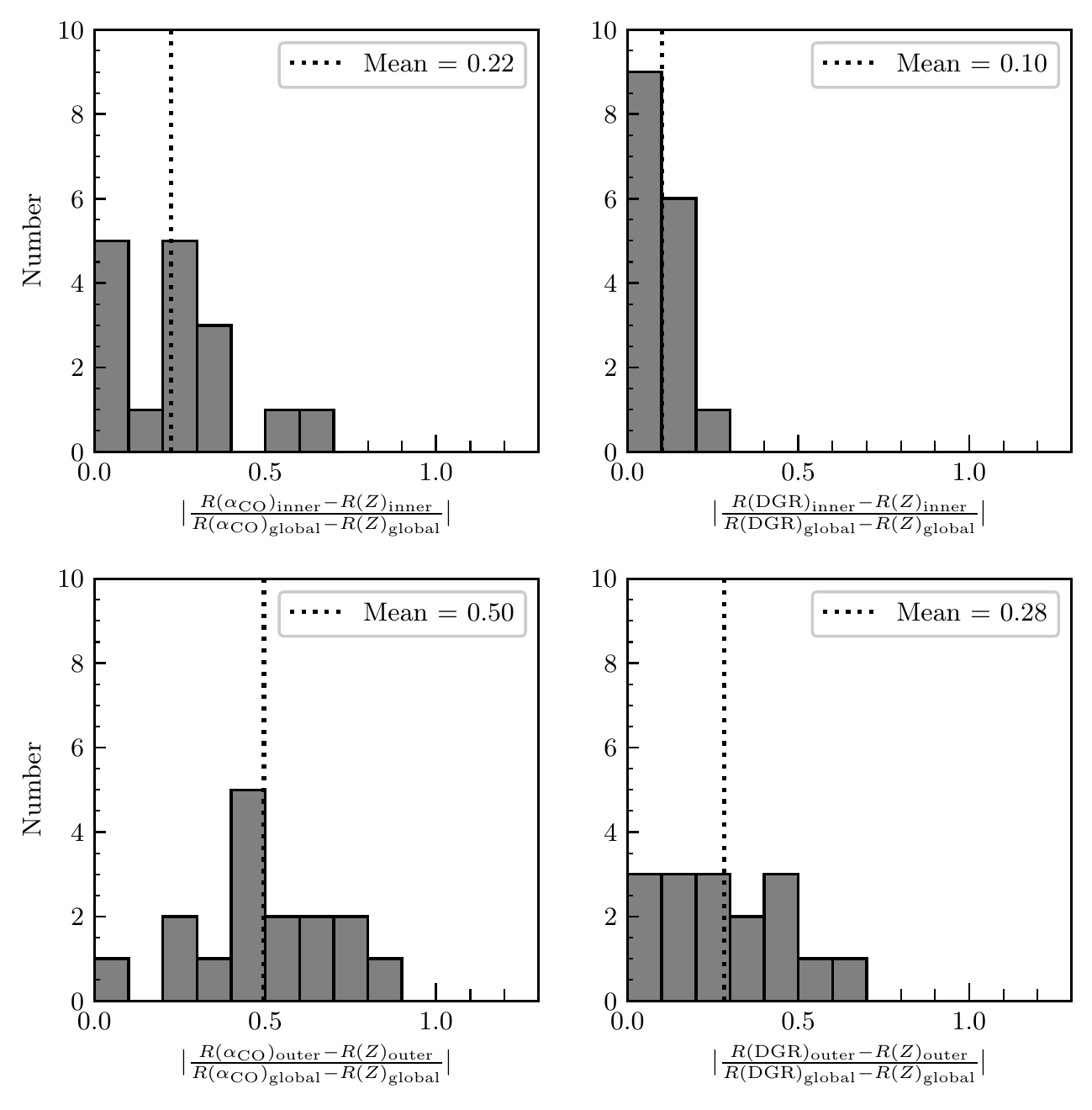}
    \end{center}
 \caption{Histograms showing the absolute value of ratio of difference between typical radius of \aco, DGR, and metallicity ($R(\alpha_\mathrm{CO})$, $R(\mathrm{DGR})$, and $R(Z)$) in the inner (or outer) region and that in the global region. $R(\alpha_\mathrm{CO})$, $R(\mathrm{DGR})$, and $R(Z)$ are the average radii over pixels where \aco\ and DGR are measured and weighted by \ICO, \sigdust, and metallicity, respectively. These ratios are measured in galaxies which listed in table \ref{tab:alpha_DGR_measure} and have the data of metallicity. The ranges of inner and outer regions are $\leq 0.2R_{25}$ and $> 0.25 R_{25}$, respectively. The dotted black line indicates the mean value in each panel.}
 \label{fig:hist_radius_aco_DGR_Z}
\end{figure*}

From the results of figure \ref{fig:alpha_dgr_metal}, we cannot find a significant correlation of $\overline{Z}$(global) with \aco(global) and DGR(global). This is because the diversity of \aco\ among our sample galaxies is excessively large in the narrow metallicity range. This is also consistent with the previous studies in which \aco\ was nearly constant in galaxies with $12+\rm{log(O/H)} > 8.2$. On the other hand, we can find a clearer correlation of metallicity with \aco\ and DGR in the inner and outer regions than with \aco(global) and DGR(global). We speculate that this results from the difference in the manner the values of \aco, DGR, and metallicity are obtained. In our sample galaxies, the metallicity expressed by equation (\ref{eq:metal}) decreases at a constant rate with a galactocentric radius, in contrast to the radial profiles of CO, \hi, and dust. When comparing \aco(global) (or DGR(global)) and $\overline{Z}$(global) within a galaxy, there is a discrepancy in the typical radius where their values are represented. On the other hand, when comparing \aco(inner) (or DGR(inner)) and $\overline{Z}$(inner) within a galaxy, the discrepancy of the typical radius where their radial values are represented becomes smaller because \aco(inner) (or DGR(inner)) and $\overline{Z}$(inner) are averaged within narrower regions than \aco(global) (or DGR(global)) and $\overline{Z}$(global). The comparison of \aco(outer) (or DGR(outer)) with $\overline{Z}$(outer) is also similar. To verify the speculation, we investigated the differences of typical radii of \aco, DGR, and metallicity ($R(\alpha_\mathrm{CO})$, $R(\mathrm{DGR})$, and $R(Z)$) in the global, inner ($r \leq 0.2R_{25}$), and outer ($r > 0.2R_{25}$) regions for galaxies which are listed in table \ref{tab:alpha_DGR_measure} and have the data of metallicity. $R(\alpha_\mathrm{CO})$, $R(\mathrm{DGR})$, and $R(Z)$ are the average radii over pixels where \aco\ and DGR are measured and weighted by \ICO, \sigdust, and metallicity, respectively. We measured $R(\alpha_\mathrm{CO}) - R(Z)$ and $R(\mathrm{DGR}) - R(Z)$ in the global, inner, and outer regions and compared them. Figure \ref{fig:hist_radius_aco_DGR_Z} shows the histogram showing the absolute value of the ratio between $R(\alpha_\mathrm{CO}) - R(Z)$ (or $R(\mathrm{DGR}) - R(Z)$) in the inner region and that in the global region. All ratios are smaller than the unity, indicating that the difference of typical radius between \aco\ (or DGR) and metallicity in the inner region is smaller than that in the global region. This trend is the same as in the case of the outer region. Therefore, we can find the correlation of \aco\ (or DGR) with the metallicity by separating them into the inner and outer regions.

Note that the three galaxies (IC\,342, NGC\,5236, and NGC\,6946) have a large metallicity gradient within their galactic disk among our samples. The gradient of metallicity is determined by $C_\mathrm{O/H}$ in equation (\ref{eq:metal}). Among our samples where \aco(global) was measured, the three galaxies have the large values of $C_\mathrm{O/H}$ while there are some galaxies with small $C_\mathrm{O/H}$ like NGC\,4321, NGC\,4736, NGC\,5248, and NGC\,7331. If we can also obtain \aco\ and DGR in the inner and outer regions of galaxies with small metallicity gradient, it is expected that the correlations between \aco, DGR and metallicity in the inner and outer regions shown in the right panels of figure \ref{fig:alpha_dgr_metal} are weak.

From figure \ref{fig:alpha_DGR_SFR}, we can also find a clearer correlation of $\overline{\Sigma_{\rm{SFR}}}$ with \aco\ and DGR in the inner and outer regions than with \aco(global) and DGR(global). The decrease in the CO-to-\hii\ conversion factor in high SFR regions is suggested by the numerical simulations. \citet{Narayanan2012} investigated the variations of \xco\ in merging galaxies. They suggested that \xco\ decreases in high SFR regions because a higher kinetic temperature and larger velocity dispersion increases the CO integrated intensity. There are also observational studies supporting this prediction (e.g., \cite{Accurso2017}; \cite{Morokuma2020}). This trend is consistent with the expectation from the numerical simulations of \aco.

In our samples, both \sigSFR\ and metallicity increase with a decreasing radius. Therefore, it is difficult to distinguish which metallicity or \sigSFR\ affect the radial variations of \aco\ and DGR within a galaxy in this study. To solve this problem, the distributions of \aco, DGR, metallicity, and \sigSFR\ with smaller spatial scales are necessary.

\subsection{Comparisons with the previous studies}\label{comp_aco}

\subsubsection{\aco\ in Sandstrom et al. (2013)} \label{comp_S13}
We compared our \aco(global) with that in S13 for the 13 galaxies in common with both our study and S13 as shown in figure \ref{fig:comp_S13_aco}. Although our \aco(global) is within the standard deviation of \aco\ in S13, our \aco(global) is lower than that in S13 for 10 out of 13 galaxies as listed in table \ref{tab:alpha_R21}. We discuss the cause of the difference considering the following: (1) total dust mass, (2) CO($J=2-1$)/CO($J=1-0$) line ratio $R_{21}$, and (3) difference in the data sampling area.

\noindent
(1) Total dust mass\\
The SED model to derive a dust mass is different between our study and S13. We discuss the difference of dust mass caused by the different SED models and the effect of its difference on the value of \aco. We obtained \sigdust\ by using the C12 SED model with a single dust temperature using infrared data (70, 100, 160, and 250 $\mu$m) from the DustPedia. S13 obtained \sigdust\ using the SED model of \citet{Draine2007a} (hereafter DL07) and infrared data (IRAC 3.6, 4.5, 5.8, and 8.0 $\mu$m, MIPS 24 and 70 $\mu$m, PACS 70, 100, and 160 $\mu$m; SPIRE 250 and 350 $\mu$m) from Key Insights on Nearby Galaxies: A Far-Infrared Survey with Herschel (KINGFISH, \cite{Kennicutt2011}). Although there is no information regarding the dust mass in S13, \citet{Aniano2020} (hereafter A20) performed the dust modeling to the full KINGFISH samples by using the DL07 model and the infrared data of IRAC, MIPS, PACS, and SPIRE, which were the same as S13. They also provided the archival data\footnote{\url{https://dataspace.princeton.edu/handle/88435/dsp01hx11xj13h}} of \sigdust\ for galaxies listed in table \ref{tab:alpha_R21}, whose \aco\ was measured by both this study and S13. Therefore, we investigated the difference of dust mass between this study using the C12 SED model and A20 as an indicator of dust mass of S13. After background subtraction, A20 convolved each image with a common point spread function and resampled to match a common pixel size. They provided a map of \sigdust\ whose FWHM and grid size are (\timeform{18.2"}, \timeform{6.0"}). Except for NGC\,4254 and NGC\,4321, the 11 galaxies listed in table \ref{tab:alpha_R21} have similar spatial resolution (beam size = \timeform{18.0"} and pixel size = \timeform{6.0"}) to that of \sigdust\ provided by A20. Therefore, we compared \sigdust\ of A20 with \sigdust\ of 11 galaxies modeling with the C12 SED model.

\begin{figure}[htbp]
 \begin{center}
  \includegraphics[scale=1.3]{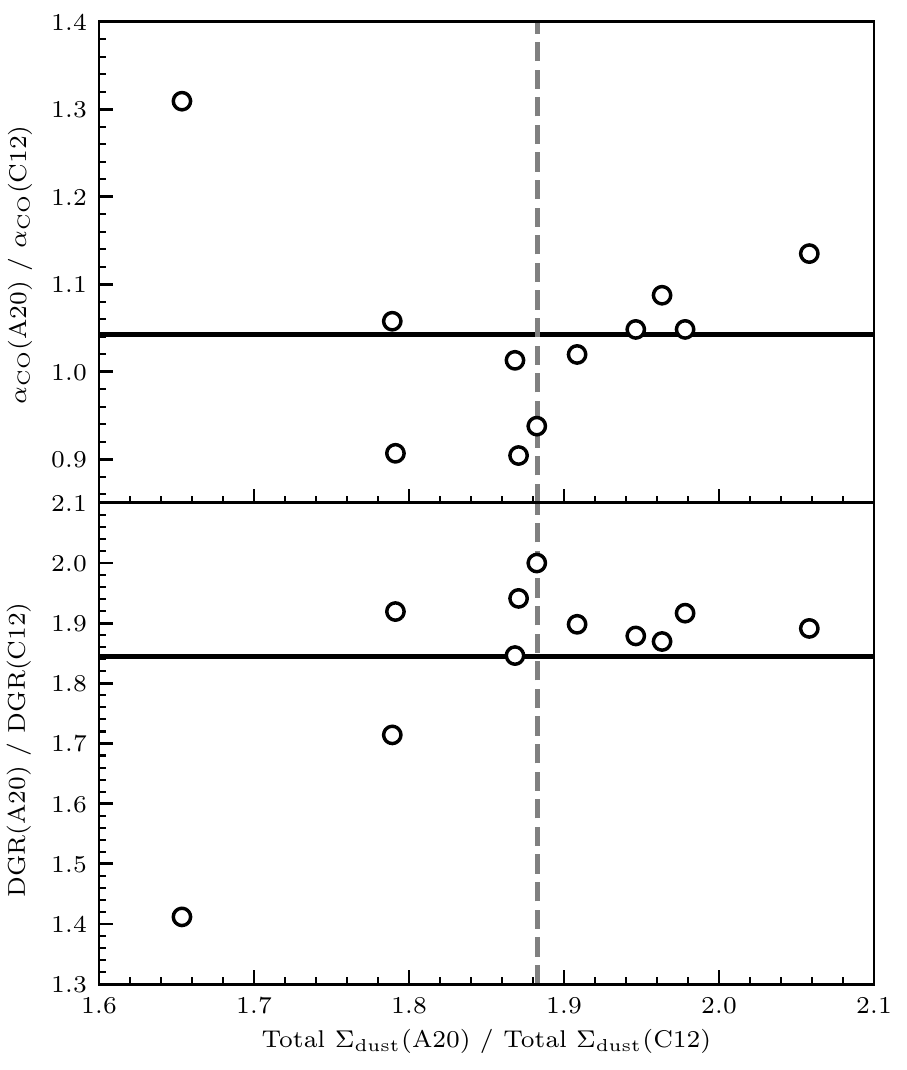}
 \end{center}
 \caption{Comparison of \aco\ (top) and DGR (bottom) measured by using the SED model of \citet{Casey2012} (C12) and \sigdust\ from \citet{Aniano2020} (A20). The horizontal axis and the dashed vertical gray line indicate the ratio of the total \sigdust\ from A20 to that estimated by the C12 model and the mean of the ratio, respectively. The vertical axis indicates the ratio of global \aco\ using \sigdust\ from A20 ($=$\aco(A20)) to that using \sigdust\ estimated by the C12 SED model ($=$\aco(C12)) (top), and the ratio of global DGR using \sigdust\ from A20 ($=$DGR(A20)) to that using \sigdust\ estimated by the C12 SED model ($=$DGR(C12)) (bottom). The solid horizontal black lines indicate the mean values of the \aco(A20)/\aco(C12) ratio and the DGR(A20)/DGR(C12) ratio, respectively.}
 \label{fig:dust_mass_alpha_DGR}
\end{figure}

\begin{figure}
 \begin{center}
  \includegraphics[scale=1.5]{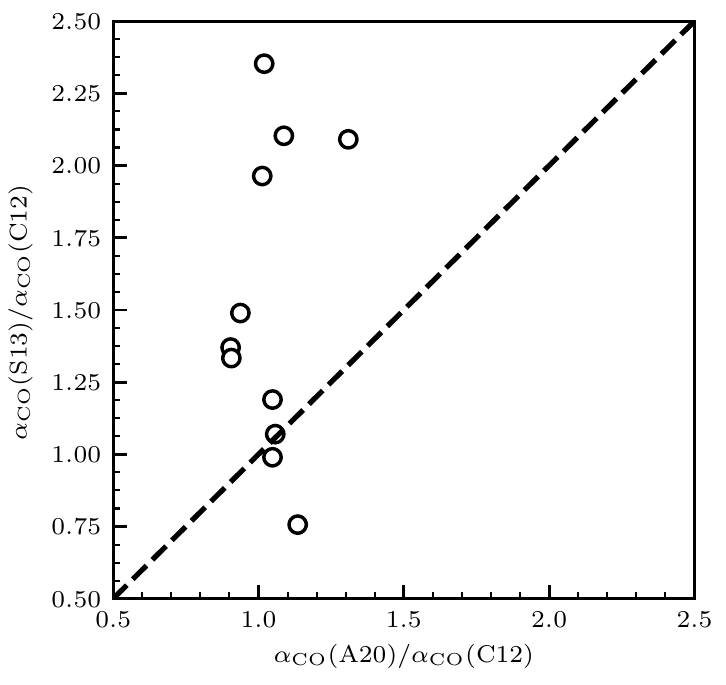}
 \end{center}
 \caption{Comparison of \aco\ measured by S13 \aco(S13)) and that measured by using the SED model of C12 (\aco(C12)) and \sigdust\ from A20 (\aco(A20)). The horizontal axis idicates the ratio of \aco(A20) to \aco(C12). The vertical axis indicates the ratio of \aco(S13) to \aco(C12). The dashed black line indicates \aco(A20)/\aco(C12) = \aco(S13)/\aco(C12).}
 \label{fig:comp_three_alpha}
\end{figure}

We measured \aco(global) and DGR(global) by using \sigdust\ from A20, which were defined as \aco(A20) and DGR(A20), in the same region where \aco(global) and DGR(global) were measured. In this discussion, we define \aco(global) and DGR(global) using \sigdust\ estimated by the C12 SED model as \aco(C12) and DGR(C12). Figure \ref{fig:dust_mass_alpha_DGR} presents the ratio of \aco(A20) to \aco(C12) and that of DGR(A20) to DGR(C12) as a function of the ratio of total \sigdust\ from A20 to that estimated by the C12 SED model for the 11 galaxies. The means of the \aco(A20)/\aco(C12) ratio and the DGR(A20)/DGR(C12) ratio, indicated by the solid horizontal black lines in figure \ref{fig:dust_mass_alpha_DGR}, are 1.04 and 1.84, respectively. Moreover, the mean of the ratio of total \sigdust\ from A20 to that estimated by the C12 SED model, indicated by the dashed vertical gray lines in figure \ref{fig:dust_mass_alpha_DGR}, is 1.88. The variation of the \aco(A20)/\aco(C12) ratio is within 30\%. In this comparison, \sigdust\ increases, while \sigatom\ and \ICO\ are not changed. Considering equation (\ref{eq:alpha_DGR}) of $\Sigma_\mathrm{dust} = \mathrm{DGR} \times (\Sigma_\mathrm{atom} + \alpha_\mathrm{CO} I_\mathrm{CO})$, increasing \sigdust\ affects increasing DGR more than \aco. If \aco\ increases and DGR does not change to compensate for the increase in the total dust mass, the \hi/\hii\ ratio is changed. This causes a change in the contrast of the total gas mass surface density and increases the variation of the DGR, and the $\chi^{2}_{\mathrm{min}}$ value becomes larger. We also compared the ratio of the mean \aco\ from S13 and listed in table \ref{tab:alpha_R21} ($=$\aco(S13)) to \aco(C12) with the \aco(A20)/\aco(C12) ratio as shown in figure \ref{fig:comp_three_alpha}. The variation of \aco(A20)/\aco(C12) ratios tends to be smaller than that of \aco(S13)/\aco(C12) ratios. Therefore, the difference in total dust mass is not the main cause of the difference between our \aco(global) and \aco\ of S13.

\noindent
(2) CO($J=2-1$)/CO($J=1-0$) line ratio $R_{21}$\\
S13 used the integrated intensity of $^{12}$CO($J=2-1$) ($I_{\mathrm{CO(J=2-1)}}$) assuming a constant $R_{21}$ ($=0.7$) to estimate the integrated intensity of $^{12}$CO($J=1-0$) ($I_{\mathrm{CO(J=1-0)}}$). However, if the true $R_{21}$ is larger than 0.7, the integrated intensity of $^{12}$CO($J=1-0$) is overestimated, and \aco\ is underestimated. On the other hand, if true $R_{21}$ is lower than 0.7, the integrated intensity of $^{12}$CO($J=1-0$) is underestimated and \aco\ is overestimated.

We investigated the effect of the variation of $R_{21}$ on the derivation of \aco(global). As a reference to true $R_{21}$ in each galaxy, we used $\overline{R_{21}}$ derived by \citet{Yajima2021}. For the 13 galaxies, we obtained the relation between our \aco(global) and \aco(S13)$\times (\overline{R_{21}}/0.7)$ shown in figure \ref{fig:comp_S13_aco}. Some galaxies have \aco(S13)$\times (\overline{R_{21}}/0.7)$ which is similar to our \aco(global) (NGC\,3184, NGC\,3521, NGC\,3627, NGC\,4254, NGC\,5055, and NGC\,7331). There are also large differences between our \aco(global) and \aco(S13)$\times (\overline{R_{21}}/0.7)$ in other galaxies (NGC\,628, NGC\,4321, NGC\,4536, NGC\,4736, NGC\,5457, NGC\,5713, and NGC\,6946). However, even if \aco(S13)\ is corrected for $R_{21}$, the variations of \aco(S13)$\times (\overline{R_{21}}/0.7)$ are consistent with the uncertainties of our \aco(global) and \aco(S13). Therefore, the assumption of the constant $R_{21}$ cannot be the main cause of the difference between the results obtained in our study and those of S13.

\begin{figure}
 \begin{center}
  \includegraphics[width=\linewidth]{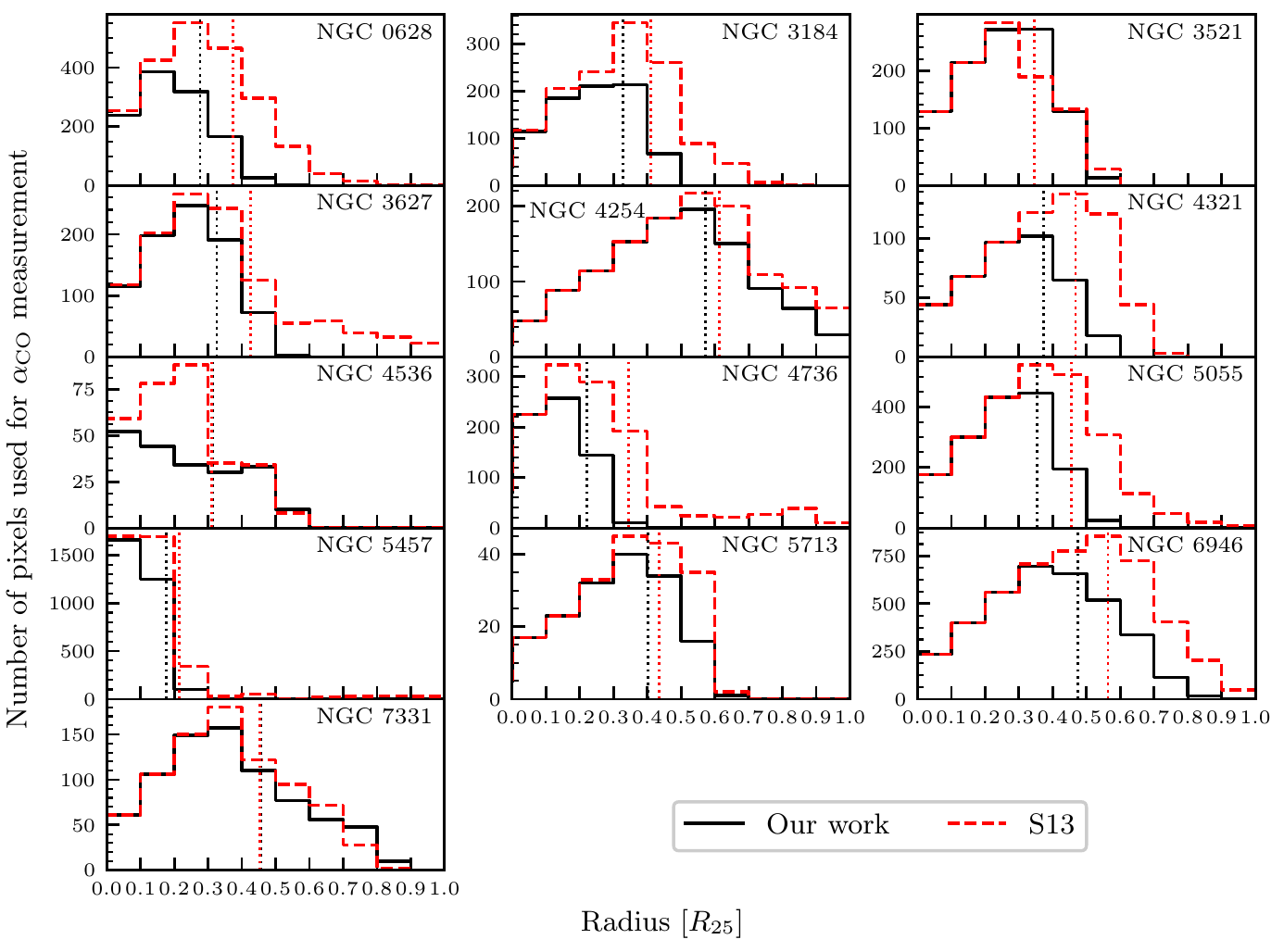}
 \end{center}
 \caption{The radial variation in the number of pixels used for the measurement of \aco\ in each annulus ring with a width = 0.1 $R_{25}$ between our study (solid black) and S13 (dashed red). The dotted black line and the dotted red line indicate the mean radii of the measurement regions of our \aco(global) and \aco\ of S13, respectively.}
 \label{fig:Npix_radial_CO}
\end{figure}

\noindent
(3) Difference in the data sampling area\\
Considering the previous studies that reported the radial variation of \aco, it is expected that \aco\ in the inner region of a galaxy tends to be smaller than that of the outer region. While S13 measured \aco\ in the region where $I_{\mathrm{CO(J=2-1)}}/0.7>1$ K km s$^{-1}$, we measured \aco(global) in the region where $I_{\mathrm{CO(J=1-0)}}>3 \Delta I_{\mathrm{CO(J=1-0)}}$, $I_{\mathrm{H \textsc{i}}}>3 \Delta I_{\mathrm{H \textsc{i}}}$, and an infrared flux density $> 3 \times \rm{RMS}$, where RMS was estimated in the emission-free region of an infrared map. The median and mean of $\Delta I_{\mathrm{CO(J=1-0)}}$ in the CO observation region of our samples listed in table \ref{tab:alpha_DGR_measure} are 0.61 and 0.69 K km s$^{-1}$, respectively. Furthermore, the achieved sensitivity for the S13 samples was better than that of our samples. Thus, it is possible that our measurements of \aco\ are biased toward the inner region where \aco\ tends to be smaller. To confirm the presumption, we compared the data sampling area for the measurements of \aco\ in our study and S13. Figure \ref{fig:Npix_radial_CO} presents the radial variation of data sampling area used for the measurement of our \aco(global) and \aco\ of S13. The vertical axis indicates the radial variation in the number of pixels used for the measurement of \aco\ in each annulus ring with a width $=0.1R_{25}$. While the dotted red line indicates the number of pixels, where $I_{\mathrm{CO(J=2-1)}}/0.7>1$ K km s$^{-1}$, the dotted black line indicates and $I_{\mathrm{CO(J=1-0)}}>3 \Delta I_{\mathrm{CO(J=1-0)}}$, $I_{\mathrm{H \textsc{i}}}>3 \Delta I_{\mathrm{H \textsc{i}}}$, and an infrared flux density $> 3 \times \rm{RMS}$. We used the $I_{\mathrm{CO(J=2-1)}}$ map obtained from the HERACLES project\footnote{\url{https://www.iram.fr/ILPA/LP001/}} and the map of each galaxy was convolved with a Gaussian kernel and matched to the spatial resolution listed in table \ref{tab:HI_CO_list}. The mean radii of the regions used for the measurement of our \aco(global) and \aco\ of S13 for the 13 galaxies are 0.35$R_{25}$ and 0.41$R_{25}$, respectively. This implies that we may measure \aco\ within the radius, which is 0.85 times smaller than the typical radius, where S13 measured \aco. As indicated in subsection \ref{radial_alpha_dgr}, \aco\ tends to be smaller in the inner region of a galactic disk in many galaxies. S13 also reported that \aco\ in the inner regions of their sample galaxies is generally lower than the remaining of the disk. Therefore, the difference in the sampling area is one of the possibilities causing the difference in \aco\ between our study and S13, that is, the value of \aco\ in S13, which is derived from a larger area than that used in our study, can be larger than our value of \aco.

\subsubsection{\aco\ in the Milky Way}\label{comp_with_MW}
In addition to our value of \aco(global) being lower than that of S13, we also found that \aco(global) of the 20 galaxies is lower than the standard value in the Milky Way, $\alpha_{\mathrm{CO}}(\mathrm{MW})=4.35$ \acounit (e.g., \cite{Bolatto2013}). \acomw\ as the typical value of the solar neighborhood has been measured by various methods; virial masses (e.g., \cite{Solomon1987}; \cite{Scoville1987}), CO isotopologues (\cite{Goldsmith2008}), extinction (e.g., \cite{Pineda2008}; \cite{Paradis2012}), dust emission (\cite{Dame2001}; \cite{Planck2011}), and $\gamma$-ray radiation (e.g., \cite{Strong1996}; \cite{Ackermann2012}).

Because our measurements are biased towards the inner region of galaxies, as indicated in subsection \ref{comp_S13}, it can also cause the difference between the value of \aco\ obtained in this study and \acomw. To confirm our presumption, we compared the typical radius where \aco(global) was measured in this study with that of \acomw. We assumed that $R_{25}$ in the Milky Way is twice the distance from the Galactic center to the solar neighborhood $R_{\odot}=8$ kpc (\cite{Bigiel2012}): the radius from the Galactic center to the solar neighborhood is $0.5R_{25}$. We obtained the mean radius of the regions used for the measurement of \aco(global) in the 22 galaxies ($=0.38 R_{25}$) in subsection \ref{result_alpha_dgr}. This result implies that the typical radius where \aco(global) was measured is 0.76 times smaller than the typical radius where \acomw\ has been measured. Previous studies have reported the radial variation of the CO-to-\hii\ conversion factor in the Milky Way. \citet{Strong2004} measured the radial variation of \xco\ in the Milky Way by the $\gamma$-ray observations and found $X_{\mathrm{CO}}/10^{20} \approx 0.4$ \xcounit\ ($\alpha_{\mathrm{CO}} \approx 0.9$ \acounit) at $R \sim 2-3\ \mathrm{kpc}$ and $X_{\mathrm{CO}}/10^{20} \approx 10$ \xcounit\ ($\alpha_{\mathrm{CO}} \approx 21.7$ \acounit) at $R > 10\ \mathrm{kpc}$ in the Milky Way. \citet{Sodroski1995} obtained $X_{\mathrm{CO}}/10^{20} \sim (0.2-0.7)$ \xcounit\ ($\alpha_{\mathrm{CO}} \sim 0.4 - 1.5$ \acounit) from 140 $\mu$m, 240 $\mu$m, and $^{12}$CO($J=1-0$) observations within approximately 400 pc of the galactic center and found that their \xco\ is 3--10 times lower than that in the galactic disk at $R=4.5-13$ kpc obtained from previous studies. They also obtained the least-square fit of \xco\ as a function of galactocentric distance $D_\mathrm{G}$ in the unit of kpc by fitting their observations and literature measurements as follows:

\begin{equation}
\left( \frac{X_\mathrm{CO}}{\mathrm{cm^{-2}(K\ km\ s^{-1})^{-1}}} \right) = 10^{0.12D_\mathrm{G}-0.34} \times 10^{20}.
\label{eq:xco_sodroski1995}
\end{equation}

\noindent
These results indicate that \aco\ in the center region of the Milky Way is smaller than \acomw\ in the solar neighborhood. Therefore, our \aco\ measurements are biased towards the inner region compared to \acomw. This is one of the possibilities leading to the lower value of our \aco\ than \acomw. Furthermore, the use of \acomw\ within a region where CO emission is detected in our samples leads to an overestimation of molecular gas mass surface density.

We also compare \aco(inner) listed in table \ref{tab:alpha_DGR_radial} ($= 0.25-0.45$ \acounit) with that in the Milky Way. We investigated the typical radius of \aco(inner) in the inner region by averaging the radii over the pixels where \aco(inner) is measured and weighting with \ICO\ in subsection \ref{alpha_dgr_SFR_metal}. The mean typical radius of \aco(inner) for four galaxies is $0.10R_{25}$. Therefore, we compare \aco(inner) with that at $0.10R_{25}$ ($=0.10\times8.0/0.5=1.60$ kpc) in the Milky Way. \xco\ of equation (\ref{eq:xco_sodroski1995}) of \citet{Sodroski1995} at 1.60 kpc is $7.1\times10^{19}$ \xcounit, which corresponds to $\alpha_\mathrm{CO} = 1.5$ \acounit. \citet{Pineda2013} surveyed the Galactic plane in the [C\,{\sc ii}] 158 $\mu$m line with the $Herschel$ and obtained the radial variation of \xco\ by using the \hii\ column density traced by CO and [C\,{\sc ii}]. From figure 20 of \citet{Pineda2013}, we can find the plots of $X_\mathrm{CO} \approx 10^{19.7-19.8}$ \xcounit, which corresponds to $\alpha_\mathrm{CO} \approx 1.1-1.4$ \acounit, at $R_\mathrm{gal}=1-2$ kpc. Therefore, our \aco(inner) is smaller than \aco\ in the Milky Way even if we compare them at the similar distance from the galactic center.

\subsubsection{DGR in the Milky Way}
As described in subsection \ref{comp_with_MW}, the typical radius at which \aco(global) and DGR(global) were measured in our study in the 22 nearby galaxies ($=0.38R_{25}$) is different from that of the solar neighborhood ($=0.5R_{25}$, \cite{Bigiel2012}). Using equation (\ref{eq:gdr_giannetti}) and defining the radius of the solar neighborhood (= 8 kpc = $0.5R_{25}$), the estimated DGR of the Milky Way at 0.38$R_{25}$ is $0.0107^{+0.0049}_{-0.0011}$. Furthermore, a linear fit of the DGR as a function of a galactocentric radius for nearby star-forming galaxies as shown in figure 12 of S13 indicates a larger DGR at $0.38R_{25}$ than the typical value of DGR for the solar neighborhood ($=0.010$). These values are larger than the average of DGR(global) obtained in our study in the 22 nearby galaxies ($= 0.0052 \pm 0.0026$). Note that the absolute value of DGR highly depends on the model adopted for SED fitting, whereas the trend of the radial variation is not affected. In subsection \ref{comp_S13}, we found that the mean and standard deviation of DGR(global) using \sigdust\ from A20 in the 11 galaxies shown in figure \ref{fig:dust_mass_alpha_DGR} is $0.0090 \pm 0.0045$, which is closer to the results of S13 and \citet{Giannetti2017} than the mean of DGR(global) using \sigdust\ estimated by the C12 SED model.

\section{Conclusions}
We simultaneously measured the spatially-resolved CO-to-\hii\ conversion factor \aco\ and DGR in nearby spiral galaxies on a kiloparsec scale by using the method developed by \citet{Leroy2011} and \citet{Sandstrom2013} (S13) with the spatially-resolved $^{12}$CO($J=1-0$) mapping data from the COMING and the CO Atlas projects. The conclusions of this study are summarized as follows:
\begin{enumerate}
\item We obtained global averaged values of \aco\ and DGR, \aco(global) and DGR(global), for 22 nearby spiral galaxies. The averages and standard deviations of \aco(global) and DGR(global) in the 22 spiral galaxies are $2.66 \pm 1.36$ \acounit\ and $0.0052 \pm 0.0026$, respectively. The average of \aco(global) is lower than not only the standard value in the Milky Way $\alpha_{\mathrm{CO}}(\mathrm{MW})=4.35$ \acounit\ (e.g., \cite{Bolatto2013}), but also the average ($= 3.1$ \acounit) of \aco\ in S13 (\aco(S13)). We divided the measurement regions for \aco\ and DGR into the inner and outer regions with a boundary at $0.2R_{25}$, and obtained the radial variations of \aco\ and DGR in the following four barred spiral galaxies: IC\,342, NGC\,3627, NGC\,5236, and NGC\,6946. In these four galaxies, the averages and standard deviations of \aco\ and DGR in the inner region ($\leq 0.2R_{25}$) are $0.36 \pm 0.08$ \acounit\ and $0.0199 \pm 0.0058$, while those in the outer region ($> 0.2R_{25}$) are $1.49 \pm 0.76$ \acounit\ and $0.0084 \pm 0.0037$, respectively. We found that \aco\ in the outer region is 2.3 to 5.3 times larger than that in the inner region.
\item We found a clearer correlation of \aco\ and DGR with metallicity and SFR surface density using the data separated into inner and outer regions than that of the global region. The correlation between \aco\ and SFR surface density is expected by theoretical studies in which the decrease of \aco\ is attributed to the increase in the CO integrated intensity owing to a high kinetic temperature and large velocity dispersion in high SFR regions.
\item  Using the mean $^{12}$CO($J=2-1$)/$^{12}$CO($J=1-0$) line ratio ($R_{21}$) from \citet{Yajima2021}, we corrected \aco(S13), which was derived using the constant $R_{21}(=0.7)$. However, the variations of \aco(S13) corrected for $R_{21}$ are consistent with the uncertainties of our \aco(global) and \aco(S13). We conclude that the constant $R_{21}$ cannot be the main cause of the difference between \aco(global) and \aco(S13). The mean radius of the regions used for the measurement of \aco(global) in the 22 nearby galaxies is $0.38R_{25}$, which is smaller than the radius of the solar neighborhood ($=0.5R_{25}$, \cite{Bigiel2012}). On the other hand, owing to the difference in the data sampling condition between our study and S13, the mean radii of the regions used for the measurement of \aco\ in the 13 galaxies in common with our study and S13 are $0.35R_{25}$ and $0.41R_{25}$, respectively. These results imply that we may measure \aco\ within a radius smaller than that of S13 and \acomw. This is one of the possibilities leading to the smaller \aco(global) in our study compared to S13 and \acomw. In our samples, the use of \acomw\ within a region where CO emission is detected leads to an overestimation of molecular gas mass surface density.
\end{enumerate}

\section*{Acknowledgments}
We are grateful to Nobeyama Radio Observatory staff for setting up and operating the Nobeyama 45-m telescope system and for supporting our COMING project. The Nobeyama 45-m radio telescope is operated by the Nobeyama Radio Observatory, a branch of the National Astronomical Observatory of Japan. This work utilized CO Atlas; Nobeyama CO Atlas of Nearby Spiral Galaxies (\cite{Kuno2007}), HERACLES; the HERA CO-Line Extragalactic Survey (\cite{Leroy2009}), THINGS; The \hi\ Nearby Galaxies Survey (\cite{Walter2008}), VIVA; VLA Imaging of Virgo in Atomic gas (\cite{Chung2009}), DustPedia (\cite{Davies2017}), and the Local Volume Legacy (\cite{Dale2009}). This work was supported by the Japan Society for the Promotion of Science KAKENHI (Grant Nos. 18K13593 and 20K04008) and MEXT Leading Initiative for Excellent Young Researchers (HJH02007). This research has also made use of the NASA/IPAC Extragalactic Database, which is operated by the Jet Propulsion Laboratory, California Institute of Technology, under contract with the National Aeronautics and Space Administration, and the NRAO Science Data Archive.

\clearpage
\begin{table*}[p!]
\begin{center}
\tbl{Parameters of the sample galaxies.}{
\begin{tabular}{llccccccccc} \hline
Galaxy & Morphology & Distance [Mpc] & PA [$^{\circ}$] & $i$ [$^{\circ}$] & $R_{25}$ [$^{\prime}$]\\ \hline
\multicolumn{6}{c}{Isolated galaxies}\\ \hline
NGC\,337 & SB(s)d & 18.90 & 119.6 & 44.5 & 1.44\\
NGC\,628 & SA(s)c & 9.02 & 20 & 7 & 5.24\\
IC\,342 & SAB(rs)cd & 3.3 & 48 & 31 & 10.69\\
NGC\,1569 & Ibm & 3.25 & 112 & 63 & 1.82\\
NGC\,2841 & SA(r)b? & 14.60 & 152.6 & 73.7 & 4.06\\
NGC\,2976 & SAc pec & 3.63 & -25.5 & 64.5 & 2.94\\
NGC\,3077 & I0 pec & 3.81 & 63.8 & 38.9 & 2.69\\
NGC\,3147 & SA(rs)bc & 39.30 & 142.79 & 35.19 & 1.95\\
NGC\,3184 & SB(s) & 8.7 & 6 & 21 & 3.71\\
NGC\,3198 & SB(rs)c & 13.40 & -145.0 & 71.5 & 4.26\\
Mrk\,33 & Im pec? & 24.9 & 124.0 & 42.6 & 0.50\\
NGC\,3351 & SB(r)b & 10.7 & -168 & 41 & 3.71\\
NGC\,3521 & SAB(rs)bc & 14.20 & -19 & 63 & 5.48\\
NGC\,3627 & SAB(s)b & 9.04 & 176 & 52 & 4.56\\
NGC\,3938 & SA(s)c & 17.9 & -154.0 & 20.9 & 2.69\\
NGC\,4030 & SA(s)bc & 29.90 & 29.6 & 39.0 & 2.08\\
NGC\,4192 & SAB(s)ab & 16.1 & 155 & 79 & 4.89\\
NGC\,4254 & SA(s)c & 16.1 & 66 & 42 & 2.69\\
NGC\,4321 & SAB(s)bc & 16.1 & 158 & 27 & 3.71\\
NGC\,4501 & SA(rs)b & 16.1 & 139 & 68 & 3.46\\
NGC\,4527 & SAB(s)bc & 16.5 & 69.5 & 70 & 3.08\\
NGC\,4535 & SAB(s)c & 16.1 & 2 & 44 & 3.54\\
NGC\,4536 & SAB(rs)bc & 16.5 & -54.5 & 64.2 & 3.79 \\
NGC\,4569 & SAB(rs)ab & 16.1 & 22 & 64 & 4.77 \\
NGC\,4579 & SAB(rs)b & 16.5 & 92.1 & 41.7 & 2.94\\
NGC\,4654 & SAB(rs)cd &  16.1 & 124 & 51 & 0.24 \\
NGC\,4689 & SA(rs)bc & 16.1 & 164 & 36 & 2.13\\
NGC\,4736 & (R)SA(r)ab & 4.3 & 119 & 40 & 5.61 \\
NGC\,5055 & SA(rs)bc & 9.04 & 98 & 61 & 6.29\\
NGC\,5236 & SAB(s)c & 4.5 & 45 & 24 & 6.44\\
NGC\,5248 & SAB(rs)bc & 13.00 & 103.9 & 38.6 & 3.08\\
NGC\,5364 & SA(rs)bc pec & 18.2 & -144.4 & 47.9 & 3.38\\
NGC\,5457 & SAB(rs)cd & 7.2 & 42 & 18 & 14.42\\
\hline
\end{tabular}}
\end{center}
\end{table*}

\clearpage
\setcounter{table}{0}
\begin{table*}[p!]
\begin{center}
\tbl{(Continued).}{
\begin{tabular}{llcccccccc} \hline
Galaxy & Morphology & Distance [Mpc] & PA [$^{\circ}$] & $i$ [$^{\circ}$] & $R_{25}$ [$^{\prime}$]\\ \hline
\multicolumn{6}{c}{Isolated galaxies}\\ \hline
NGC\,5713 & SAB(rs)bc pec & 19.5 & -157 & 33 & 1.38\\
NGC\,6946 & SAB(rs)cd & 5.5 & 242 & 40 & 5.74\\
NGC\,7331 & SA(s)b & 13.90 & 167.7 & 75.8 & 5.24\\ \hline
\multicolumn{6}{c}{Interacting galaxies}\\ \hline
NGC\,4298/NGC\,4302 & & & & & \\
\,\,NGC\,4298 & SA(rs)c & 16.5 & -48.9 & 54.8 & 1.62\\
VV 219 & & & & & \\
\,\,NGC\,4567 & SA(rs)bc & 16.5 & 80 & 46 & 1.48\\
\,\,NGC\,4568 & SA(rs)bc & 16.5 & 23 & 64 & 2.29\\
Arp\,116 & & & & & \\
\,\,NGC\,4647 & SAB(rs)c & 16.5 & 98.5 & 39.3 & 1.44\\ \hline
\end{tabular}}
\label{tab:galaxy_list}
\begin{tabnote}
Note. Morphology, distance, position angle (PA), and inclination angle $i$ were obtained from \citet{Sorai2019} and \citet{Kuno2007}. $R_{25}$ of COMING and CO Atlas samples was obtained from \citet{Sorai2019} and NED Database, respectively.
\end{tabnote}
\end{center}
\end{table*}

\clearpage
\begin{table*}[p!]
\begin{center}
\tbl{Summary of $^{12}$CO($J=1-0$) and \hi\ data.}{
\begin{tabular}{lccccc} \hline
  \multirow{2}{*}{Galaxy} & \multicolumn{2}{c}{Reference} & \multicolumn{2}{c}{Adopted spatial resolution} & \multirow{2}{*}{\begin{tabular}[c]{@{}l@{}}Spatial scale of \\ beam size {[}kpc{]}\end{tabular}} \\ \cline{2-5} & $^{12}$CO($J=1-0$) & \hi & Beam size [$^{\prime \prime}$] & Pixel size [$^{\prime \prime}$] \\  \hline
\multicolumn{6}{c}{Isolated galaxies}\\ \hline
NGC\,337  & COMING & N & 29.6 $\times$ 21.1 & 6.0 & 2.7 $\times$ 1.9 \\
NGC\,628  & COMING & T & 18.0 $\times$ 18.0 & 6.0 & 0.8 $\times$ 0.8 \\
IC\,342 & CO Atlas & C00 & 38.0 $\times$ 37.0 & 6.1 & 0.6 $\times$ 0.6 \\
NGC\,1569 & COMING & T & 18.0 $\times$ 18.0 & 6.0 & 0.3 $\times$ 0.3 \\
NGC\,2841 & COMING & T & 18.0 $\times$ 18.0 & 6.0 & 1.3 $\times$ 1.3 \\
NGC\,2976 & COMING & T & 18.0 $\times$ 18.0 & 6.0 & 0.3 $\times$ 0.3 \\
NGC\,3077 & COMING & T & 18.0 $\times$ 18.0 & 6.0 & 0.3 $\times$ 0.3 \\
NGC\,3147 & COMING & N & 25.3 $\times$ 22.0 & 6.0 & 4.8 $\times$ 4.8 \\
NGC\,3184 & CO Atlas & T & 18.0 $\times$ 18.0 & 6.0 & 0.8 $\times$ 0.8 \\
NGC\,3198 & COMING & T & 18.0 $\times$ 18.0 & 6.0 & 1.2 $\times$ 1.2 \\
Mrk\,33 & COMING & N & 18.0 $\times$ 18.0 & 6.0 & 2.2 $\times$ 2.2 \\
NGC\,3351 & COMING & T & 18.0 $\times$ 18.0 & 6.0 & 0.9 $\times$ 0.9 \\
NGC\,3521 & CO Atlas & T & 18.0 $\times$ 18.0 & 6.0 & 1.2 $\times$ 1.2 \\
NGC\,3627 & COMING & T & 18.0 $\times$ 18.0 & 6.0 & 0.8 $\times$ 0.8 \\
NGC\,3938 & COMING & N & 18.8 $\times$ 18.2 & 6.0 & 1.6 $\times$ 1.6 \\
NGC\,4030 & COMING & N & 18.0 $\times$ 18.0 & 6.0 & 2.6 $\times$ 2.6 \\
NGC\,4192 & CO Atlas & V & 28.0 $\times$ 25.9 & 10.0 & 2.2 $\times$ 2.0 \\
NGC\,4254 & CO Atlas & V & 26.8 $\times$ 24.5 & 6.0 & 2.1 $\times$ 1.9 \\
NGC\,4321 & CO Atlas & V & 31.1 $\times$ 28.1 & 10.0 & 2.4 $\times$ 2.2 \\
NGC\,4501 & CO Atlas & V & 18.0 $\times$ 18.0 & 6.0 & 1.4 $\times$ 1.4 \\
NGC\,4527 & COMING & N & 18.0 $\times$ 18.0 & 6.0 & 1.4 $\times$ 1.4 \\
NGC\,4535 & CO Atlas & V & 25.0 $\times$ 24.1 & 10.0 & 2.0 $\times$ 1.9 \\
NGC\,4536 & COMING & N & 18.0 $\times$ 18.0 & 6.0 & 1.4 $\times$ 1.4 \\
NGC\,4569 & CO Atlas & V & 18.0 $\times$ 18.0 & 6.0 & 1.4 $\times$ 1.4 \\
NGC\,4579 & COMING & V & 42.4 $\times$ 34.5 & 10.0 & 3.4 $\times$ 2.8 \\
NGC\,4654 & CO Atlas & V & 18.0 $\times$ 18.0 & 6.0 & 1.4 $\times$ 1.4 \\
NGC\,4689 & CO Atlas & V & 18.0 $\times$ 18.0 & 6.0 & 1.4 $\times$ 1.4 \\
NGC\,4736 & CO Atlas & T & 18.0 $\times$ 18.0 & 6.0 & 0.4 $\times$ 0.4 \\
NGC\,5055 & COMING & T & 18.0 $\times$ 18.0 & 6.0 & 0.8 $\times$ 0.8 \\
NGC\,5236 & CO Atlas & T & 18.0 $\times$ 18.0 & 6.0 & 0.4 $\times$ 0.4 \\
NGC\,5248 & COMING & N & 18.0 $\times$ 18.0 & 6.0 & 1.1 $\times$ 1.1 \\
NGC\,5364 & COMING & N & 18.0 $\times$ 18.0 & 6.0 & 1.6 $\times$ 1.6 \\
NGC\,5457 & CO Atlas & T & 18.0 $\times$ 18.0 & 6.0 & 0.6 $\times$ 0.6 \\
\hline
\end{tabular}}
\end{center}
\end{table*}

\clearpage
\setcounter{table}{1}
\begin{table*}[p!]
\begin{center}
\tbl{(Continued).}{
\begin{tabular}{lccccc} \hline
  \multirow{2}{*}{Galaxy} & \multicolumn{2}{c}{Reference} & \multicolumn{2}{c}{Adopted spatial resolution} & \multirow{2}{*}{\begin{tabular}[c]{@{}l@{}}Spatial scale of \\ beam size {[}kpc{]}\end{tabular}} \\ \cline{2-5} & $^{12}$CO($J=1-0$) & \hi & Beam size [$^{\prime \prime}$] & Pixel size [$^{\prime \prime}$] \\  \hline
\multicolumn{6}{c}{Isolated galaxies}\\ \hline
NGC\,5713 & COMING & N & 18.0 $\times$ 18.0 & 6.0 & 1.7 $\times$ 1.7 \\
NGC\,6946 & CO Atlas & T & 18.0 $\times$ 18.0 & 6.0 & 0.5 $\times$ 0.5 \\
NGC\,7331 & COMING & T & 18.0 $\times$ 18.0 & 6.0 & 1.2 $\times$ 1.2 \\ \hline
\multicolumn{6}{c}{Interacting galaxies}\\ \hline
NGC\,4298 & COMING & V & 18.0 $\times$ 18.0 & 6.0 & 1.4 $\times$ 1.4\\
NGC\,4567/NGC\,4568 & COMING & V & 18.0 $\times$ 18.0 & 6.0 & 1.4 $\times$ 1.4\\
NGC\,4647 & COMING & N & 18.0 $\times$ 18.0 & 6.0 & 1.4 $\times$ 1.4\\ \hline
\end{tabular}}
\label{tab:HI_CO_list}
\begin{tabnote}
  Note. \hi\ data are obtained from THINGS (T, \cite{Walter2008}), VIVA (V, \cite{Chung2009}) and NRAO Science Data Archive (N, https://archive.nrao.edu/archive/advquery.jsp). \hi\ data of IC 342 is used for \citet{Crosthwaite2000} (C00) obtained from NED Database.
\end{tabnote}
\end{center}
\end{table*}

\begin{table}[hbt]
\begin{center}
\tbl{Results of \aco, DGR, and properties.}{
\begin{tabular}{lccccc}
\hline
Galaxy & \aco\ ($\Delta$\aco)  & DGR ($\Delta$DGR) & $R_{\mathrm{max}}$ & $\overline{\Sigma_{\mathrm{SFR}}}$ & $\overline{Z}$ \\
 & [\acounit] &  & [$\times R_{25}$] & [$10^{-2}\ M_{\odot}\ \mathrm{yr^{-1}\ kpc^{-2}}$] &  \\
 & (1) & (2) & (3) & (4) & (5) \\ \hline
NGC\,628 & 3.94 (1.50 -- 17.44) & 0.0033 (0.0010 -- 0.0061) & 0.61 & 0.69 & 8.64\\
IC\,342 & 1.72 (1.33 -- 2.30) & 0.0057 (0.0047 -- 0.0066) & 0.88 & 1.03 & 8.63\\
NGC\,3147 & 2.91 (1.05 -- 11.59) & 0.0058 (0.0019 -- 0.0114) & 0.79 & 0.68 & ---$^{**}$\\
NGC\,3184 & 2.52 (0.70 -- 12.17) & 0.0046 (0.0015 -- 0.0083) & 0.58 & 0.50 & 8.52\\
NGC\,3521 & 3.87 (1.84 -- 9.56) & 0.0026 (0.0013 -- 0.0041) & 0.66 & 0.80 & 8.52\\
NGC\,3627 & 0.51 (0.20 -- 1.21) & 0.0108 (0.0070 -- 0.0150) & 0.61 & 1.89 & ---$^{**}$ \\
NGC\,4030 & 3.69 (1.49 -- 15.20) & 0.0039 (0.0012 -- 0.0074) & 0.81 & 1.34 & 8.65 \\
NGC\,4254 & 4.55 (2.34 -- 10.15) & 0.0023 (0.0013 -- 0.0035) & 1.09 & 1.13 & 8.54\\
NGC\,4321 & 1.95 (0.64 -- 7.08) & 0.0069 (0.0028 -- 0.0124) & 0.66 & 0.97 & 8.66\\
NGC\,4501 & 3.05 (0.75 -- 10.53) & 0.0052 (0.0025 -- 0.0085) & 1.03 & 0.41 & 8.67\\
NGC\,4527 & 1.87 (0.54 -- 6.53) & 0.0037 (0.0018 -- 0.0062) & 0.82 & ---$^{*}$ & ---$^{**}$\\
NGC\,4536 & 2.43 (0.52 -- 13.65) & 0.0035 (0.0012 -- 0.0069) & 0.65 & 2.10 & ---$^{**}$\\
NGC\,4654 & 3.47 (0.68 -- 17.43) & 0.0035 (0.0012 -- 0.0067) & 0.77 & 0.93 & 8.52\\
NGC\,4736 & 0.73 (0.18 -- 2.32) & 0.0068 (0.0039 -- 0.0101) & 0.44 & 2.29 & 8.56\\
NGC\,5055 & 3.11 (1.59 -- 6.85) & 0.0048 (0.0027 -- 0.0073) & 0.72 & 0.59 & 8.70\\
NGC\,5236 & 0.84 (0.60 -- 1.23) & 0.0137 (0.0106 -- 0.0169) & 0.56 & 4.98 & 8.71\\
NGC\,5248 & 2.32 (1.03 -- 6.32) & 0.0043 (0.0021 -- 0.0070) & 0.78 & 1.26 & 8.61\\
NGC\,5457 & 3.04 (1.88 -- 5.00) & 0.0046 (0.0033 -- 0.0061) & 0.40 & 0.75 & 8.55\\
NGC\,5713 & 2.20 (0.24 -- 17.05) & 0.0034 (0.0008 -- 0.0074) & 0.71 & 4.19 & ---$^{**}$\\
NGC\,6946 & 1.50 (1.09 -- 2.13) & 0.0062 (0.0050 -- 0.0075) & 0.94 & 1.37 & 8.57\\
NGC\,7331 & 6.58 (2.62 -- 22.64) & 0.0027 (0.0010 -- 0.0049) & 0.96 & 0.40 & 8.58\\
NGC\,4568 & 1.72 (0.40 -- 9.30) & 0.0071 (0.0021 -- 0.0143) & 0.67 & 1.13 & ---$^{**}$\\ \hline
\end{tabular}}
\label{tab:alpha_DGR_measure}
\begin{tabnote}
(1)--(2) Measured global \aco\ and DGR. $\Delta \alpha_{\mathrm{CO}}$ and $\Delta$DGR indicate the range between the minimum and maximum values of \aco\ and DGR at $\chi^{2} = \chi^{2}_{\mathrm{min}}+2.3$. (3) The maximum radius within which \aco\ and DGR in columns (1) and (2) are measured. (4)--(5) Mean of the star formation rate surface density and metallicity ($12+\log(\rm{O/H})$ in equation (\ref{eq:metal})) over pixels where global \aco\ and DGR are measured.\\
$^{*}$ Archived data of 24 $\mu$m is unavailable.\\
$^{**}$ The lack of metallicity data in \citet{Pilyugin2014}.
\end{tabnote}
\end{center}
\end{table}

\begin{table}[hbt]
\begin{center}
\tbl{Radial variation of \aco, DGR, and properties.}{
\begin{tabular}{lccccc}
\hline
Galaxy & region & \aco\ ($\Delta$\aco)  & DGR ($\Delta$DGR) & $\overline{\Sigma_{\mathrm{SFR}}}$ & $\overline{Z}$ \\
 &  & [\acounit] &  & [$10^{-2}\ M_{\odot}\ \mathrm{yr^{-1}\ kpc^{-2}}$] & \\
 &  & (1) & (2) & (3) & (4) \\ \hline
IC\,342 & inner & 0.45 (0.23 -- 1.23) & 0.0171 (0.0074 -- 0.0278) & 4.50 & 8.76 \\
       & outer & 2.39 (1.61 -- 3.65) & 0.0047 (0.0035 -- 0.0060) & 0.53 & 8.61 \\
NGC\,3627 & inner & 0.25 (0.03 -- 2.11) & 0.0157 (0.0039 -- 0.0302) & 3.18 & ---$^{*}$ \\
         & outer & 0.69 (0.16 -- 2.32) & 0.0096 (0.0049 -- 0.0149) & 1.60 & ---$^{*}$ \\
NGC\,5236 & inner & 0.34 (0.18 -- 0.80) & 0.0299 (0.0151 -- 0.0471) & 15.35 & 8.75 \\
         & outer & 0.78 (0.43 -- 1.42) & 0.0139 (0.0098 -- 0.0183) & 2.85 & 8.70 \\
NGC\,6946 & inner & 0.41 (0.17 -- 1.77) & 0.0167 (0.0051 -- 0.0298) & 6.17 & 8.68 \\
         & outer & 2.09 (1.27 -- 3.49) & 0.0053 (0.0039 -- 0.0069) & 0.91 & 8.56 \\ \hline
\end{tabular}}
\label{tab:alpha_DGR_radial}
\begin{tabnote}
(1)--(2) The measured \aco\ and DGR in the inner ($r \leq 0.2R_{25}$) and outer ($0.2R_{25} < r \leq R_{\mathrm{max}}$, where $ R_{\mathrm{max}}$ is the same as column (3) in table \ref{tab:alpha_DGR_measure} region. $\Delta \alpha_{\mathrm{CO}}$ and $\Delta$DGR indicate the range between the minimum and maximum values of \aco\ and DGR at $\chi^{2} = \chi^{2}_{\mathrm{min}}+2.3$. (3)-(4) Mean of star formation rate surface density and metallicity ($12+\log(\rm{O/H})$ in equation (\ref{eq:metal})) over pixels where radial \aco\ and DGR are measured in the inner and outer regions, respectively.\\
$^{*}$ The lack of metallicity data in \citet{Pilyugin2014}.
\end{tabnote}
\end{center}
\end{table}

\begin{table}[hbt]
\begin{center}
\tbl{\aco\ and $R_{21}$ in galaxies which are consistent with S13.}{
\begin{tabular}{lcccc}
\hline
Galaxy & \aco\ ($\Delta \alpha_{\mathrm{CO}}$) & \aco(S13) (Std. Dev.) & $\overline{R_{21}}$ & \aco(S13)$\times (\overline{R_{21}}/0.7)$ \\
 & [\acounit] & [\acounit] & & [\acounit] \\
 & (1) & (2) & (3) & (4) \\ \hline
NGC\,628 & 3.94 (1.50 -- 17.44) & 3.9 (2.1 -- 7.4) & $0.54 \pm 0.14$ & $3.00 \pm 0.78$ \\
NGC\,3184 & 2.52 (0.70 -- 12.17) & 5.3 (2.9 -- 9.5) & $0.56 \pm 0.14$ & $4.24 \pm 1.06$ \\
NGC\,3521 & 3.87 (1.84 -- 9.56) & 7.6 (4.9 -- 11.8) & $0.63 \pm 0.12$ & $6.84 \pm 1.30$ \\
NGC\,3627 & 0.51 (0.20 -- 1.21) & 1.2 (0.4 -- 3.3) & $0.46 \pm 0.10$ & $0.79 \pm 0.17$ \\
NGC\,4254 & 4.55 (2.34 -- 10.15) & 3.4 (2.1 -- 5.7) & $0.72 \pm 0.17$ & $3.50 \pm 0.83$ \\
NGC\,4321 & 1.95 (0.64 -- 7.08) & 2.2 (1.1 -- 4.6) & $0.83 \pm 0.18$ & $2.61 \pm 0.57$ \\
NGC\,4536 & 2.43 (0.52 -- 13.65) & 2.6 (1.0 -- 6.7) & $0.84 \pm 0.24$ & $3.12 \pm 0.89$ \\
NGC\,4736 & 0.73 (0.18 -- 2.32) & 1.0 (0.5 -- 2.0) & $0.88 \pm 0.24$ & $1.26 \pm 0.34$ \\
NGC\,5055 & 3.11 (1.59 -- 6.85) & 3.7 (1.9 -- 7.4) & $0.56 \pm 0.11$ & $2.96 \pm 0.58$ \\
NGC\,5457 & 3.04 (1.88 -- 5.00) & 2.3 (1.1 -- 4.8) & $0.64 \pm 0.18$ & $2.10 \pm 0.59$ \\
NGC\,5713 & 2.20 (0.24 -- 17.05) & 4.6 (1.7 -- 12.6) & $0.80 \pm 0.24$ & $5.25 \pm 1.58$ \\
NGC\,6946 & 1.50 (1.09 -- 2.13) & 2.0 (0.9 -- 4.4) & $0.71 \pm 0.17$ & $2.02 \pm 0.49$ \\
NGC\,7331 & 6.59 (2.62 -- 22.64) & 9.8 (6.2 -- 15.3) & $0.55 \pm 0.17$ & $7.70 \pm 2.38$ \\ \hline
\end{tabular}}
\label{tab:alpha_R21}
\begin{tabnote}
(1) Global \aco\ in this study. $\Delta \alpha_{\mathrm{CO}}$ is the range between the minimum and maximum values of \aco\ at $\chi^{2} = \chi^{2}_{\mathrm{min}}+2.3$. (2) Mean \aco\ and the standard deviation listed in table 4 of \citet{Sandstrom2013} (S13). (3) Mean of $R_{21}$ weighted by the integrated intensity of $^{12}$CO($J=1-0$) and standard deviation of $R_{21}$ measured in \citet{Yajima2021}. (4) Corrected \aco(S13) by multiplying the ratio of $\overline{R_{21}}$ to $R_{21} = 0.7$, which is the constant value S13 adopted.
\end{tabnote}
\end{center}
\end{table}

\clearpage

\appendix

\section{Maps used for the measurement of \aco(global) and DGR(global) and results of $\chi^{2}$ fitting for \aco(global) and DGR(global)} \label{maps_PP}
This appendix presents maps of the integrated intensity of $^{12}$CO($J=1-0$) (\ICO), the atomic gas mass surface density (\sigatom), the characteristic cold dust temperature ($T_{\mathrm{dust}}$) derived from the SED model of \citet{Casey2012}, the dust mass surface density (\sigdust), and the star formation rate surface density (\sigSFR). Furthermore, the result of $\chi^{2}$ fitting for \aco(global) and DGR(global) in each galaxy is also shown.

\begin{figure*}
 \begin{center}
  \includegraphics[width=\linewidth]{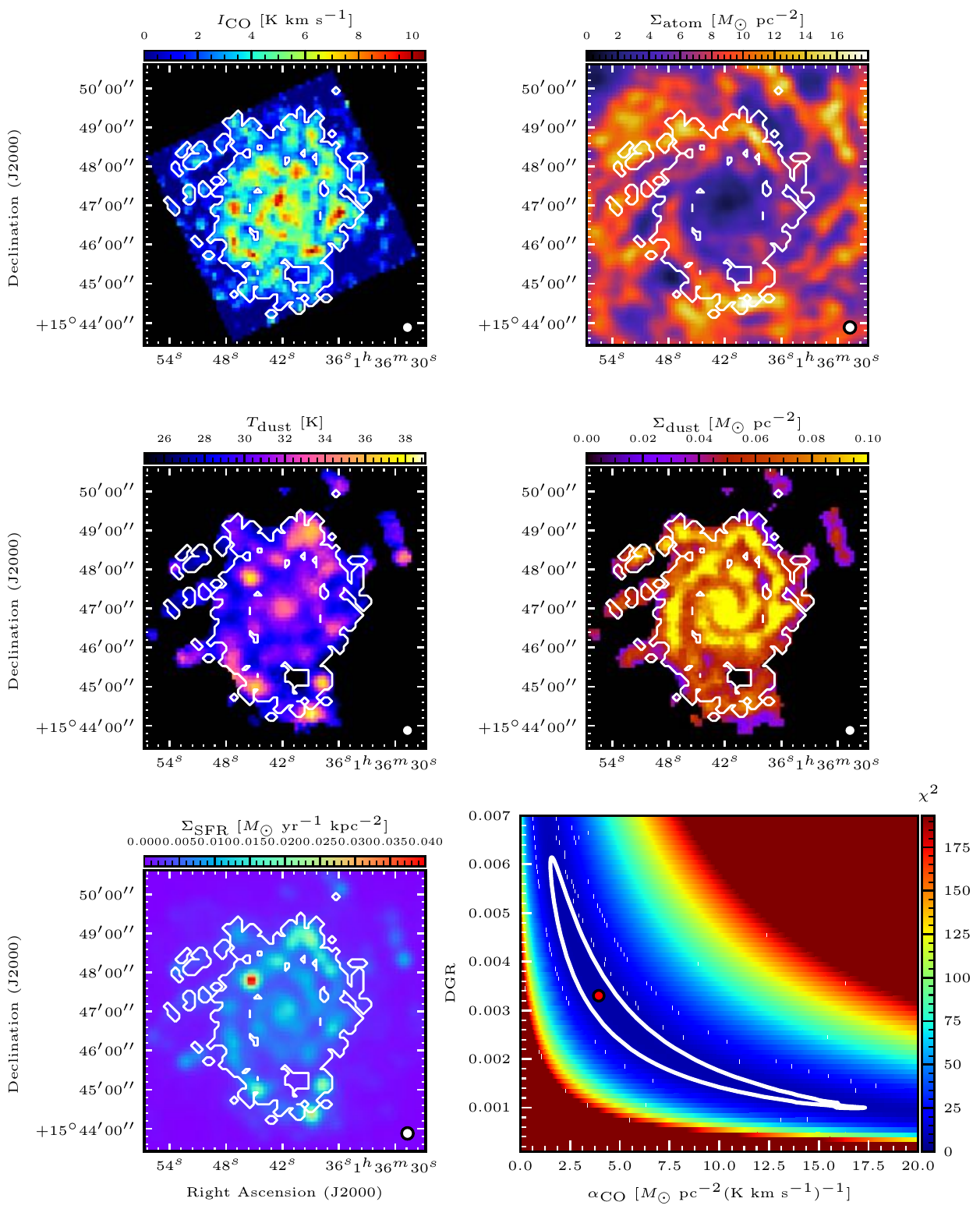}
 \end{center}
 \caption{Same as figure \ref{fig:alpha_dgr_search} including characteristic cold dust temperature ($T_{\rm{dust}}$) derived from the SED model of \citet{Casey2012} (middle left) and star formation rate surface density (\sigSFR) (bottom left) but for NGC\,628.}
\label{fig:n0628}
\end{figure*}

\begin{figure*}
 \begin{center}
  \includegraphics[width=\linewidth]{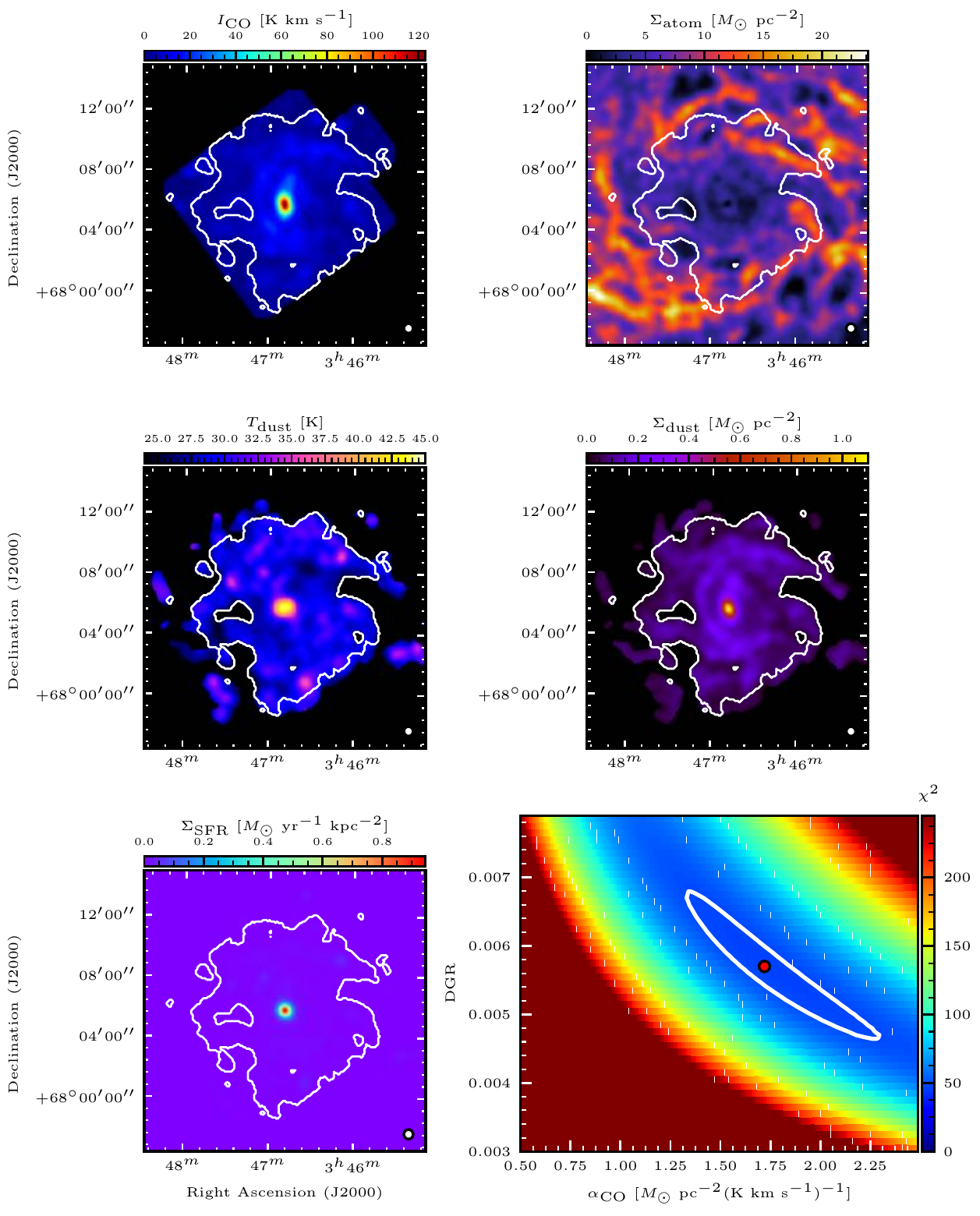}
 \end{center}
 \caption{Same as figure \ref{fig:n0628} but for IC\,342.}
\label{fig:IC342}
\end{figure*}

\clearpage
\begin{figure*}
 \begin{center}
  \includegraphics[width=\linewidth]{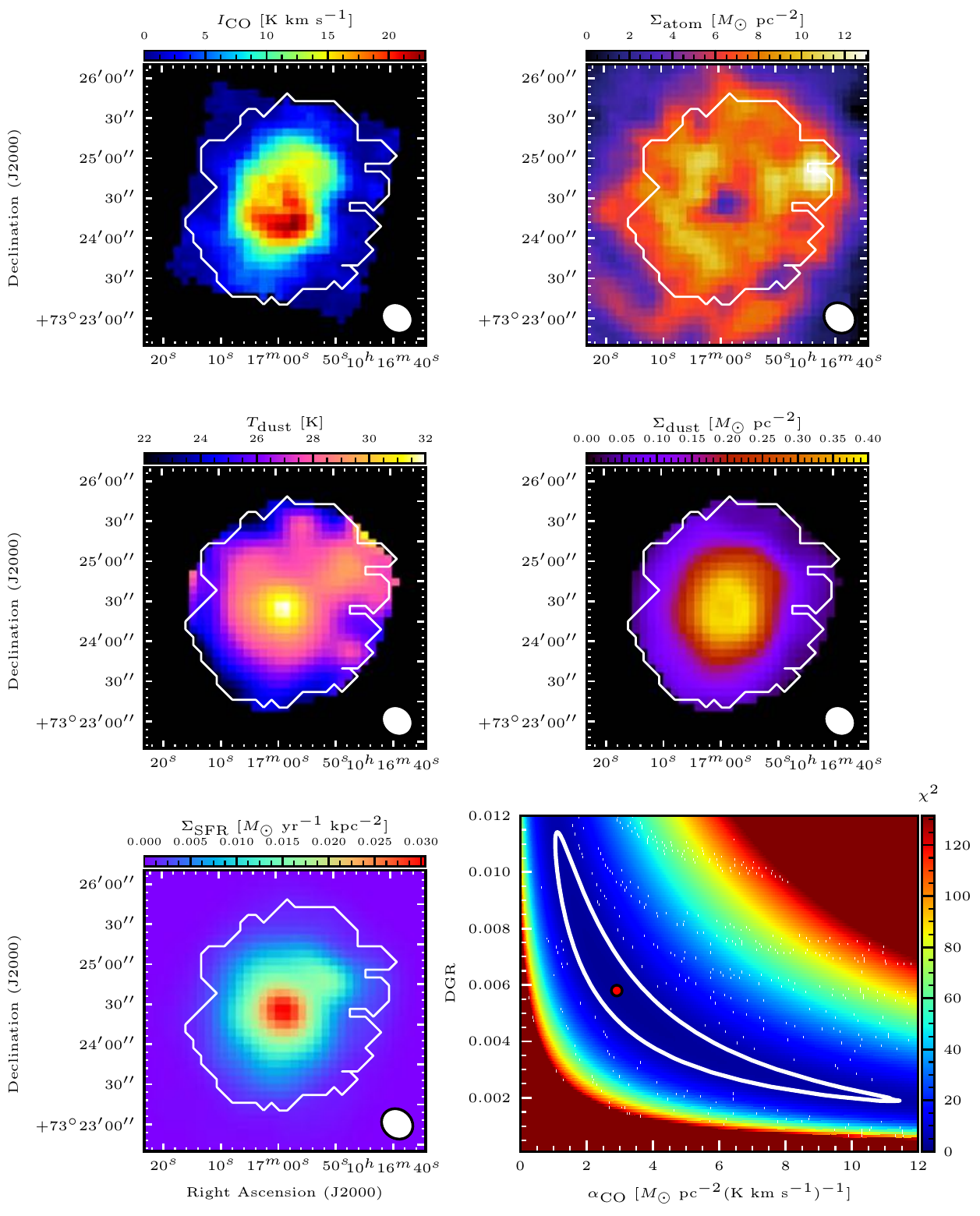}
 \end{center}
 \caption{Same as figure \ref{fig:n0628} but for NGC\,3147.}
 \label{fig:n3147}
\end{figure*}

\clearpage
\begin{figure*}
 \begin{center}
  \includegraphics[width=\linewidth]{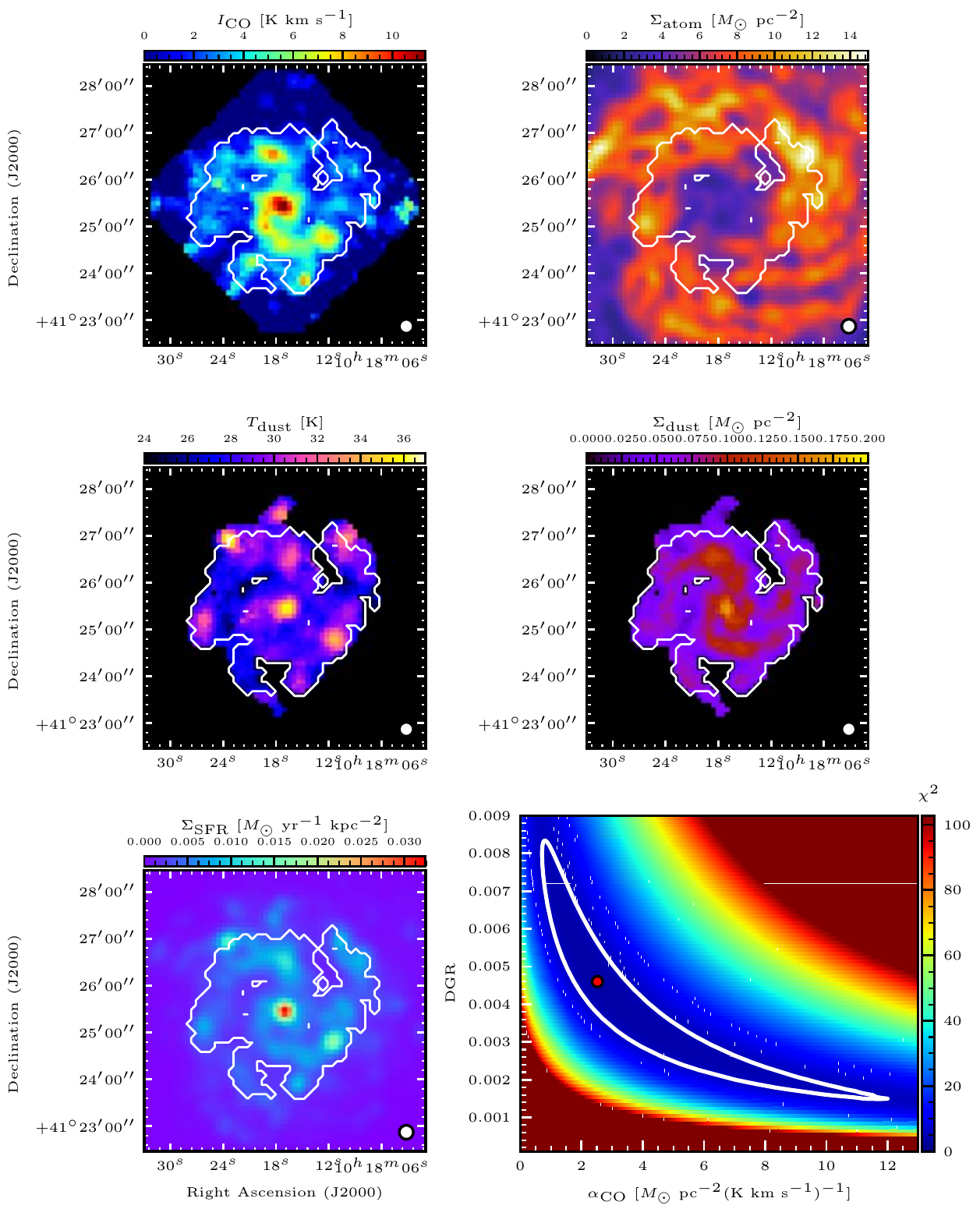}
 \end{center}
 \caption{Same as figure \ref{fig:n0628} but for NGC\,3184.}
 \label{fig:n3184}
\end{figure*}

\clearpage
\begin{figure*}
 \begin{center}
  \includegraphics[width=\linewidth]{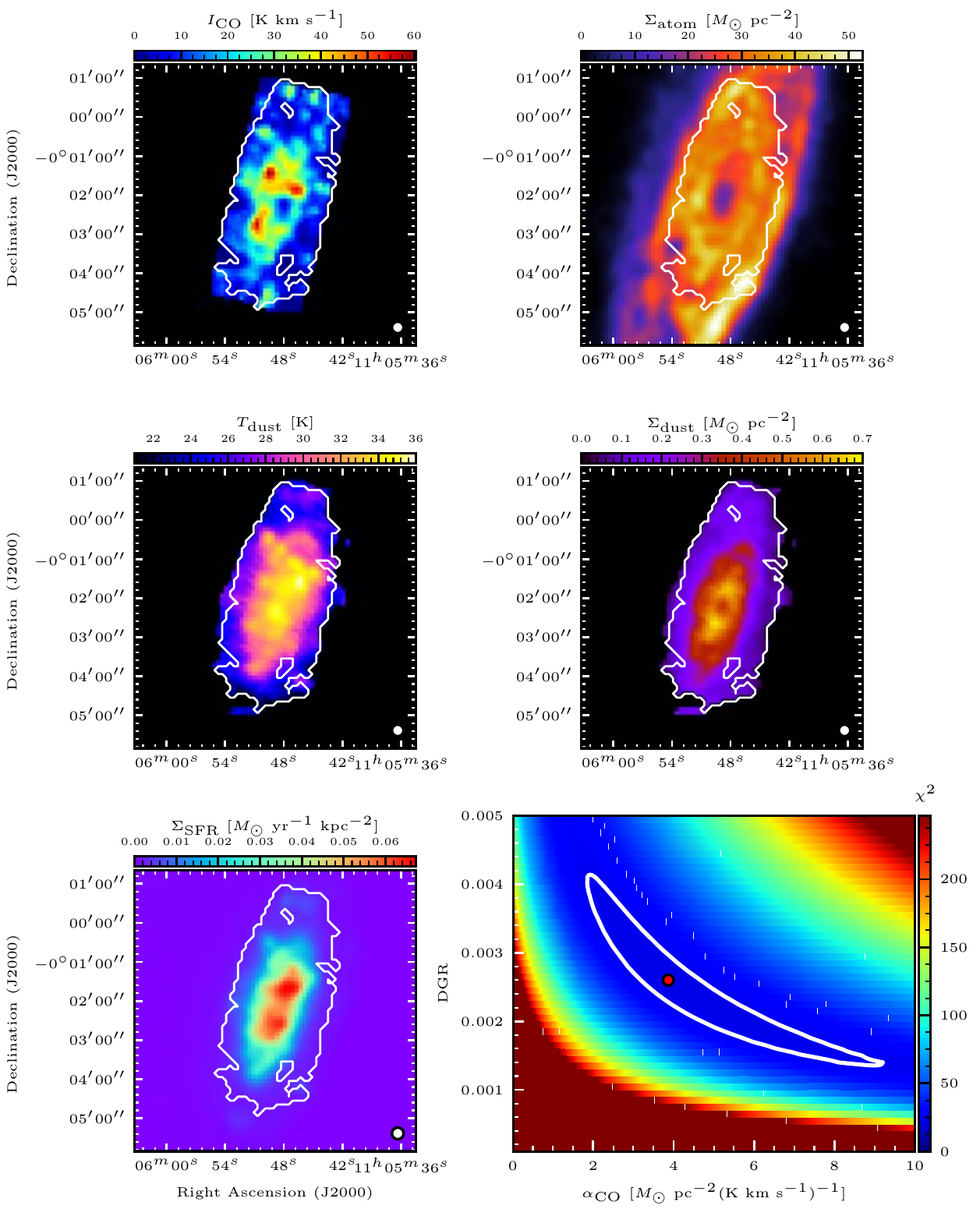}
 \end{center}
 \caption{Same as figure \ref{fig:n0628} but for NGC\,3521.}
 \label{fig:n3521}
\end{figure*}

\clearpage
\begin{figure*}
 \begin{center}
  \includegraphics[width=\linewidth]{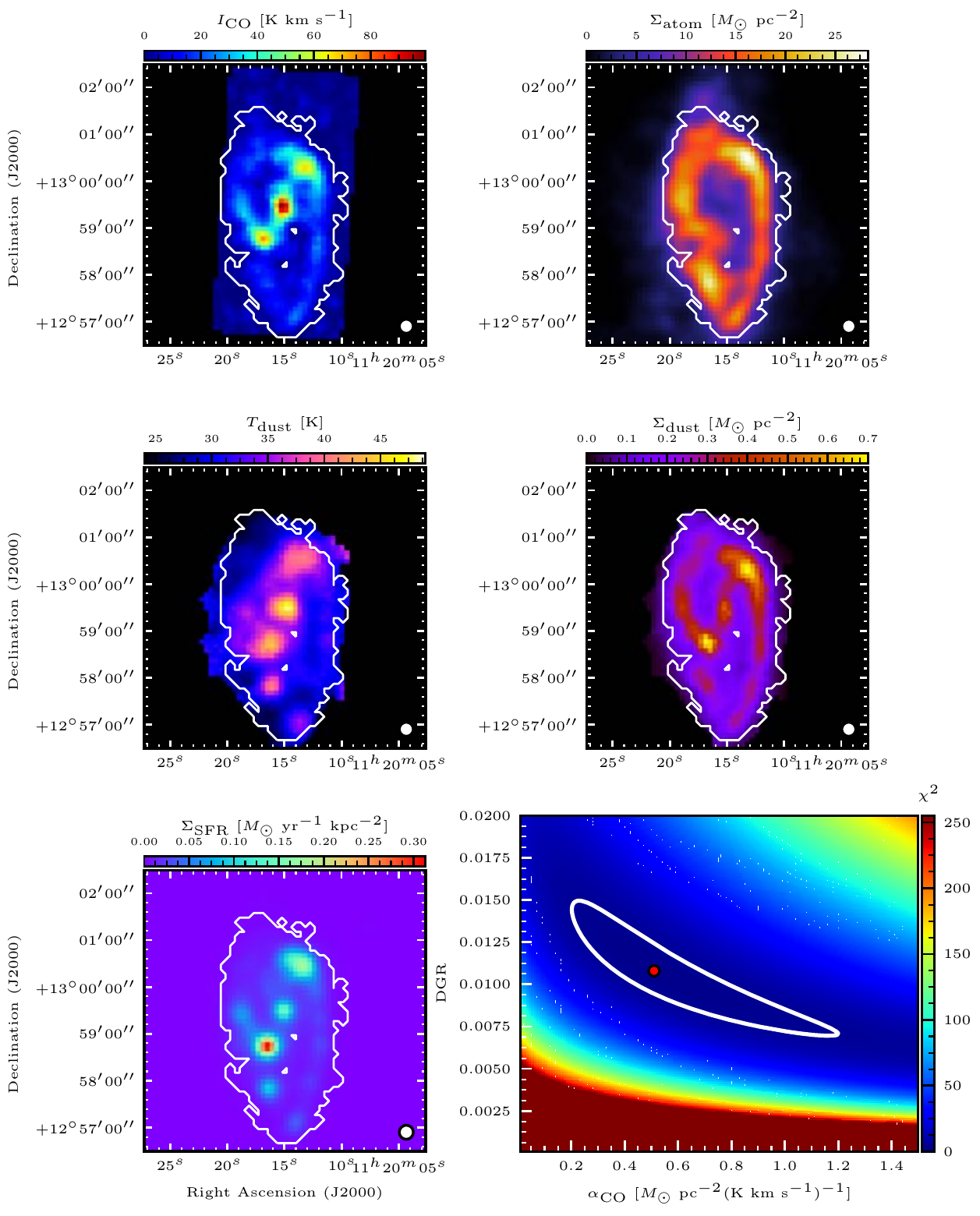}
 \end{center}
 \caption{Same as figure \ref{fig:n0628} but for NGC\,3627.}
 \label{fig:n3627}
\end{figure*}

\clearpage
\begin{figure}
 \begin{center}
  \includegraphics[width=\linewidth]{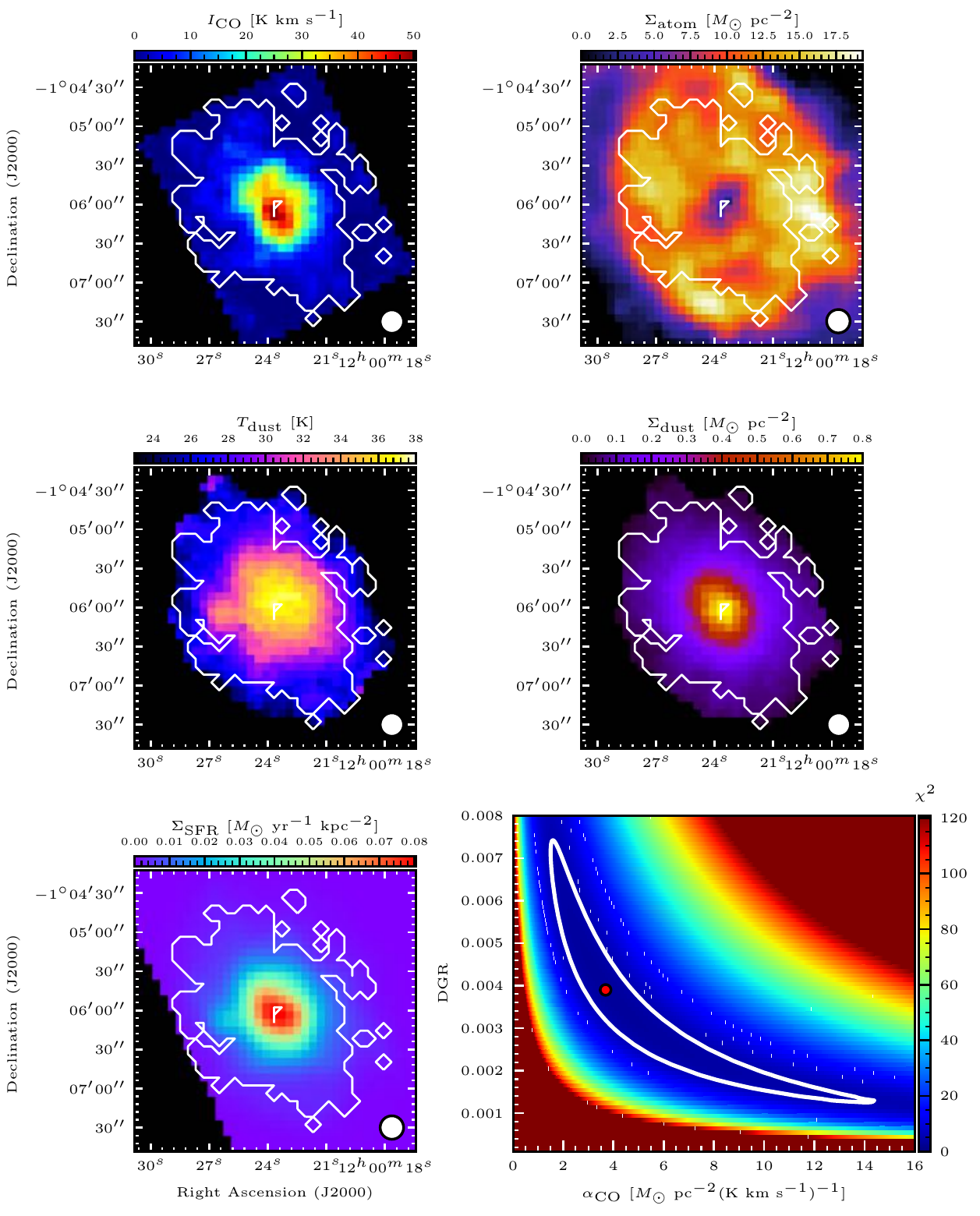}
 \end{center}
 \caption{Same as figure \ref{fig:n0628} but for NGC\,4030.}
 \label{fig:n4030}
\end{figure}

\clearpage
\begin{figure}
 \begin{center}
  \includegraphics[width=\linewidth]{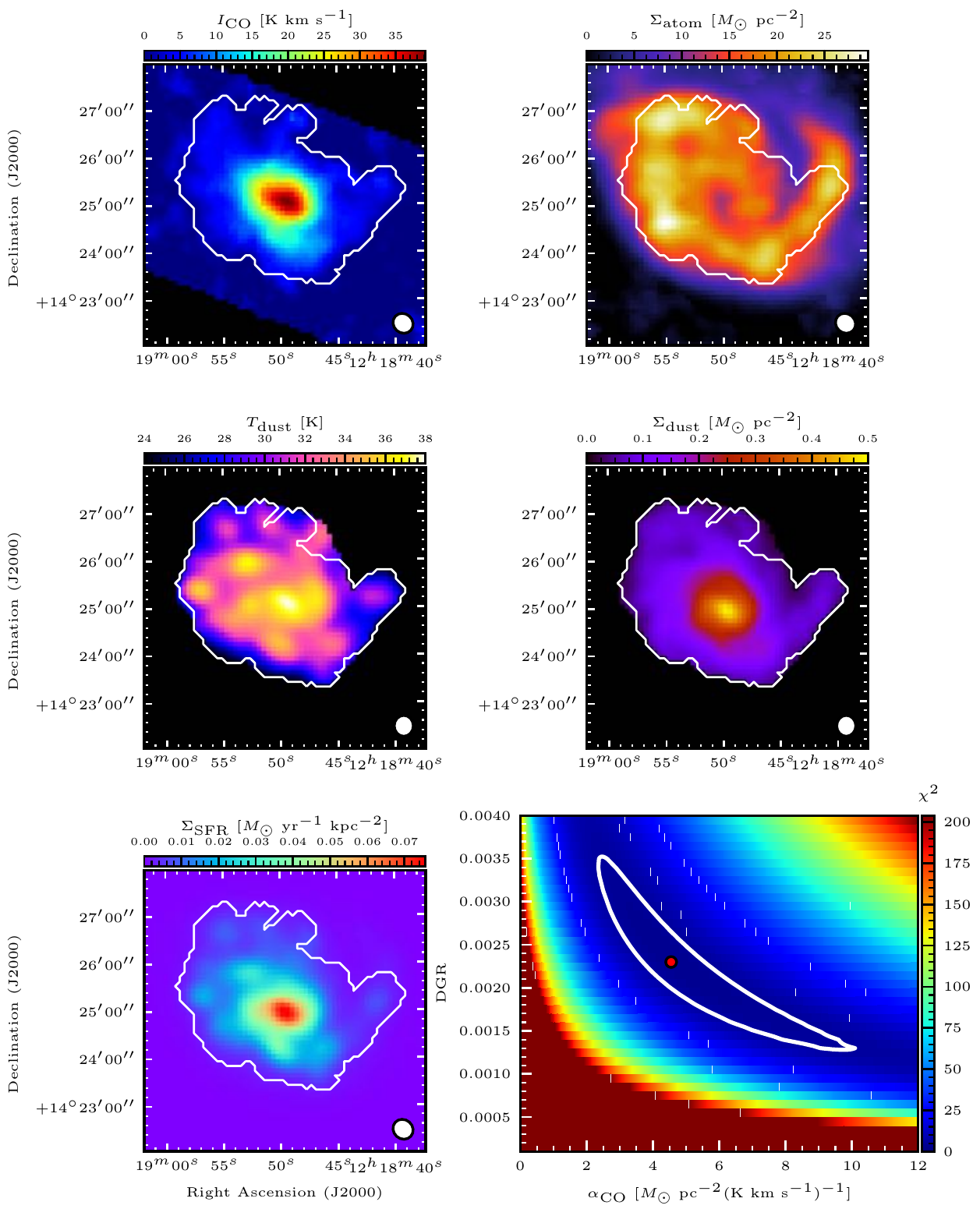}
 \end{center}
 \caption{Same as figure \ref{fig:n0628} but for NGC\,4254.}
 \label{fig:n4254}
\end{figure}

\clearpage
\begin{figure*}
 \begin{center}
  \includegraphics[width=\linewidth]{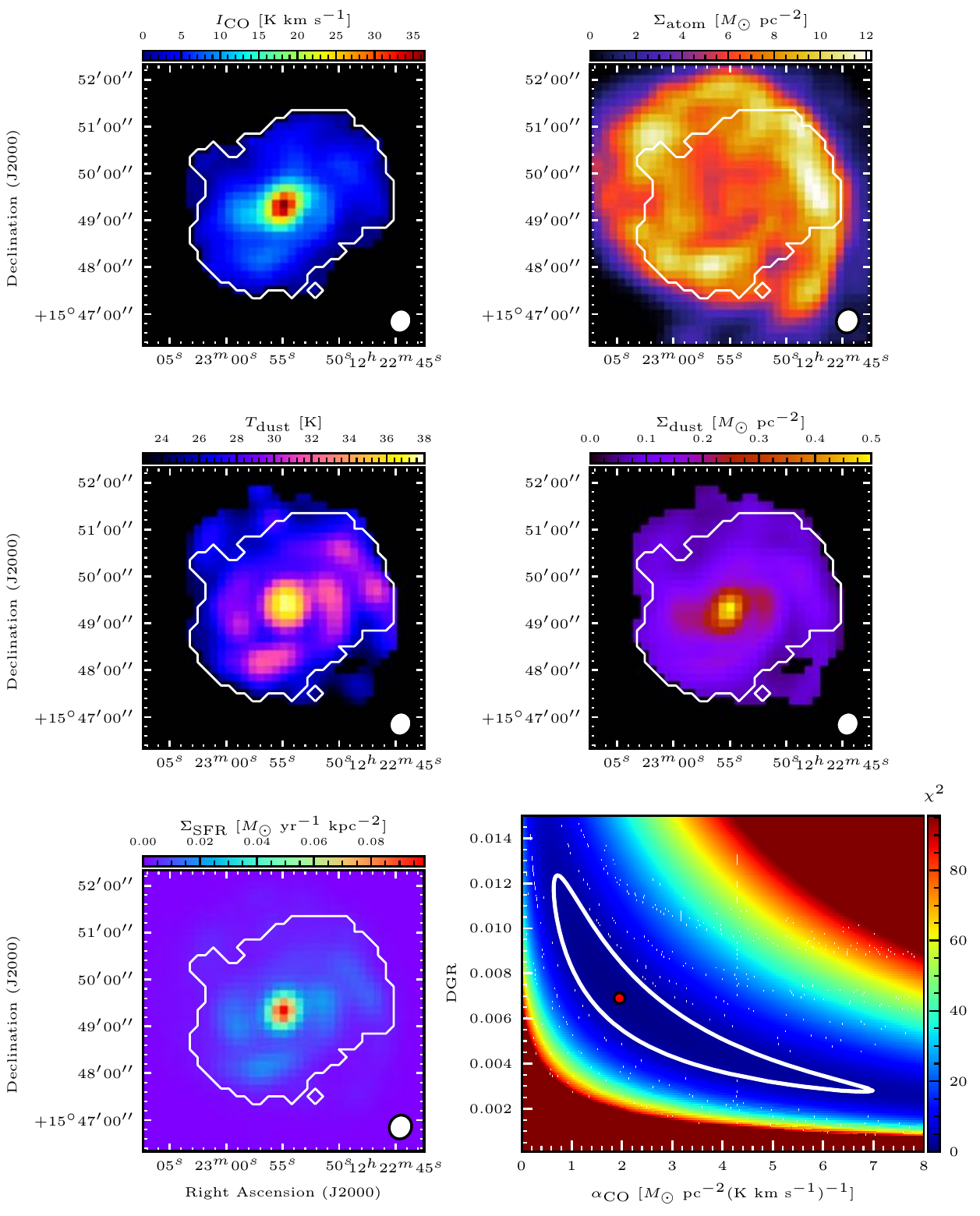}
 \end{center}
 \caption{Same as figure \ref{fig:n0628} but for NGC\,4321.}
 \label{fig:n4321}
\end{figure*}

\clearpage
\begin{figure*}
 \begin{center}
  \includegraphics[width=\linewidth]{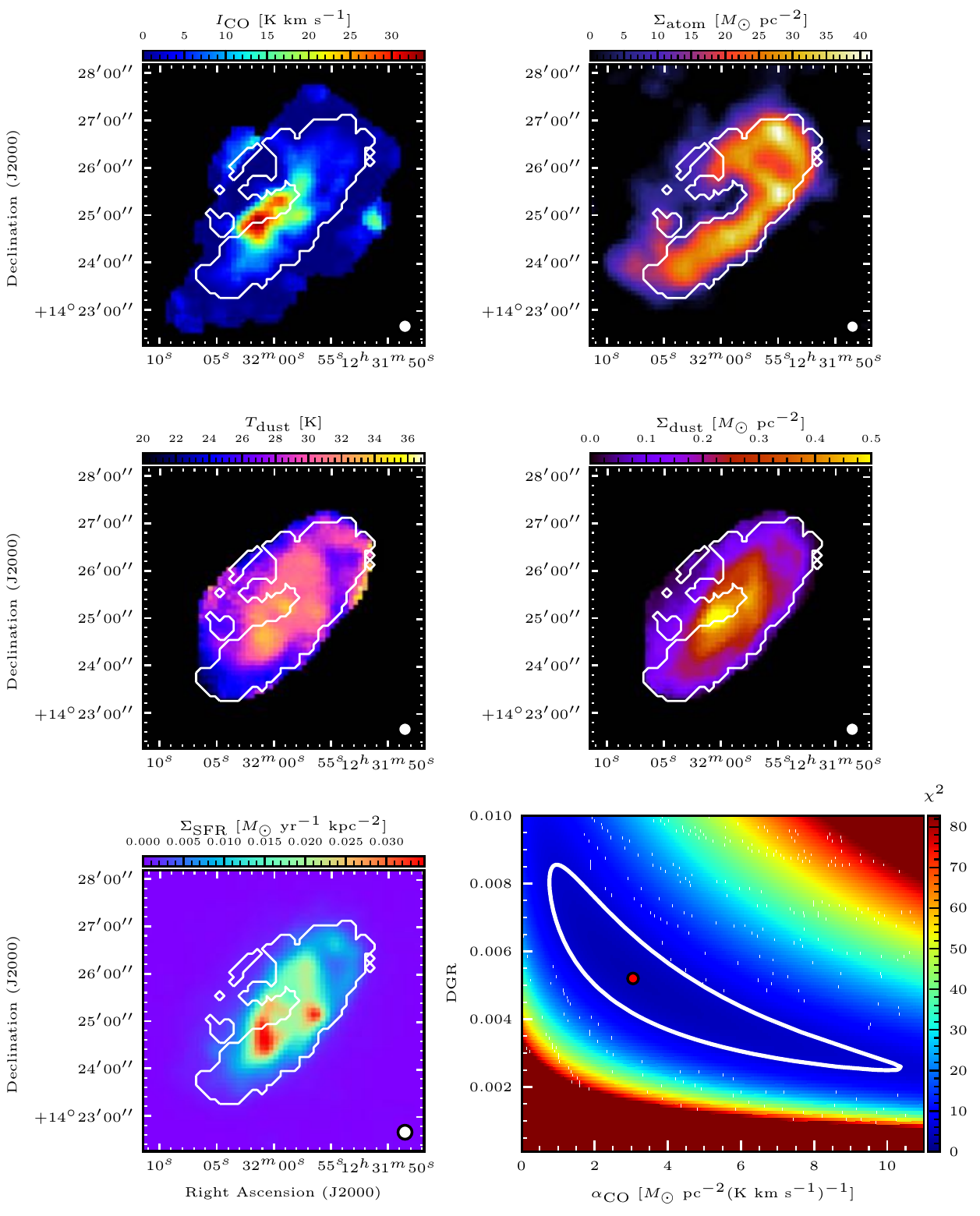}
 \end{center}
 \caption{Same as figure \ref{fig:n0628} but for NGC\,4501.}
 \label{fig:n4501}
\end{figure*}

\clearpage
\begin{figure*}
 \begin{center}
  \includegraphics[width=\linewidth]{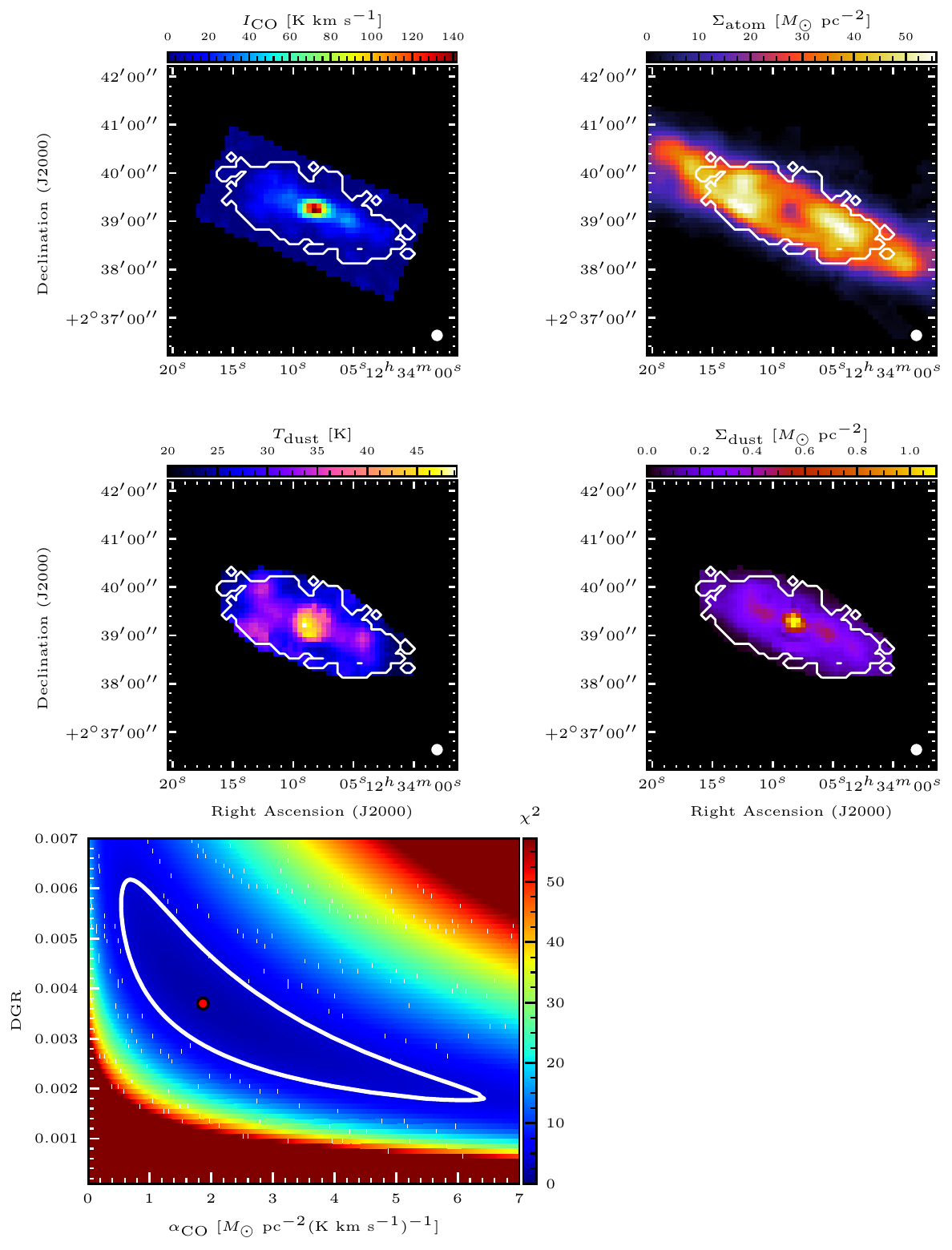}
 \end{center}
 \caption{Same as figure \ref{fig:n0628} but for NGC\,4527.}
 \label{fig:n4527}
\end{figure*}

\clearpage
\begin{figure*}
 \begin{center}
  \includegraphics[width=\linewidth]{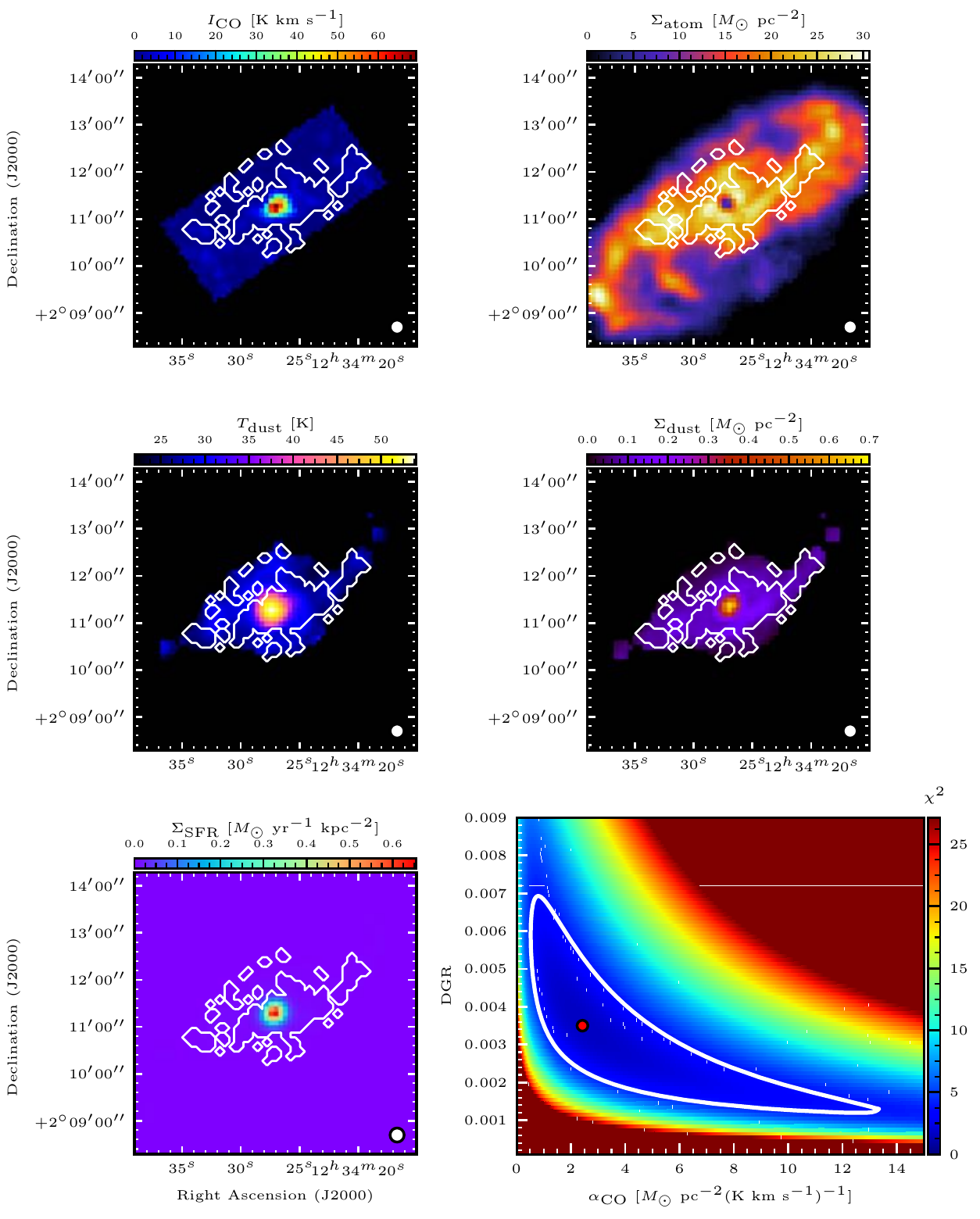}
 \end{center}
 \caption{Same as figure \ref{fig:n0628} but for NGC\,4536.}
 \label{fig:n4536}
\end{figure*}

\clearpage
\begin{figure*}
 \begin{center}
  \includegraphics[width=\linewidth]{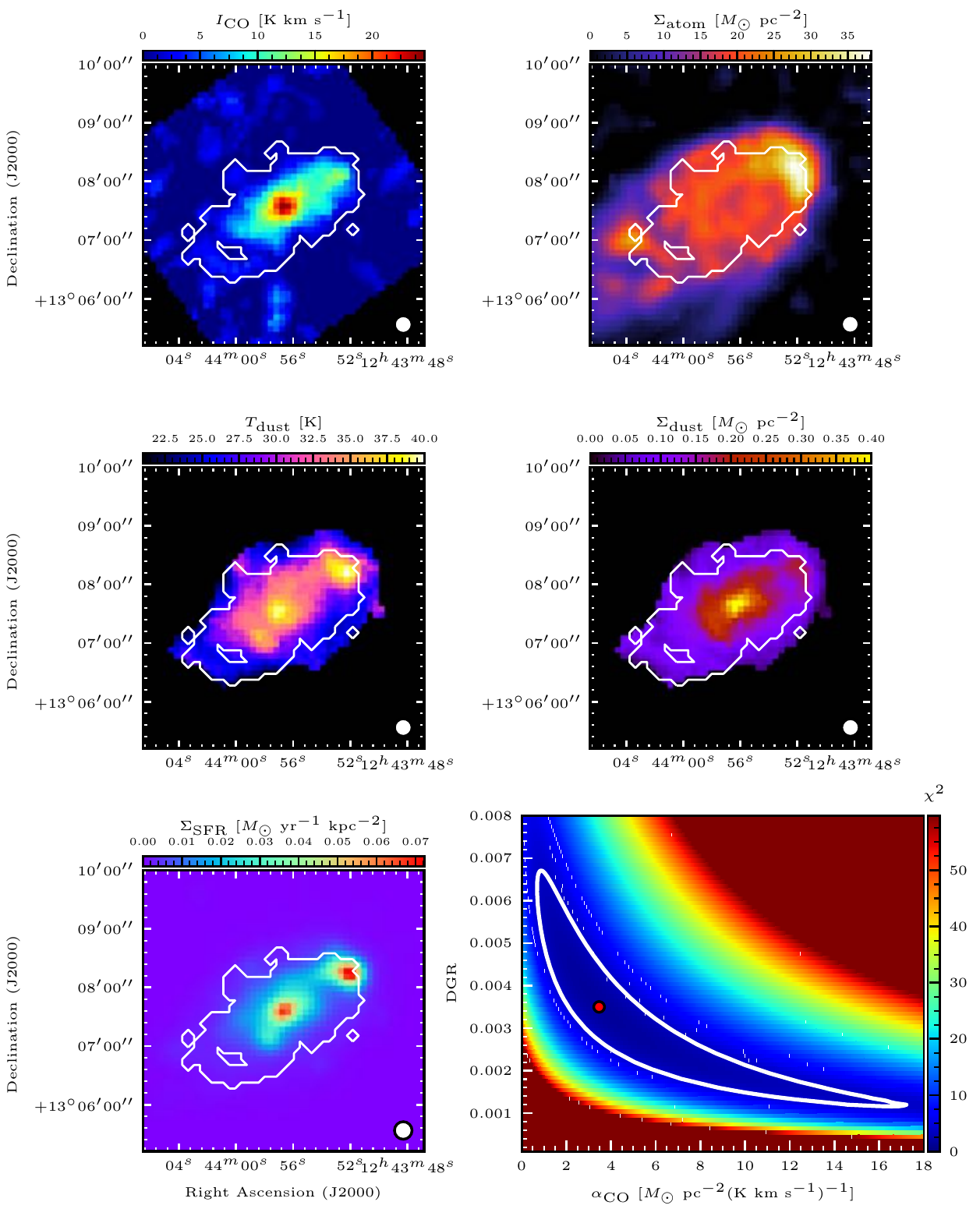}
 \end{center}
 \caption{Same as figure \ref{fig:n0628} but for NGC\,4654.}
 \label{fig:n4654}
\end{figure*}

\clearpage
\begin{figure*}
 \begin{center}
  \includegraphics[width=\linewidth]{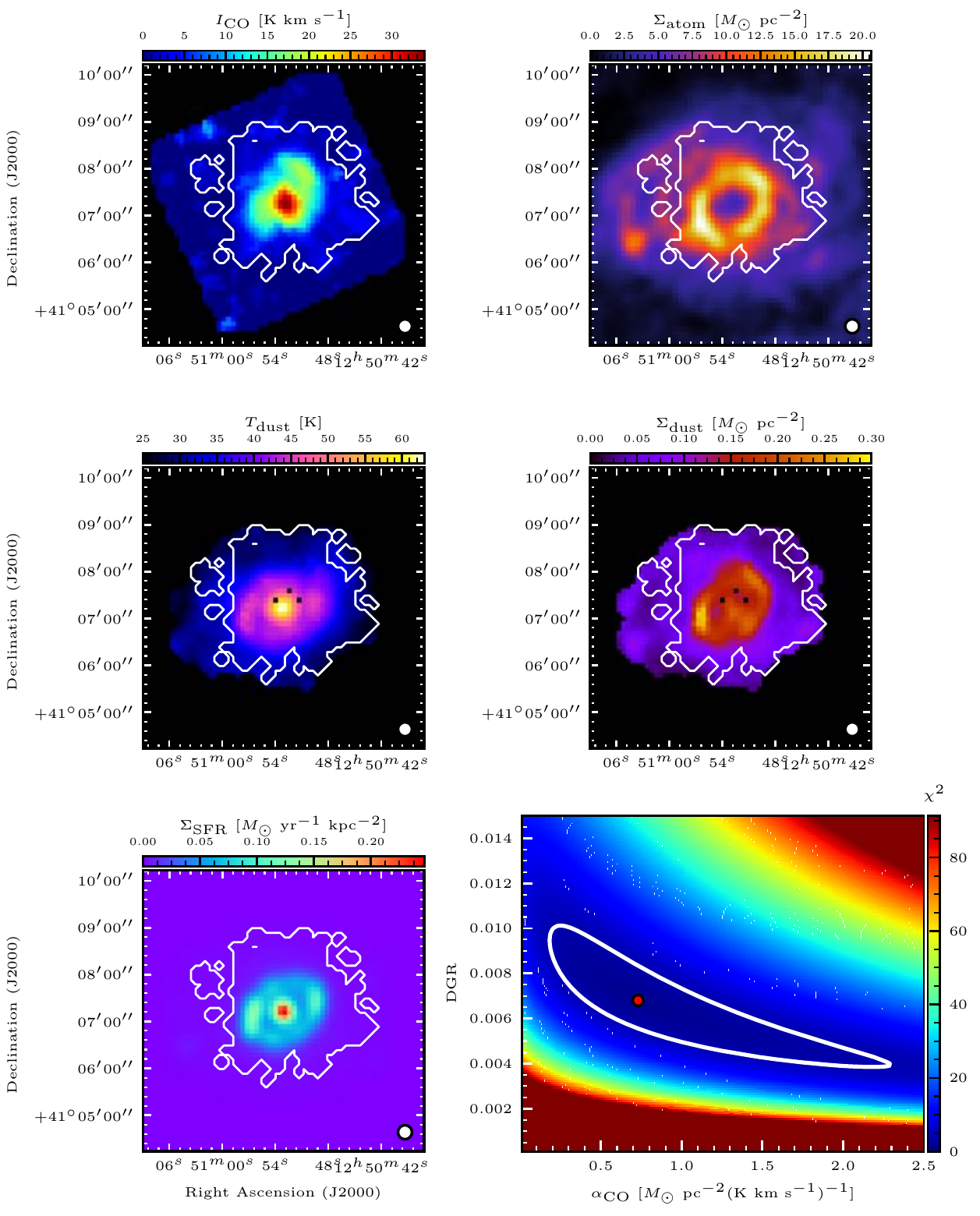}
 \end{center}
 \caption{Same as figure \ref{fig:n0628} but for NGC\,4736.}
 \label{fig:n4736}
\end{figure*}

\clearpage
\begin{figure*}
 \begin{center}
  \includegraphics[width=\linewidth]{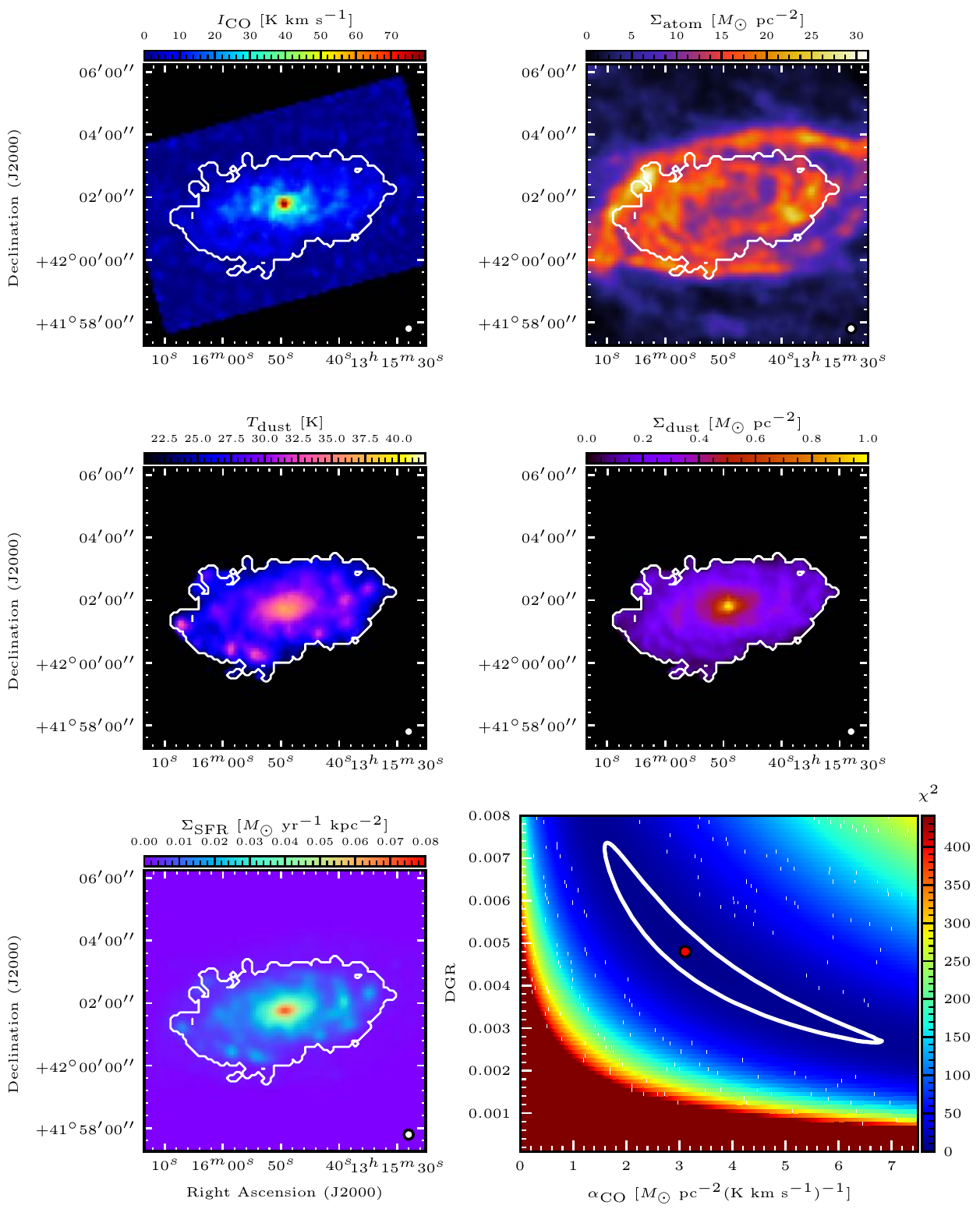}
 \end{center}
 \caption{Same as figure \ref{fig:n0628} but for NGC\,5055.}
 \label{fig:n5055}
\end{figure*}

\clearpage
\begin{figure*}
 \begin{center}
  \includegraphics[width=\linewidth]{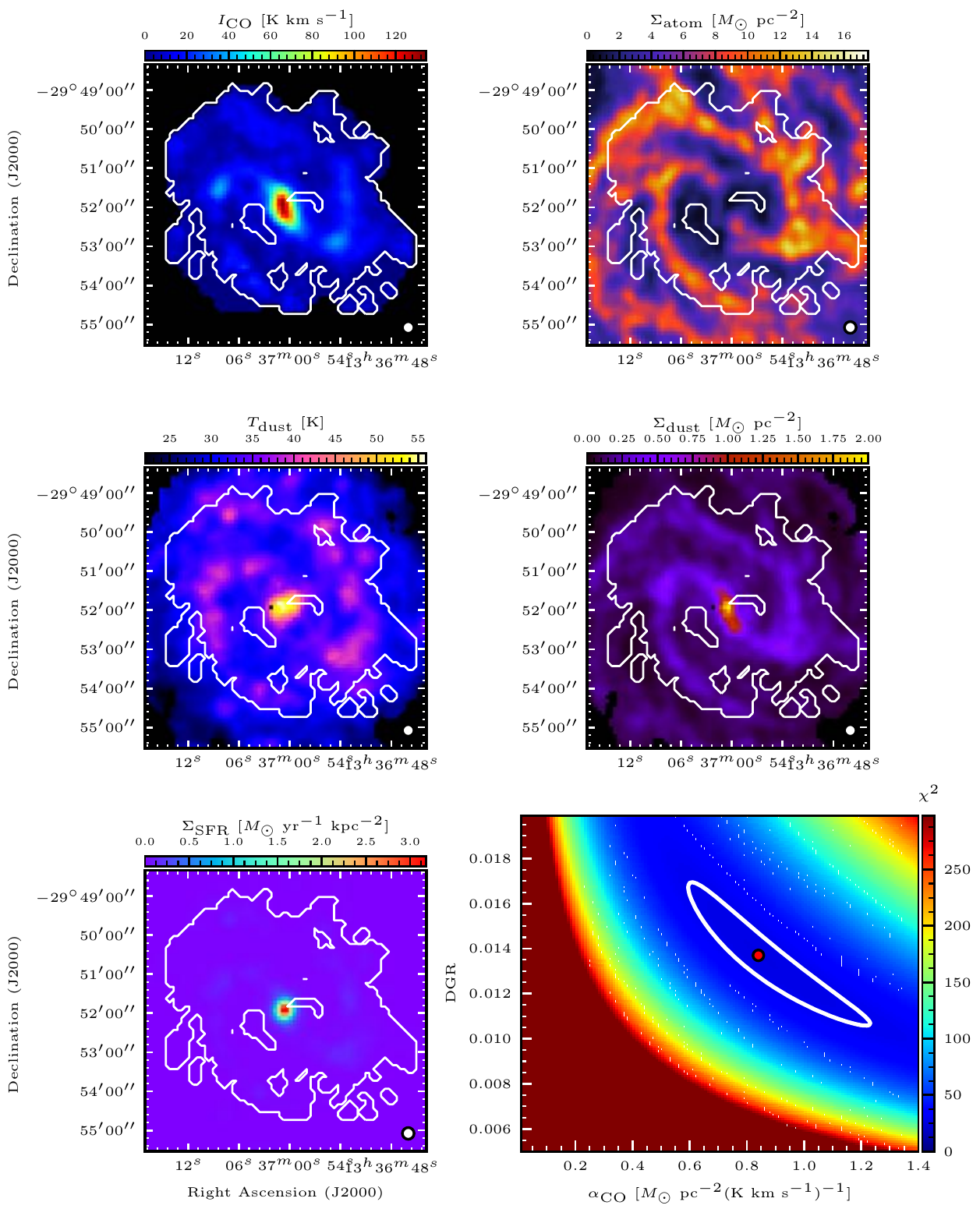}
 \end{center}
 \caption{Same as figure \ref{fig:n0628} but for NGC\,5236.}
 \label{fig:n5236}
\end{figure*}

\clearpage
\begin{figure*}
 \begin{center}
  \includegraphics[width=\linewidth]{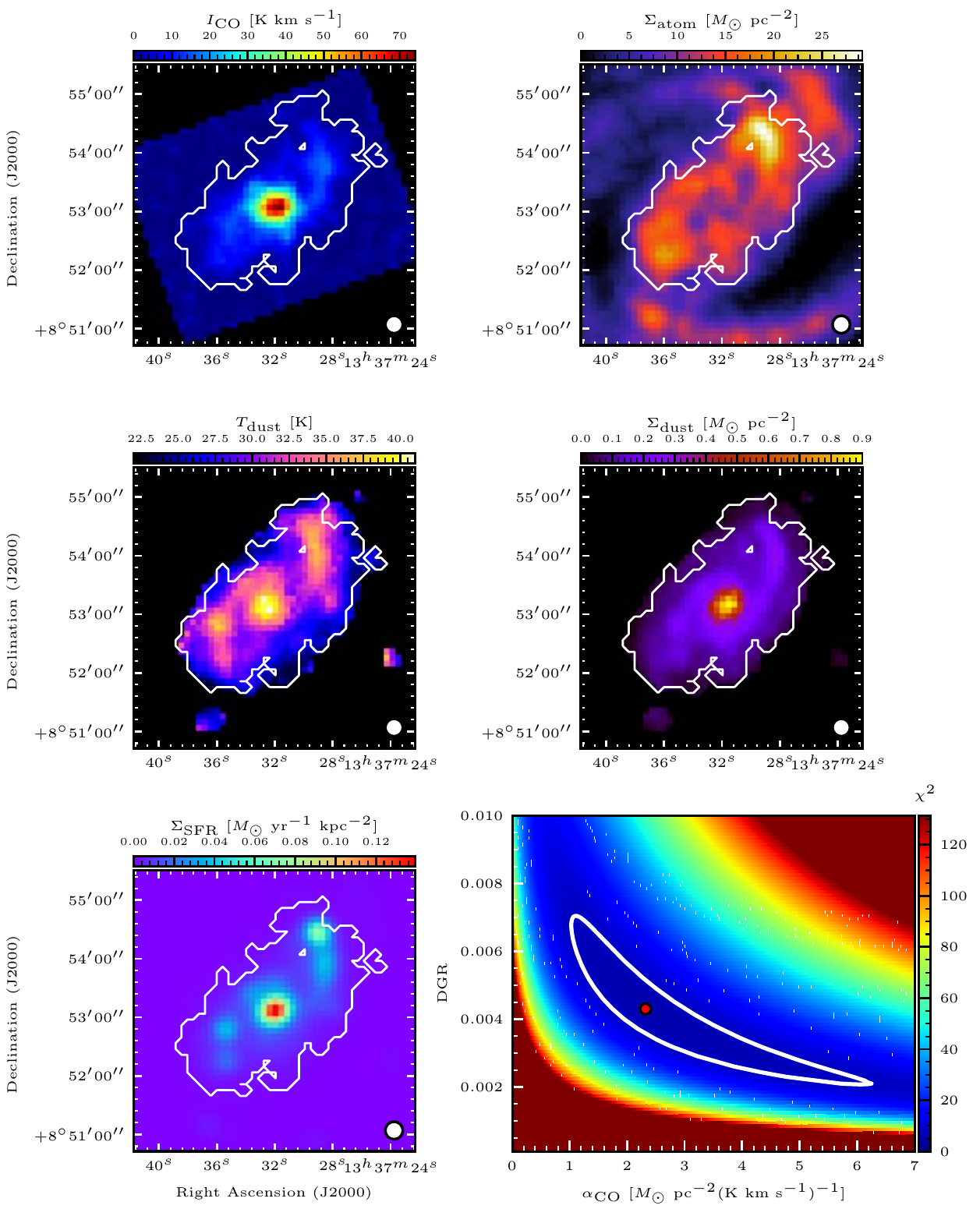}
 \end{center}
 \caption{Same as figure \ref{fig:n0628} but for NGC\,5248.}
 \label{fig:n5248}
\end{figure*}

\clearpage
\begin{figure*}
 \begin{center}
  \includegraphics[width=\linewidth]{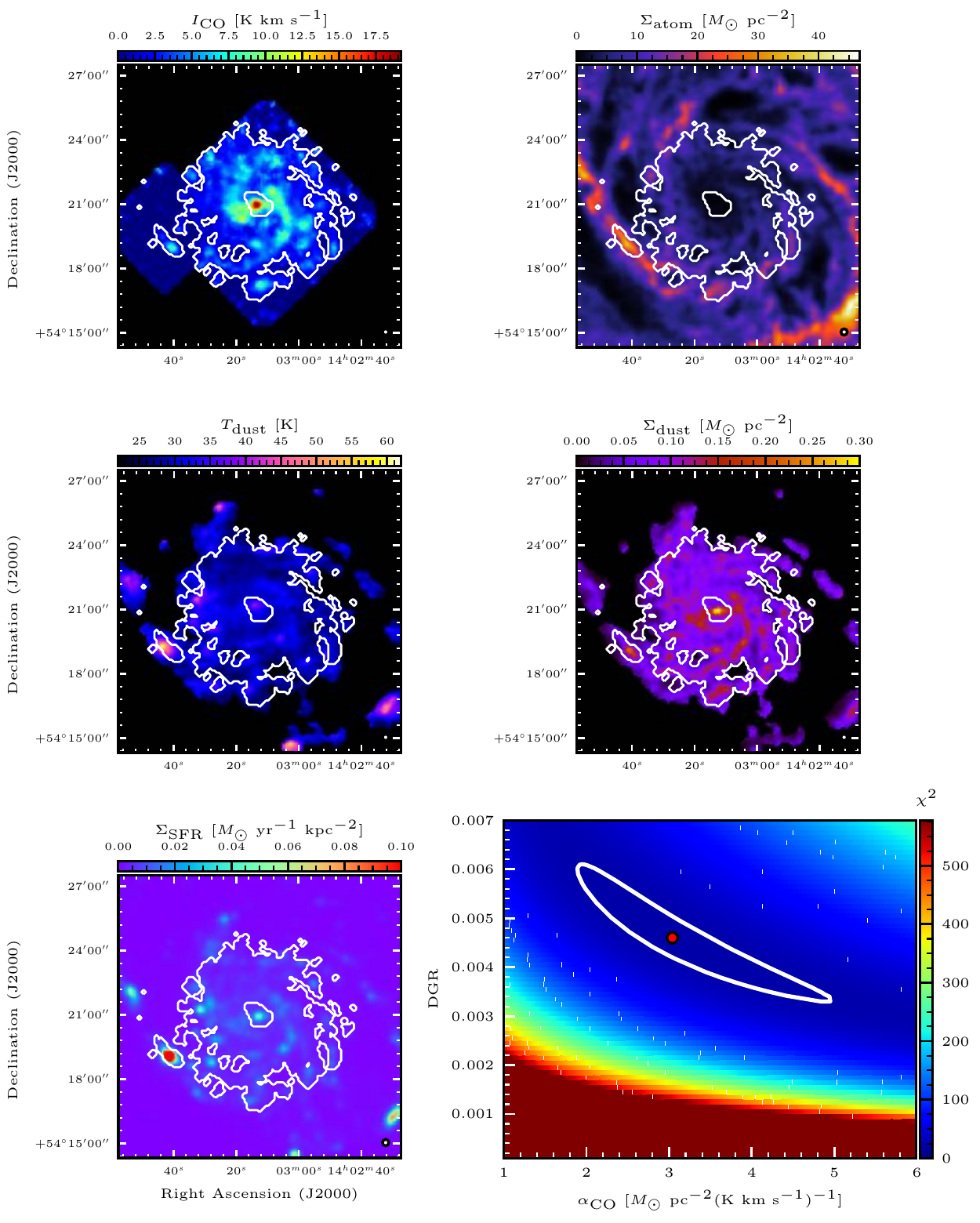}
 \end{center}
 \caption{Same as figure \ref{fig:n0628} but for NGC\,5457.}
 \label{fig:n5457}
\end{figure*}

\clearpage
\begin{figure*}
 \begin{center}
  \includegraphics[width=\linewidth]{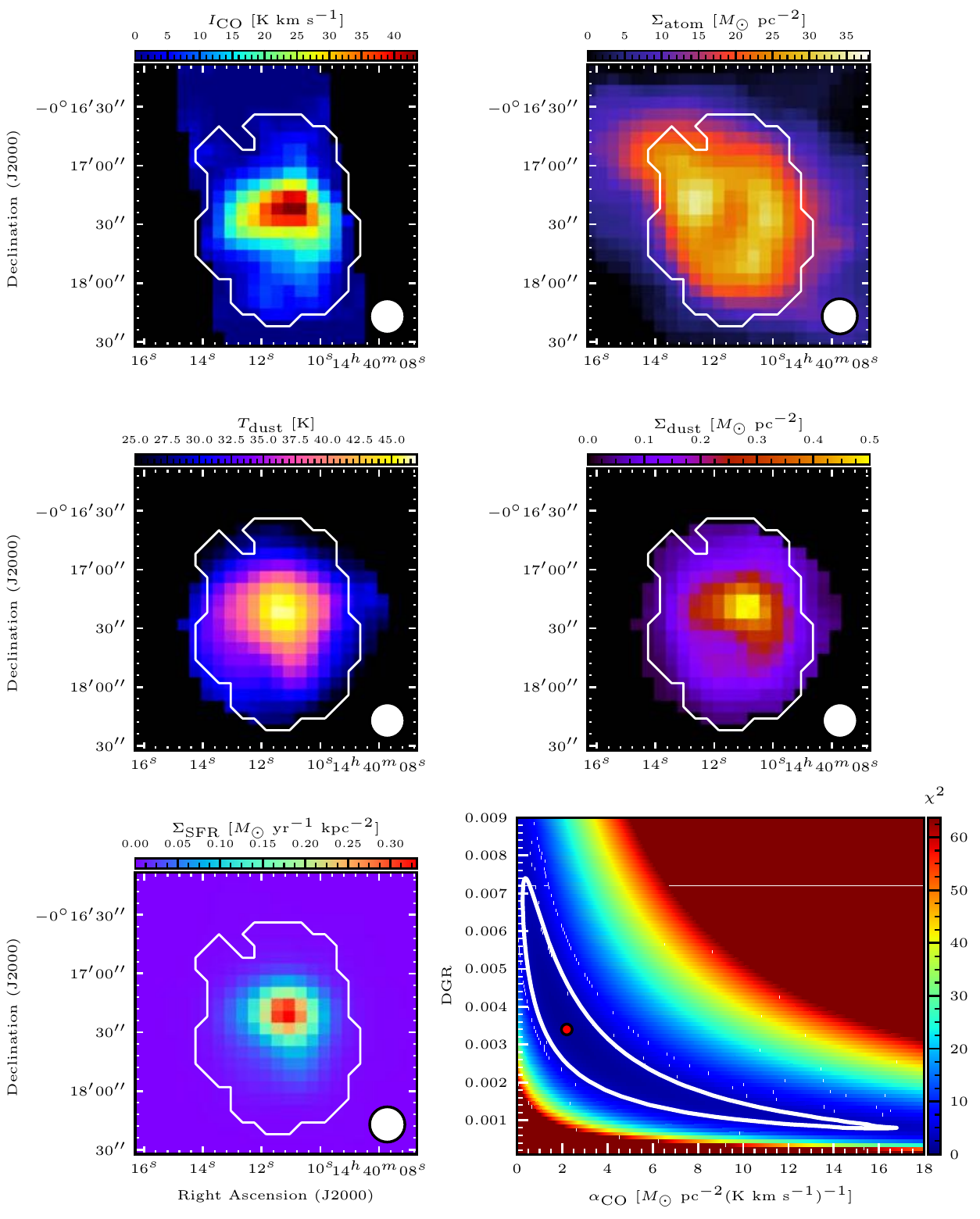}
 \end{center}
 \caption{Same as figure \ref{fig:n0628} but for NGC\,5713.}
 \label{fig:n5713}
\end{figure*}

\clearpage
\begin{figure*}
 \begin{center}
  \includegraphics[width=\linewidth]{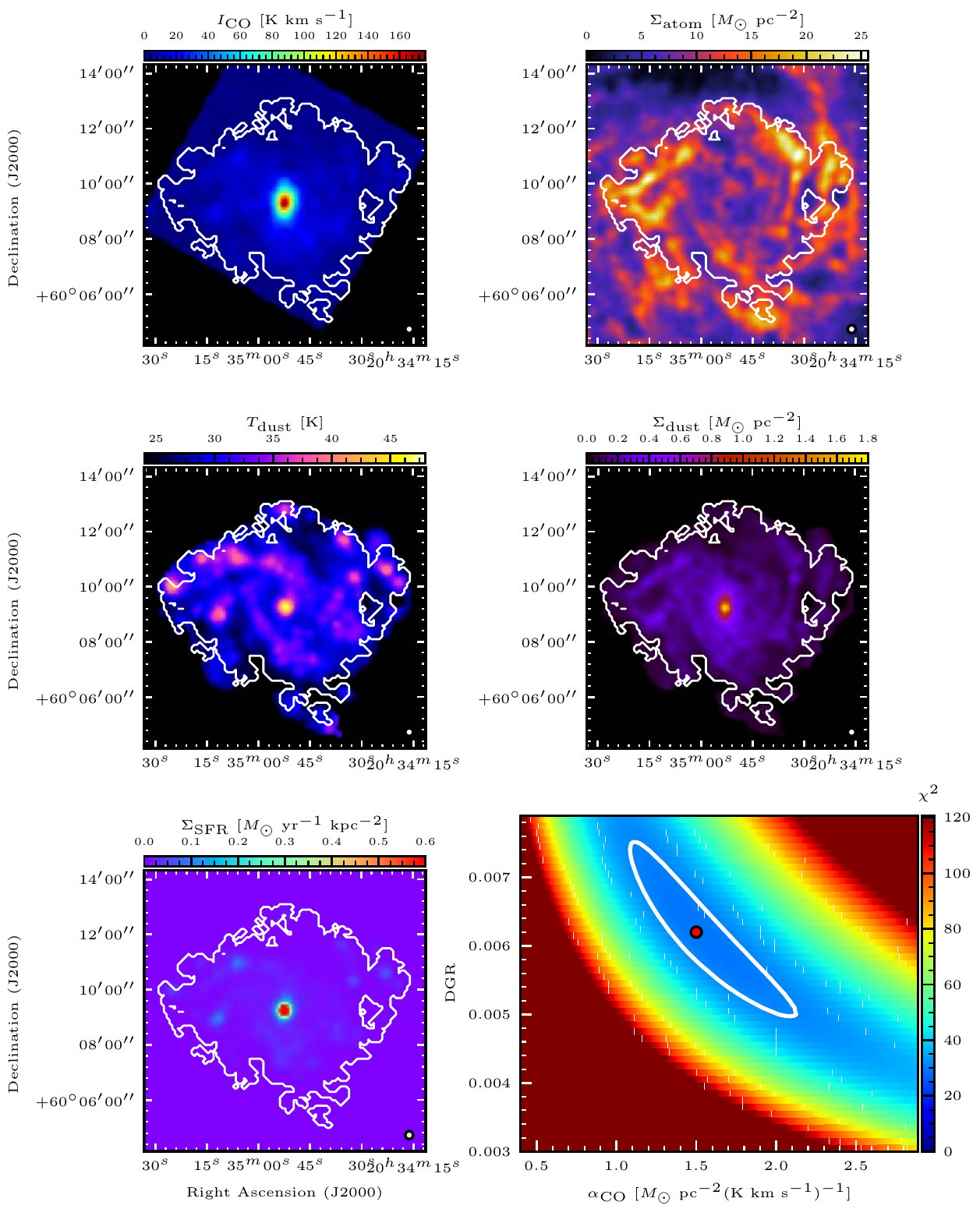}
 \end{center}
 \caption{Same as figure \ref{fig:n0628} but for NGC\,6946.}
 \label{fig:n6946}
\end{figure*}

\clearpage
\begin{figure*}
 \begin{center}
  \includegraphics[width=\linewidth]{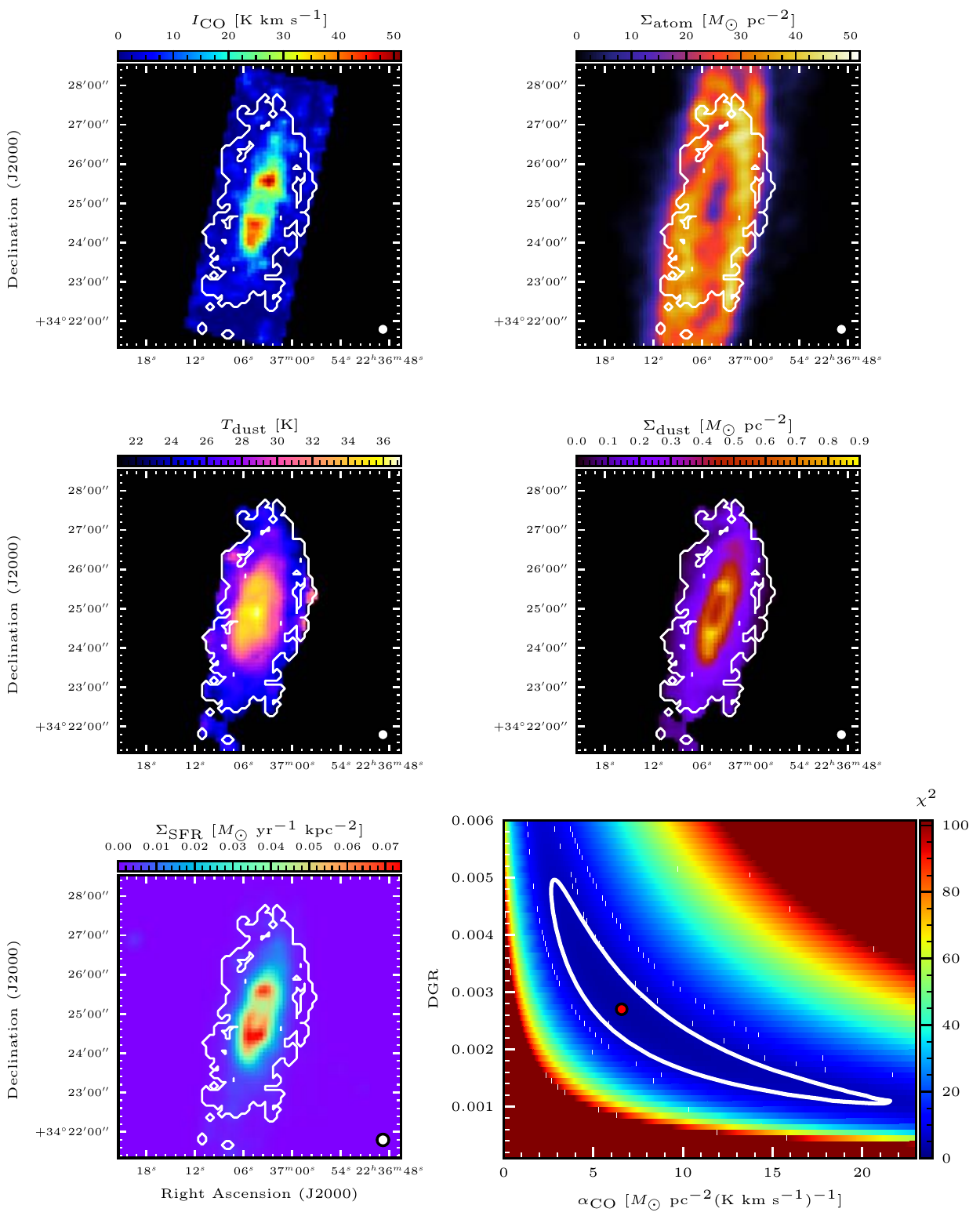}
 \end{center}
 \caption{Same as figure \ref{fig:n0628} but for NGC\,7331.}
 \label{fig:n7331}
\end{figure*}

\clearpage
\begin{figure*}
 \begin{center}
  \includegraphics[width=\linewidth]{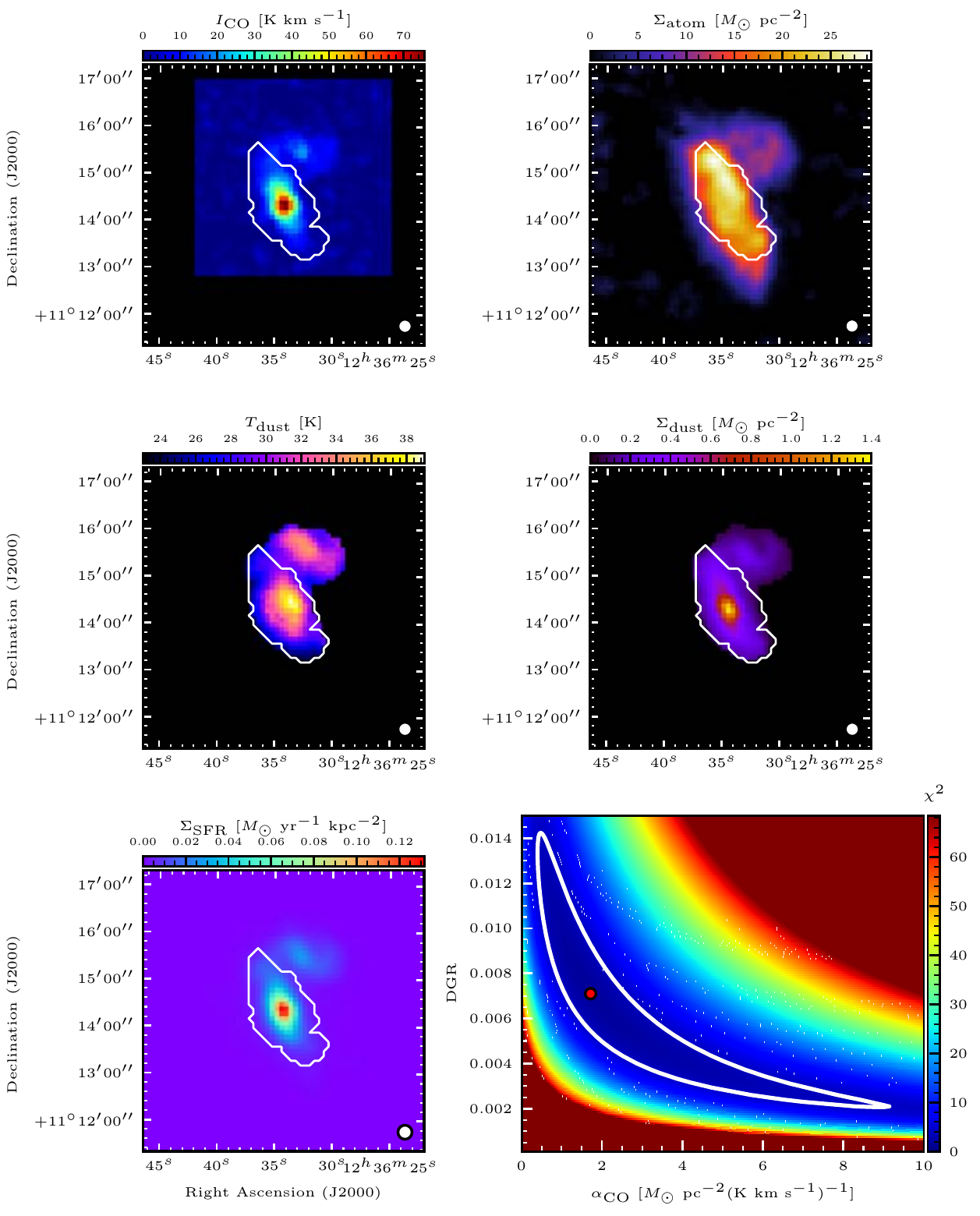}
 \end{center}
 \caption{Same as figure \ref{fig:n0628} but for NGC\,4568}
 \label{fig:n4568}
\end{figure*}

%%%%%%%%%%%%%%%%%%%%
%% Reference
%%%%%%%%%%%%%%%%%%%%

\end{document}